\documentclass[12pt]{iopart}

\usepackage{iopams}  
\usepackage{hyperref,todonotes}
\usepackage{subcaption}
\usepackage{overpic}

\usepackage{hyperref}
\usepackage{cite}
\usepackage{cleveref}

\usepackage{tikz,environ}
\usetikzlibrary{snakes}
\usetikzlibrary{decorations}
\usetikzlibrary{trees}
\usetikzlibrary{decorations.pathmorphing}
\usetikzlibrary{decorations.markings}
\usetikzlibrary{external}
\usetikzlibrary{intersections}
\usetikzlibrary{shapes,arrows}
\usetikzlibrary{arrows.meta}
\usetikzlibrary{calc}
\usetikzlibrary{shapes.misc}
\usetikzlibrary{decorations.text}
\usetikzlibrary{backgrounds}
\usetikzlibrary{fadings}
\usetikzlibrary{tikzmark,calc,arrows,shapes,decorations.pathreplacing}

\tikzset{
	graviton/.style={decorate,line width=0.1mm, decoration={snake,amplitude=.3mm, segment length=0.8mm}},
	photon/.style={decorate, decoration={snake}, draw=red},
	scalar/.style={postaction={decorate},
	},
	massive/.style={postaction={decorate},
		line width=0.75mm,
	},
	massless/.style={postaction={decorate},
	},
	masslessWithDot/.style={postaction={decorate},
		decoration={
			markings,
			mark=at position 0.5 with {\fill circle (2pt);}}
	},
	massiveWithDot/.style={postaction={decorate},
		line width=0.5mm,
		decoration={
			markings,
			mark=at position 0.5 with {\fill circle (2pt);}}
	},
	massiveWithArrow/.style={postaction={decorate},
		line width=0.75mm,
		decoration={
			markings,
			mark=at position 0.5 with {\arrow{latex}}}
	},
	massiveLin/.style={postaction={decorate},
		double,
		thick,
		fill=white
	},
	massivePhi/.style={postaction={decorate},
		line width=0.75mm,
		dashed
	},
	masslessPhi/.style={postaction={decorate},
		dashed
	},
	unitaryCut/.style={postaction={draw,densely dashed,blue,thin},
		line width = 0.2cm,white
	},
	gluon/.style={decorate, draw=magenta,
		decoration={coil,amplitude=4pt, segment length=5pt}},
	partial ellipse/.style args={#1:#2:#3}{
		insert path={+ (#1:#3) arc (#1:#2:#3)}
	},
	cross/.style={cross out, draw=black, minimum size=2*(#1-\pgflinewidth), inner sep=0pt, outer sep=0pt},
	branchCut/.style={postaction={decorate},
		snake=zigzag,
		decoration = {snake=zigzag,segment length = 2mm, amplitude = 2mm}	
	}
	cross/.default={1pt}
}

\colorlet{mred}{black!30!red}
\colorlet{mgreen}{black!30!green}
\colorlet{mblue}{black!30!blue}
\colorlet{morange}{blue!70!red}

\newcommand{\trees}{
\begin{tikzpicture}[scale=1]
	    \coordinate (e1) at (-0.5,-1.);	
		\coordinate (e2) at (-0.5,0);
		\coordinate (e3) at (.5,0);
		\coordinate (e4) at (.5,-1.);
		
		\coordinate (v1) at (-.25,-.5);
		\coordinate (v2) at (.25,-.5);
		
		\draw[massive] (e1) -- (v1) -- (e2) ;
		\draw[massive] (e3) -- (v2) -- (e4) ;
		\draw[massive] (v1) -- (v2) ;

        \node[below left]  at (e1) {$p_1$};
        \node[above left]  at (e2) {$p_2$};
        \node[above right] at (e3) {$p_3$};
        \node[below right] at (e4) {$p_4$};
\end{tikzpicture}}

\newcommand{\treet}{
\begin{tikzpicture}[scale=1]
	    \coordinate (e1) at (-0.5,-1.);	
		\coordinate (e2) at (-0.5,0);
		\coordinate (e3) at (.5,0);
		\coordinate (e4) at (.5,-1.);	
		
		\coordinate (v1) at (0,-.25);
		\coordinate (v2) at (0,-.75);
		
		\draw[massive] (e2) -- (v1) -- (e3) ;
		\draw[massive] (e1) -- (v2) -- (e4) ;
		\draw[massive] (v1) -- (v2) ;

        \node[below left]  at (e1) {$p_1$};
        \node[above left]  at (e2) {$p_2$};
        \node[above right] at (e3) {$p_3$};
        \node[below right] at (e4) {$p_4$};
\end{tikzpicture}}

\newcommand{\dblboxS}[2]{
\begin{tikzpicture}[scale=1.7]
	    \coordinate (e1) at (-1,0);	
		\coordinate (e2) at (-1,1);
		\coordinate (e3) at (1,1);
		\coordinate (e4) at (1,0);	
		
		\coordinate (v1) at (-.75,.15);
		\coordinate (v2) at (-.75,.85);
		\coordinate (v3) at (.75,.85);
		\coordinate (v4) at (.75,.15);
		\coordinate (v5) at (0,.85);
		\coordinate (v6) at (0,.15);
		\coordinate (v7) at (-.38,.5);
		\coordinate (v8) at (.38,.5);
		
		\draw[massive] (e1) -- (v1) ;
		\draw[massive] (e2) -- (v2) ;
		\draw[massive] (e3) -- (v3) ;
		\draw[massive] (e4) -- (v4) ;
		\draw[massive] (v1) -- (v2) -- (v3) -- (v4) -- (v1) ;
		\draw[massive] (v5) -- (v6) ; 

        \node[below left]  at (e1) {$1$};
        \node[above left]  at (e2) {$2$};
        \node[above right] at (e3) {$3$};
        \node[below right] at (e4) {$4$};
        
        \node at (v7) {$#1$};
        \node at (v8) {$#2$};
\end{tikzpicture}}

\newcommand{\dblboxT}[2]{
\begin{tikzpicture}[scale=1.7]
	    \coordinate (e1) at (0,-1);	
		\coordinate (e2) at (1,-1);
		\coordinate (e3) at (1,1);
		\coordinate (e4) at (0,1);	
		
		\coordinate (v1) at (.15,-.75);
		\coordinate (v2) at (.85,-.75);
		\coordinate (v3) at (.85,.75);
		\coordinate (v4) at (.15,.75);
		\coordinate (v5) at (.85,0);
		\coordinate (v6) at (.15,0);
		\coordinate (v7) at (.5,-.38);
		\coordinate (v8) at (.5,.38);
		
		\draw[massive] (e1) -- (v1) ;
		\draw[massive] (e2) -- (v2) ;
		\draw[massive] (e3) -- (v3) ;
		\draw[massive] (e4) -- (v4) ;
		\draw[massive] (v1) -- (v2) -- (v3) -- (v4) -- (v1) ;
		\draw[massive] (v5) -- (v6) ; 

        \node[below left]  at (e1) {$1$};
        \node[below right]  at (e2) {$4$};
        \node[above right] at (e3) {$3$};
        \node[above left] at (e4) {$2$};
        
        \node at (v7) {$#1$};
        \node at (v8) {$#2$};
\end{tikzpicture}}
\usepackage[export]{adjustbox}

\newcommand{\la}{\langle}
\newcommand{\ra}{\rangle}

\newcommand{\ab}[1]{\langle #1\rangle }
\newcommand{\lam}[1]{\lambda_{#1}}
\newcommand{\lamt}[1]{\widetilde{\lambda}_{#1}}

\newcommand{\Q}{\mathcal{Q}}
\newcommand{\tw}{\widetilde}
\newcommand{\twEta}{\widetilde{\eta}}
\newcommand{\MHVbar}{\overline{{\rm MHV}}}

\usepackage{etoolbox}

\makeatletter
\def\@mkboth#1#2{}
\newlength\appendixwidth
\preto\appendix{\addtocontents{toc}{\protect\patchl@section}}
\newcommand{\patchl@section}{%
  \settowidth{\appendixwidth}{\textbf{Appendix }}%
  \addtolength{\appendixwidth}{1.5em}%
  \patchcmd{\l@section}{1.5em}{\appendixwidth}{}{\ddt}%
}
\makeatother

\begin{document}
\bibliographystyle{iopart-num}
\newcommand{\eprint}[2][]{\href{https://arxiv.org/abs/#2}{\tt{#2}}}

\begin{flushright}
	SAGEX-22-08
\end{flushright}

\title[Positive Geometry]{The SAGEX Review on Scattering Amplitudes \\ 
Chapter 7: Positive Geometry of Scattering Amplitudes}

\vspace{-.5cm}
\author{Enrico Herrmann}
\address{Mani L. Bhaumik Institute for Theoretical Physics,\\
UCLA Department of Physics and Astronomy, Los Angeles, CA 90095, USA}
\ead{eh10@g.ucla.edu}

\author{Jaroslav Trnka}
\address{Center for Quantum Mathematics and Physics (QMAP), \\
Department of Physics, University of California, Davis, CA 95616, USA}
\ead{trnka@ucdavis.edu}


\begin{abstract}
Scattering amplitudes are both a wonderful playground to discover novel ideas in Quantum Field Theory and simultaneously of immense phenomenological importance to make precision predictions for e.g.~particle collider observables and more recently also for gravitational wave signals. In this review chapter, we give an overview of some of the exciting recent progress on reformulating QFT in terms of mathematical, geometric quantities, such as polytopes, associahedra, Grassmanians, and the amplituhedron. In this novel approach, standard notions of locality and unitarity are derived concepts rather than fundamental ingredients in the construction which might give us a handle on a number of open questions in QFT that have evaded an answer for decades. 
We first give a basic summary of positive geometry before discussing the associahedron---one of the simplest physically relevant geometric examples---and its relation to tree-level scattering amplitudes in bi-adjoint $\phi^3$ theory. Our second example is the amplituhedron construction for scattering amplitudes in planar maximally supersymmetric Yang-Mills theory. 
\end{abstract}

%
%
%
%
%

\newpage 
\tableofcontents

\section{Introduction}

Scattering amplitudes have been termed ``the most perfect microscopic structures in the universe'' \cite{Dixon:2011xs} and have caught researchers' attention over many decades. As is being highlighted in the various chapters of this review article \cite{Travaglini:2022uwo}, amplitudes play a key role in numerous phenomenological applications ranging from particle colliders all the way to precision gravitational wave science. On the other hand, scattering amplitudes allow us to probe deep and fundamental questions at the heart of Quantum Field Theory and to discover novel and truly surprising structures. 

Inspired by the immense progress in understanding quantum gravitational phenomena in AdS via the holographic principle in terms of a QFT living on the boundary of spacetime \cite{Maldacena:1997re}, it is natural to wonder whether related ideas extend to flat space. Is there some theory at infinity that computes S-matrices without a detailed knowledge of local and unitary evolution in the \emph{bulk} of spacetime, contrary to the standard picture of fields, Lagrangians, and the path integral? In recent years, several pictures of a putative flat-space holographic dual emerged, one based on studying amplitudes in special conformal bases as CFT correlators on the \emph{celestial sphere} at infinity along the lines of ideas pioneered by Strominger et al.~e.g.~in~\cite{Strominger:2013jfa,Strominger:2017zoo,Pasterski:2016qvg,Raclariu:2021zjz}. An alternative paradigm starts from a diametrically opposite philosophy of the 1960's S-matrix bootstrap program where people aspired to derive all properties of scattering amplitudes from the fundamental principles of locality and unitarity. Instead, one looks for different fundamental principles and novel mathematical structures where the S-matrix emerges as the natural answer to basic questions in this modified setup. Here, we will describe geometric structures either directly in the space of asymptotic scattering kinematics or in certain auxiliary spaces where the notion of \emph{positivity} ultimately gives rise to local unitary spacetime physics as a derived property. 

The concept of geometric ideas for scattering amplitudes is not entirely new. In the context of perturbative string theory~\cite{Green:1987sp,Polchinski:1998rq}, scattering amplitudes are obtained in terms of CFT correlation functions on the worldsheet. This is associated with a fundamental geometric object: the moduli space of marked points on a Riemann surface and foreshadows a common theme throughout our review: the factorizing boundary structure of the geometry gives rise to the factorization properties of scattering amplitudes. Recently, similar ideas have been extended to field theories themselves via twistor string constructions \cite{Witten:2003nn,Berkovits:2004hg,Roiban:2004yf}, and the scattering equations within the CHY formalism~\cite{Cachazo:2013gna,Cachazo:2013hca,Mason:2013sva}. 

Starting with the pioneering work of Andrew Hodges \cite{Hodges:2009hk}, a seemingly very different set of mathematical ideas has emerged in the context of scattering amplitudes in planar maximally supersymmetric Yang-Mills theory ($\mathcal{N}{=}4$ SYM). Hodges was the first to point out that certain tree-level amplitudes in $\mathcal{N}{=}4$ SYM can be interpreted as the volumes of polytopes in projective space. Although his idea did not directly generalize to more complicated amplitudes, the fundamental philosophy was morally correct. Almost simultaneously another set of ideas led to the discovery of the positive Grassmannian \cite{Arkani-Hamed:2009ljj,Arkani-Hamed:2012zlh} in an amplitudes setting which did generalize to more complicated amplitudes and ultimately culminated in the \emph{amplituhedron} construction \cite{Arkani-Hamed:2013jha,Arkani-Hamed:2013kca}. Even though these newly discovered geometric structures might be somewhat unfamiliar to physicists, at a fundamental level, their mathematical definitions are combinatorial in nature, see e.g.~\cite{Postnikov:2006kva,postnikov2008matching,williams2004enumeration} and relatively simple to state. The amplituhedron is an example of a more general structure---a \emph{positive geometry} \cite{Arkani-Hamed:2017tmz}---where the factorizing boundary structure of the geometry gives rise to all physical properties of scattering amplitudes (including locality and unitarity) and exposes novel symmetries \cite{Drummond:2006rz,Alday:2007hr,Drummond:2008vq,Drummond:2009fd} that are completely hidden in standard Feynman diagrammatic representations of amplitudes.

Originally, these geometric structures were formulated in auxiliary spaces where the relation to the physics of scattering amplitudes was at least one step removed. However, more recently, a reformulation of the amplituhedron directly in kinematic space \cite{Arkani-Hamed:2017vfh} has emerged which brings us closer to the holographic duality described above where properties of the S-matrix arise from mathematical principles imposed on scattering data at asymptotic infinity. It is this picture of geometric structures on kinematic space that led to the discovery of the even simpler \emph{associahedron} geometry underlying scattering amplitudes of models as mundane as bi-adjoint scalar $\phi^3$ theory \cite{Arkani-Hamed:2017mur} that has subsequently been generalized to various types of scalar theories. 

In these notes, we hope to give an elementary introduction to some of the novel mathematical concepts and highlight them in simple settings with illustrative examples. We aim to be pedagogical and often suppress important but otherwise distracting technical details. We had to make judicious choices on the presented material, but hopefully, the references to the original research papers serve as starting point for further individual study. Our discussion was particularly influenced by Refs.~\cite{Arkani-Hamed:2013jha,Arkani-Hamed:2017tmz,Arkani-Hamed:2017vfh,Arkani-Hamed:2017mur} and the recent, mathematically oriented review \cite{Ferro:2020ygk} of the subject. 


The remainder of this chapter is structured as follows: In section \ref{sec:pos_geom_primer} we give a succinct introduction to \emph{positive geometries} and the concept of canonical differential forms with singularities on the boundary of the geometric space. Subsequently---and historically out of order---we discuss the Arkani-Hamed, Bai, He, and Yan (ABHY) associahedron in the context of tree-level scattering amplitudes in bi-adjoint scalar $\phi^3$ theory in section \ref{sec:associahedron}. The associahedron is polytopal in nature with linear boundaries of all codimension, rendering it one of the simplest geometric structures. In section \ref{sec:amplituhedron} we turn to the amplituhedron construction of scattering amplitudes and loop integrands in planar $\mathcal{N}{=}4$ SYM. We discuss both the original construction in some auxiliary space as well as the more recent topological definition in kinematic space directly. We close our discussion with a very brief bird's-eye view over other recent contexts where positive geometries played an important role in different physical contexts ranging from cosmology to bounds on EFT coefficients. Finally, we end with an outlook on open problems for the future.

\section{A Postive Geometry Primer}
\label{sec:pos_geom_primer}

In our review, we will primarily discuss two examples of geometric structures and their relation to scattering amplitudes. First, we cover one of the simplest cases; the \emph{associahedron} (well known to mathematicians, see e.g.~\cite{Stasheff:1963I,Ziegler:1995polytopes,Ceballos:2014aaa}) and its relation to scattering amplitudes in bi-adjoint $\phi^3$ theory \cite{Arkani-Hamed:2017mur}. As a second example, we consider the \emph{amplituhedron} \cite{Arkani-Hamed:2013jha,Arkani-Hamed:2013kca,Arkani-Hamed:2017vfh} and its relation to scattering amplitudes in planar maximally supersymmetric Yang-Mills theory. Both the associahedron, as well as the amplituhedron belong to a more general class of \emph{positive geometries} \cite{Arkani-Hamed:2017tmz} (see also e.g.~\cite{Arkani-Hamed:2017mur,Ferro:2020ygk} for a concise summary and \cite{Lam:notes} for a more mathematical take on the connection to amplitudes). Discussing the relevant aspects of positive geometry in general will allow us to treat both the associahedron and amplituhedron in a uniform fashion in the remainder of these notes. As we will explain in the following sections, in physics, the relevant geometric spaces can either live directly in the kinematic space of the particle scattering setup, or in some auxiliary spaces. For our primer, we mostly follow the summary in appendix A of Ref.~\cite{Arkani-Hamed:2017mur}.

Fundamentally, a positive geometry requires the specification of two main ingredients: (1) a particular geometric space\footnote{In particular, the space is given by a complex $D$-dimensional projective variety $X$ providing the ambient space together with a subset $X_{\geq0}$ that defines a real (oriented) $D$-dimensional slice.} and (2) an associated differential form. Crucially, this geometry possesses boundaries of \emph{all} codimensions. One of the simplest examples of such spaces are polytopes which have linear boundaries (see e.g.~Fig.~\ref{fig:p2_p3_polytopes}). It is also possible to have curved boundaries that are defined by polynomial equations of higher degree \cite{Arkani-Hamed:2017tmz}. The relation between the physics of scattering amplitudes and a $D$-dimensional positive geometry $\mathcal{A}$ is encoded in the \emph{canonical differential $D$-form} on the geometric space, denoted by $\Omega(\mathcal{A})$, with the following properties: 
\begin{itemize}
  \item $\Omega(\mathcal{A})$ has simple poles on the boundaries of the space and nowhere else.
  \item For any hyper-surface $H$ containing a boundary $\mathcal{B}$ of $\mathcal{A}$, the canonical form on $\mathcal{B}$ is recursively obtained from the one on $\mathcal{A}$ by taking the residue along $H$
  \begin{equation}
  \label{eq:pos_geom_omega_boundary}
  \hspace{3cm}
      \Omega(\mathcal{B}) = {\rm Res}_H \, \Omega (\mathcal{A})
  \end{equation}
  \item If $\mathcal{A}$ is zero-dimensional (a point), then $\Omega(\mathcal{A})=\pm 1$, where the sign takes the role of an orientation of the space. 
\end{itemize}
Positive geometries naturally ``live'' in (real) projective space $\mathbb{P}^N(\mathbb{R})$ instead of the more familiar Euclidean space $\mathbb{R}^N$. However, since $\mathbb{R}^N$ can be embedded in $\mathbb{P}^N(\mathbb{R})$ via $x\to(1,y)$ (with $x\in \mathbb{P}^N(\mathbb{R})$ and $y \in \mathbb{R}^N$) one can visualize projective space as ordinary Euclidean space with a special choice of a hyperplane at infinity. 

\hspace{-.4cm}
\subsection{A First Example---Projective Polytopes}
\label{subsec:proj_polytopes}

\begin{figure}[ht!]
    \centering
    \vspace{-1cm}
    \raisebox{-10pt}{\includegraphics[scale=.25]{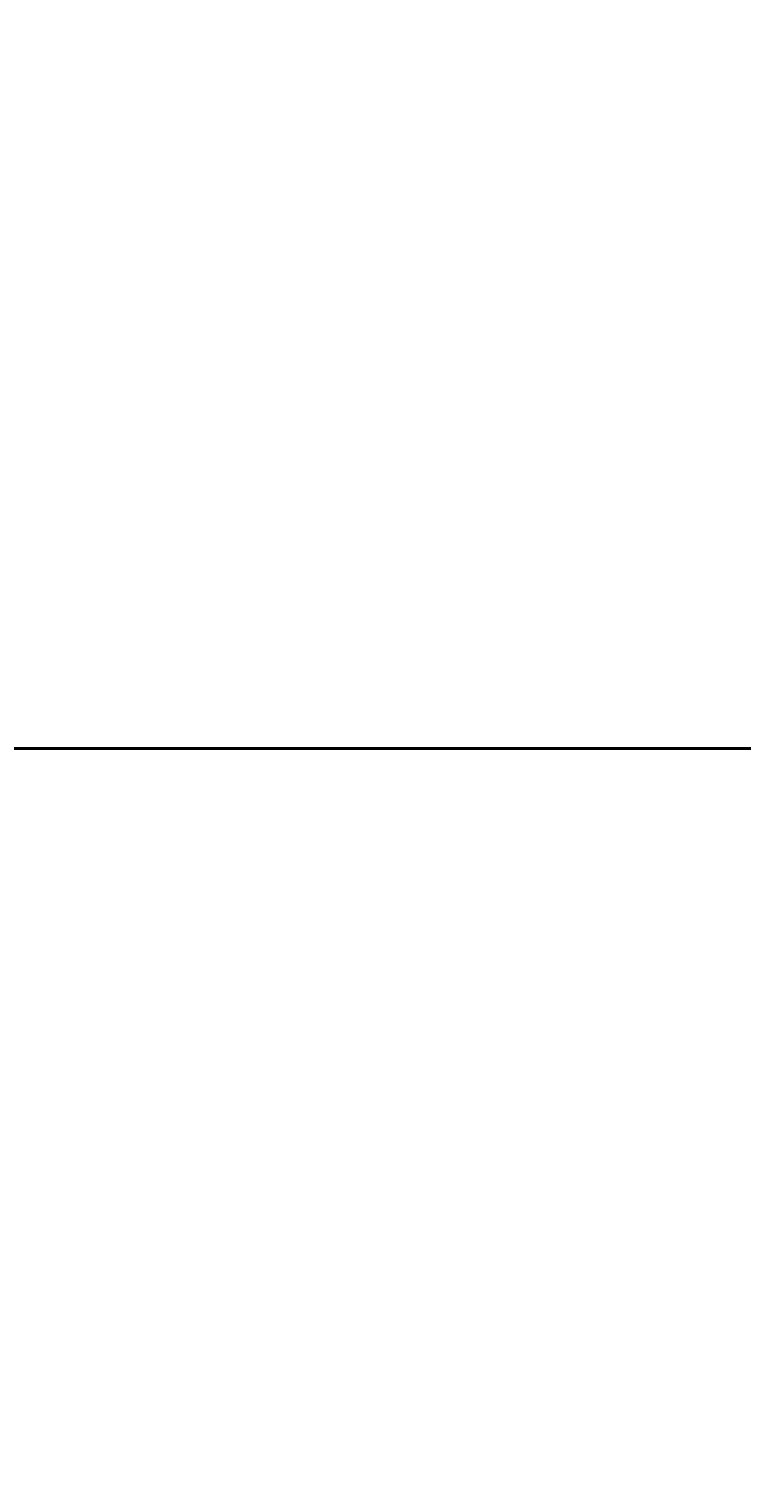}}
    \qquad
    \raisebox{20pt}{\includegraphics[scale=.15]{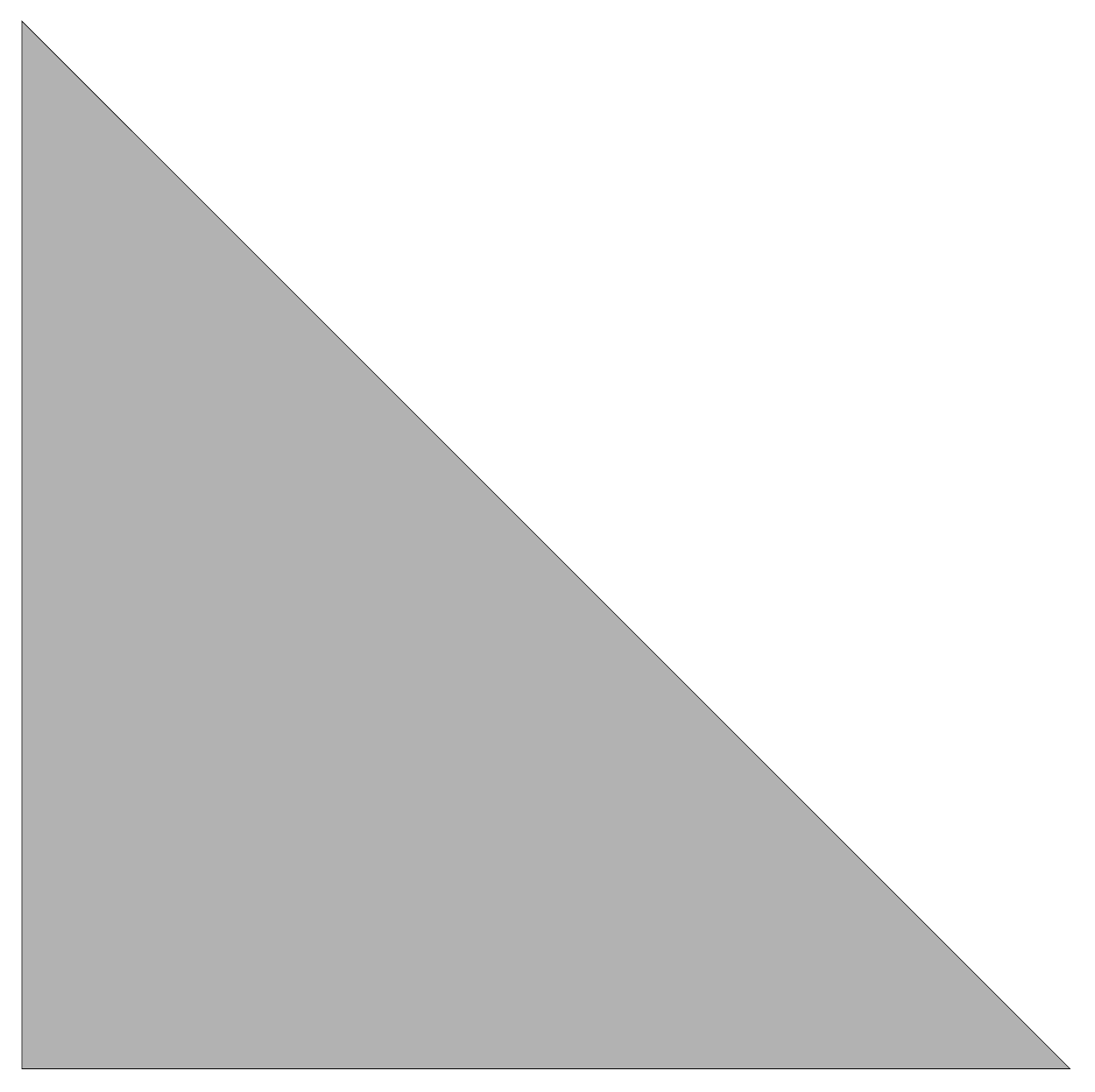}}
    \qquad
    \raisebox{00pt}{\includegraphics[scale=.45]{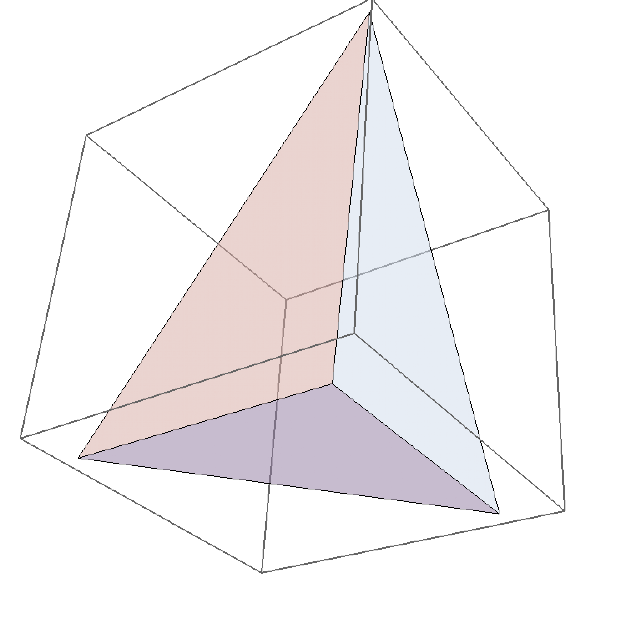}}
    \vspace{-.5cm}
    \caption{Representation of low-dimensional projective \emph{simplices} in $\mathbb{P}^N$, for $N=1,2,3$ that are carved out by $N+1$ inequalities.}
    \label{fig:proj_simplices_examples}
\end{figure}

For the following discussions, we are going to built our intuition by first studying \emph{projective polytopes} leaving more complicated positive geometries such as the positive Grassmannian and amplituhedra to later sections. Even before discussing polytopes, the simplest possible examples of positive geometries are (projective) \emph{simplices}. Mathematically, a projective $N$-simplex $\Delta$ in $\mathbb{P}^N$ is given by $N+1$ linear inequalities as follows (see e.g.~\cite{Ferro:2020ygk}): If $Y$ is a point in $\mathbb{P}^N$ with vector components $Y^A$, $A=0,1,\ldots,N$, then any linear inequality can be written as $Y\cdot W \equiv Y^A W_A \geq 0$ in terms of a dual vector $W \in \mathbb{R}^{N+1}$, so that 
\begin{equation}
\label{eq:simplex_def}
    \Delta = \left\{
    Y \in \mathbb{P}^N(\mathbb{R})\ | \ Y \cdot W_{i} \geq 0,
    \quad {\rm for}\ i=1,\ldots N{+}1 
    \right\}\,.
\end{equation}
The dual vectors $W_{i}$ correspond to the \emph{facets} of the simplex. For $N=1,2,3$, the relevant simplices are the (projective) line-segments, triangles, and tetrahedra that are carved out by $2,3,4$ inequalities, respectively (see Fig.~\ref{fig:proj_simplices_examples}). 
From the definition of the projective simplices in Eq.~(\ref{eq:simplex_def}) and the explicit low-dimensional examples of Fig.~\ref{fig:proj_simplices_examples}, it is clear that every boundary $Y\cdot W_{i}=0$ is again a projective simplex in one lower dimension and therefore satisfies all the defining conditions of a positive geometry. (The line segments in $\mathbb{P}^1$ are bounded by points. The triangles in $\mathbb{P}^2$ are bounded by line segments, which in turn are bounded by points, et cetera.) Instead of the facet-centric definition of the simplices, we can equivalently use a vertex-centric description where we denote the $N+1$ vertices of $\Delta$ by $Z_i\in \mathbb{R}^{N+1}/\{0\}$, $i\in\{1,\ldots, N+1\}$. 

\begin{figure}[ht!]
    \centering
    \vspace{-.5cm}
    \raisebox{5pt}{\includegraphics[scale=.15]{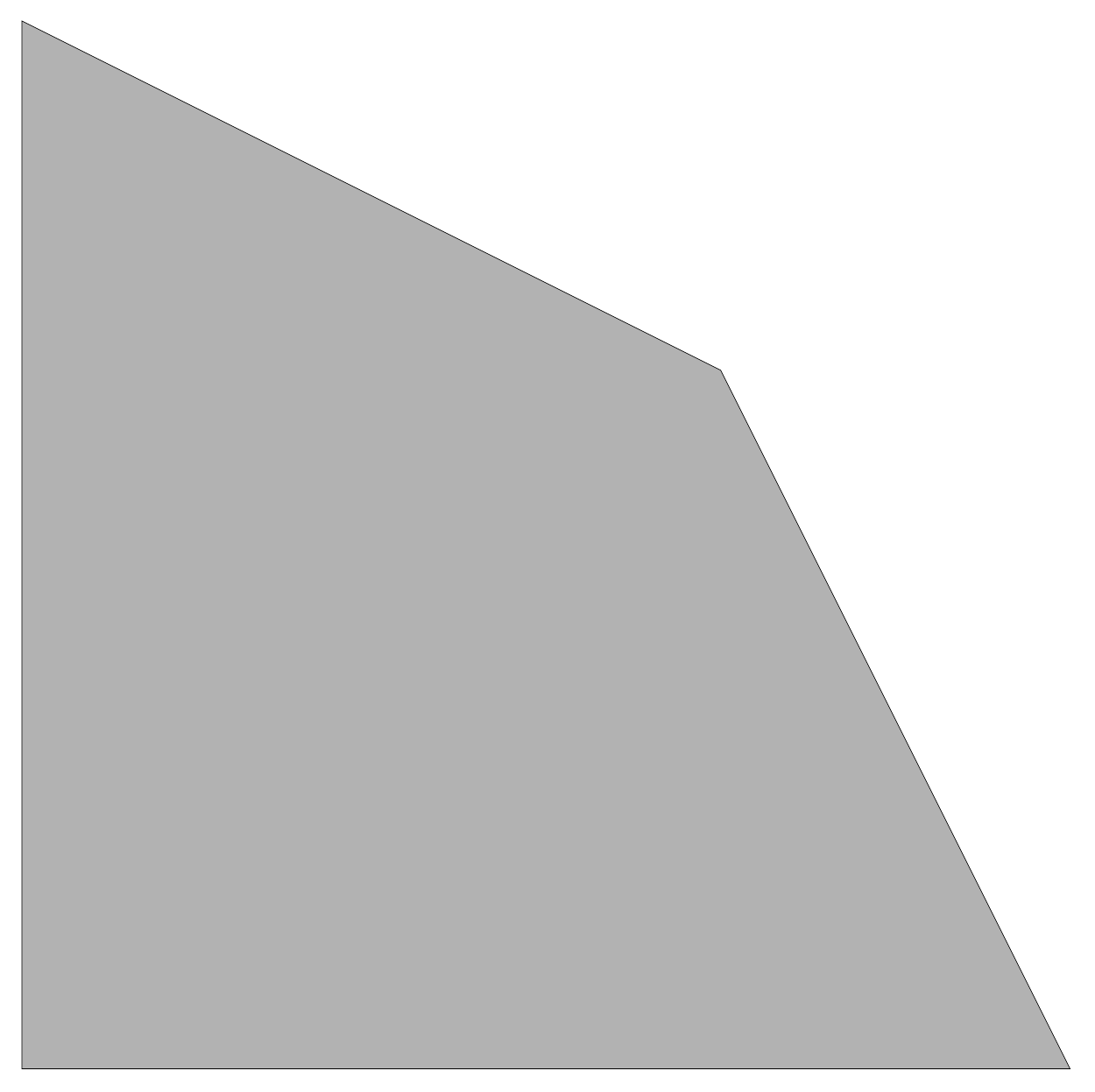}}
    \qquad
    \raisebox{5pt}{\includegraphics[scale=.15]{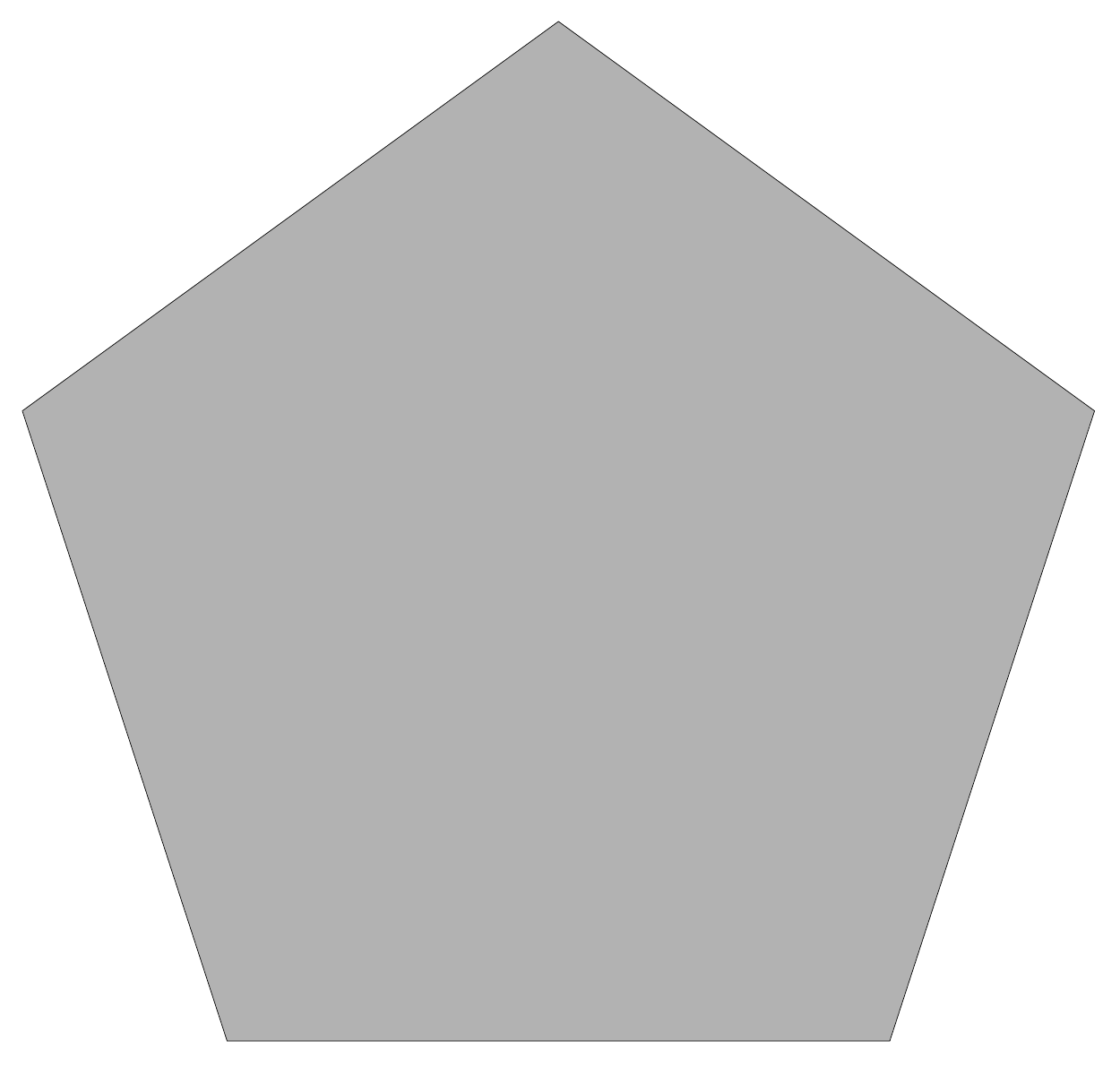}}
    \qquad
    \includegraphics[scale=.2]{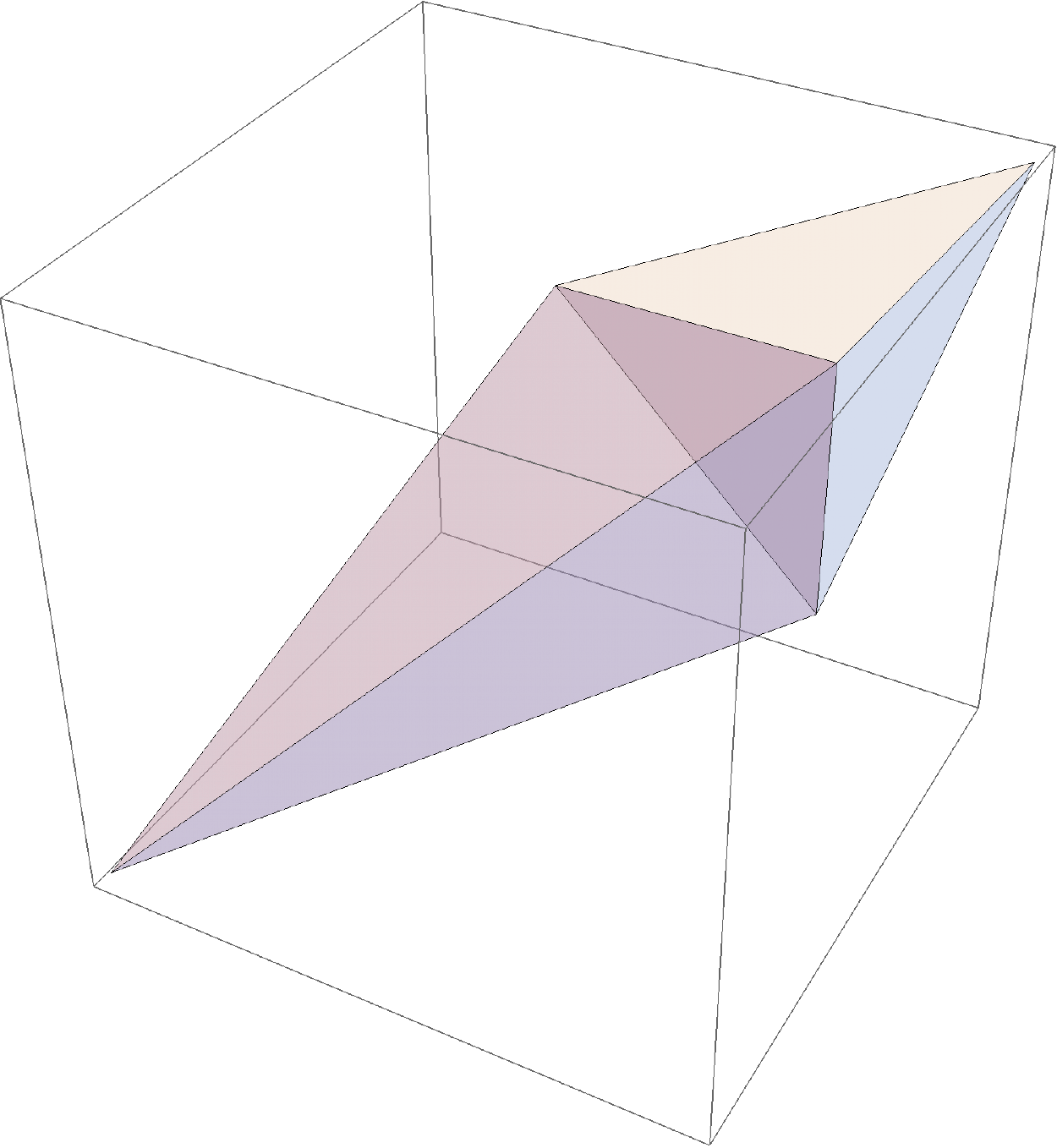}
    \caption{
    Convex projective polytopes in $\mathbb{P}^2$, for $M=4,5$, and in $\mathbb{P}^3$ for $M=5$.}
    \label{fig:p2_p3_polytopes}
    \vspace{-.5cm}
\end{figure}

\emph{Projective simplices} generalize to \emph{projective convex polytopes} by allowing more than $M>N+1$ vertices $Z_1,\ldots Z_M \in \mathbb{R}^{N+1}$ whose convex hull forms a projective polytope $\mathcal{P}\equiv \mathcal{P}(Z_1,\ldots,Z_M)\subset \mathbb{P}^{N}(\mathbb{R})$
\begin{equation}
\hspace{-1cm}
    \mathcal{P} = {\rm Hull} (Z_1,\ldots,Z_M) = 
    \left \{
    \sum^{M}_{i=1} c_i Z_i \in \mathbb{P}^{N}(\mathbb{R}) | c_i \geq 0\,,\quad  i=1,\ldots, M
    \right\}\,.
\end{equation}
Particular examples in $\mathbb{P}^2$ and $\mathbb{P}^3$ are depicted in Fig.~\ref{fig:p2_p3_polytopes}.
Let us note that the convex hull definition is insufficient to uniquely fix the polytopal space and additional conditions are necessary. One (highly redundant) definition that completely specifies the combinatorics of the facet structure of the polytope involves grouping the $M$ points $Z_i \in \mathbb{R}^{N+1}$ into a $M \times (N+1)$ matrix $Z$ and fixing the signs of \emph{all} ordered minors. 
\begin{equation}
\hspace{-.5cm}
    \ab{i_1,i_2,{\ldots}, i_{N{+}1}} \,,
    \qquad {\rm for }\ 
 1\leq i_1 < i_2 < \cdots < i_{N{+}1} \leq M\,,
\end{equation}
where, $\ab{i_1,i_2,{\ldots}, i_{N{+}1}} {\equiv} \ab{Z_{i_1},Z_{i_2},{\ldots} Z_{i_{N{+}1}}} {\equiv} \varepsilon_{A_1 A_2 \cdots A_{N+1}} Z^{A_1}_{i_1}\, Z^{A_2}_{i_2}{\cdots} Z^{A_{N+1}}_{i_{N+1}}$. Beyond $N{=}2$, the vertices of a generic polytope are not ordered. If one insists on a natural ordering, one obtains a \emph{cyclic polytope} where \emph{all} ordered minors are positive. 

\begin{figure}[ht!]
    \centering
    \raisebox{-20pt}{\includegraphics[scale=.12]{./figures/poly4_P2.pdf}}
    =
    \hspace{.1cm}
    \raisebox{-20pt}{\includegraphics[scale=.12]{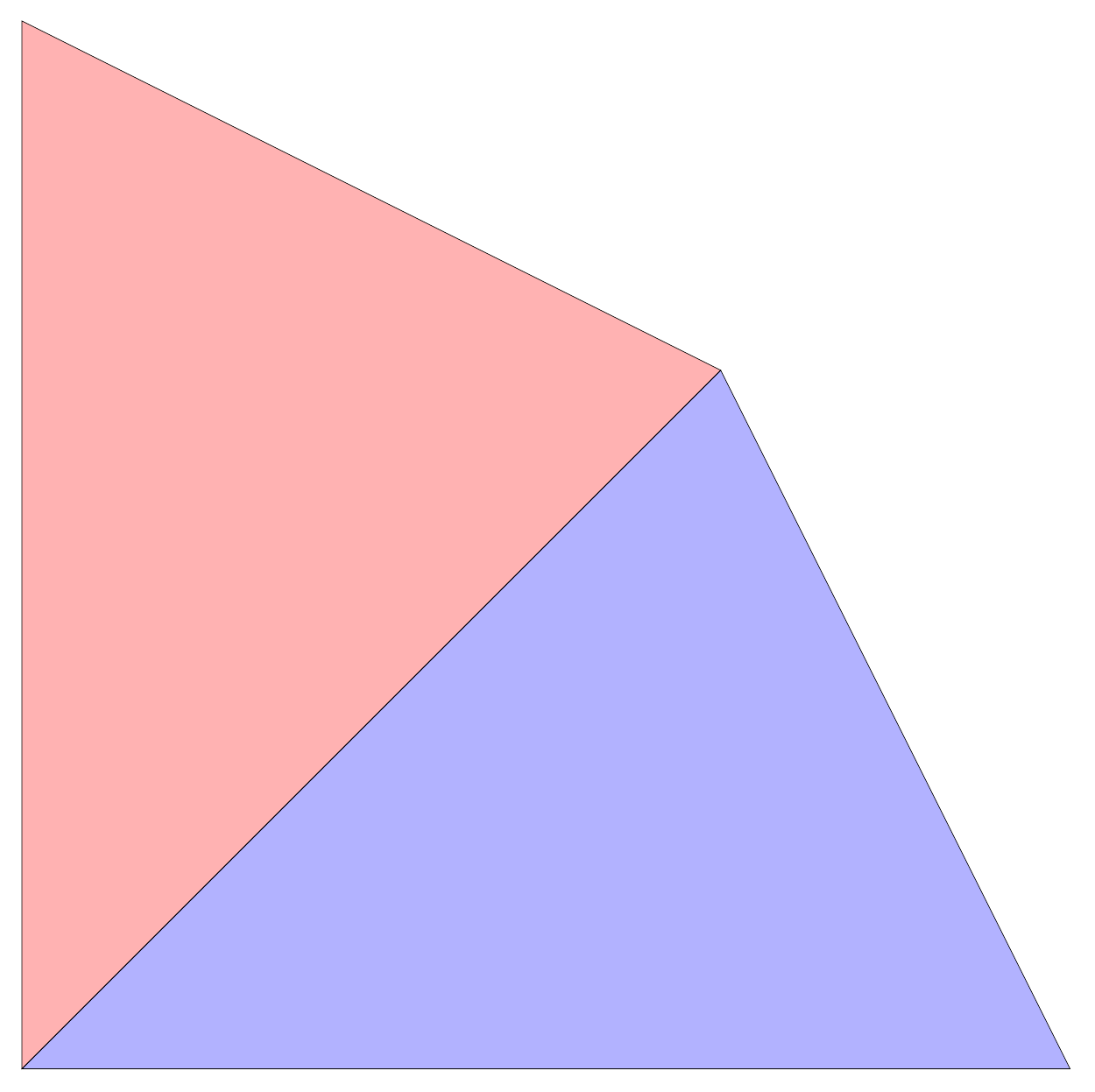}}
    =
    \hspace{.1cm}
    \raisebox{-20pt}{\includegraphics[scale=.12]{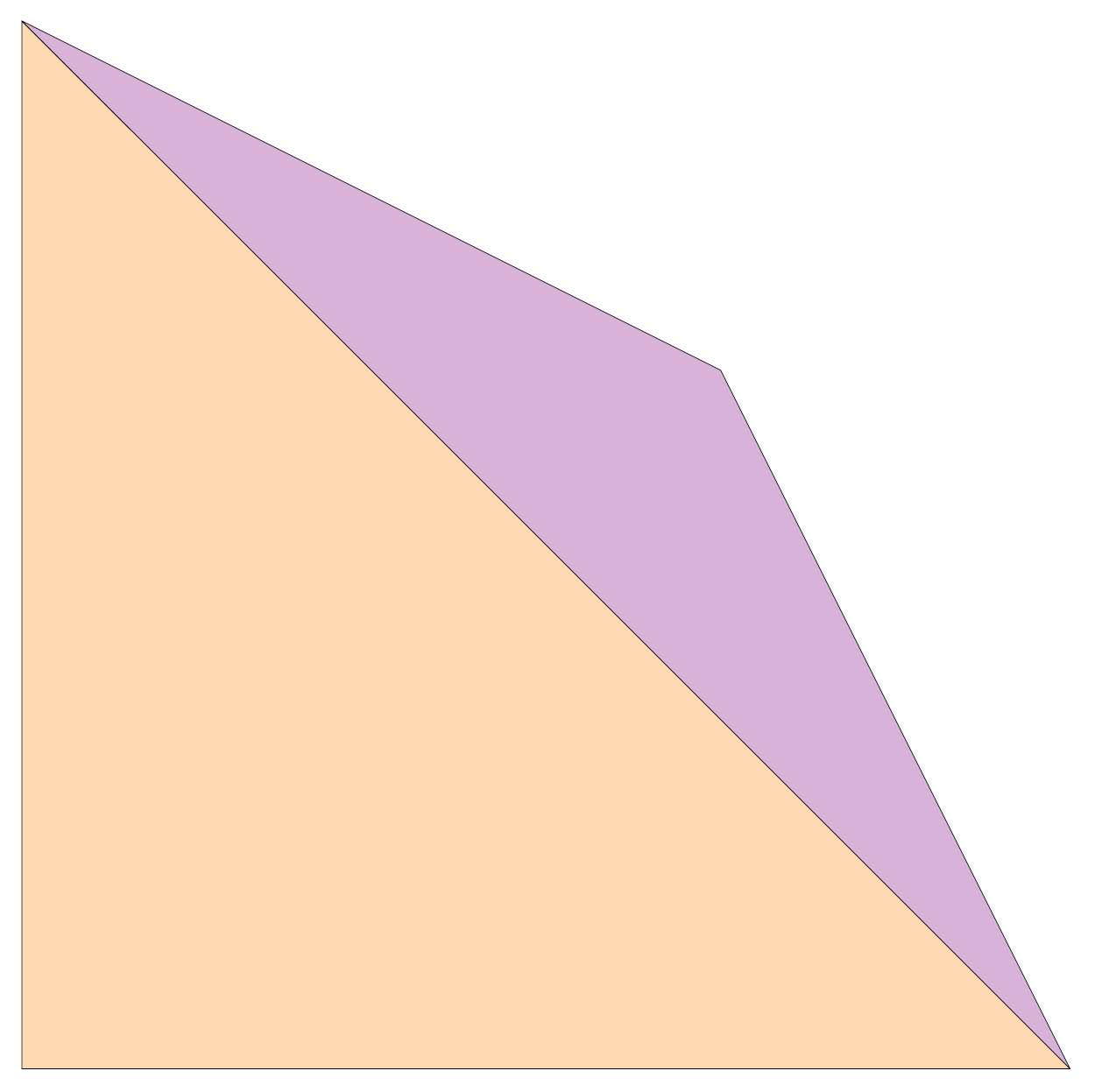}}
    =
    \hspace{.1cm}
    \raisebox{-20pt}{\includegraphics[scale=.12]{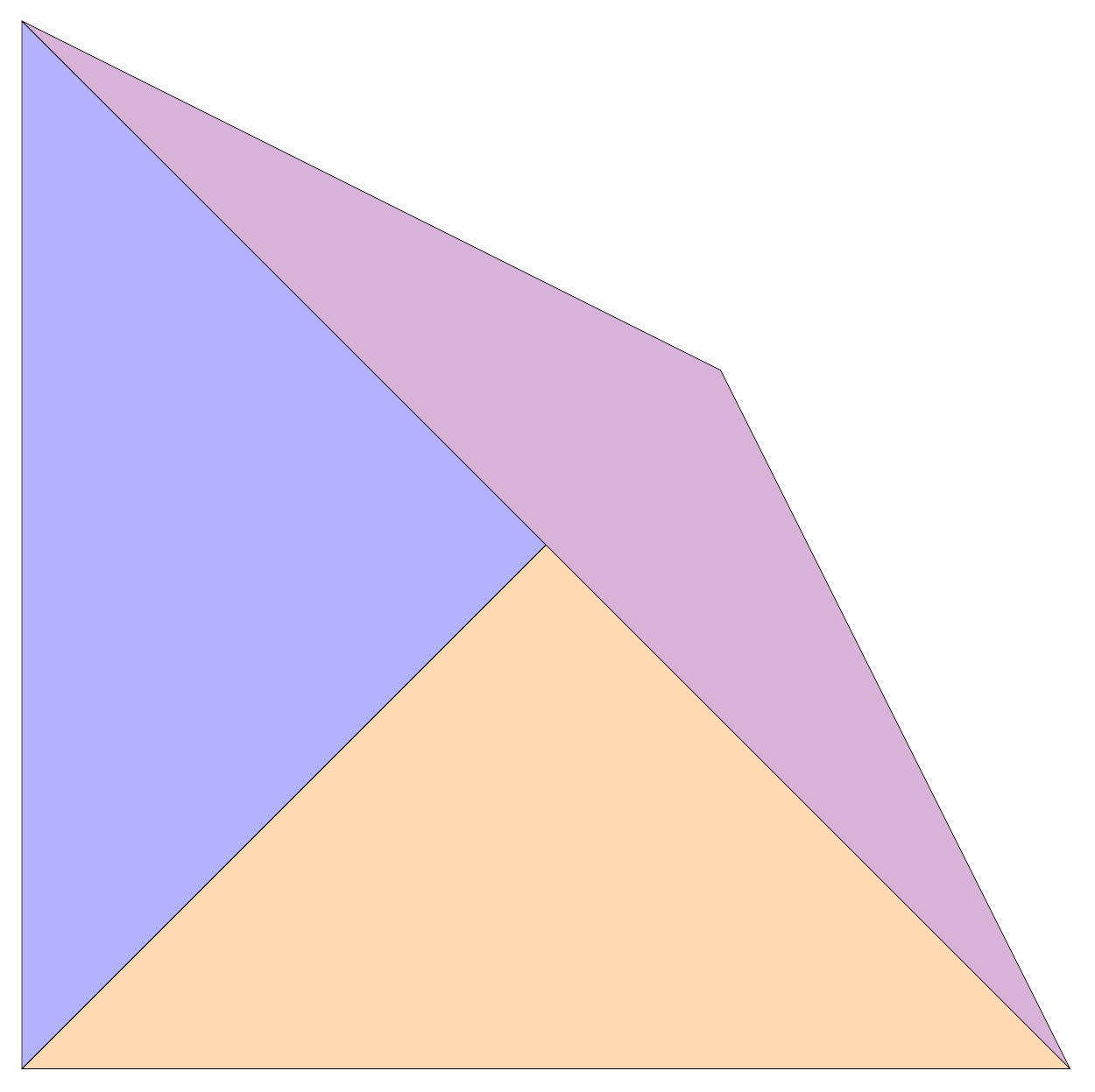}}
    =
    \hspace{.1cm}
    \raisebox{-20pt}{\includegraphics[scale=.2]{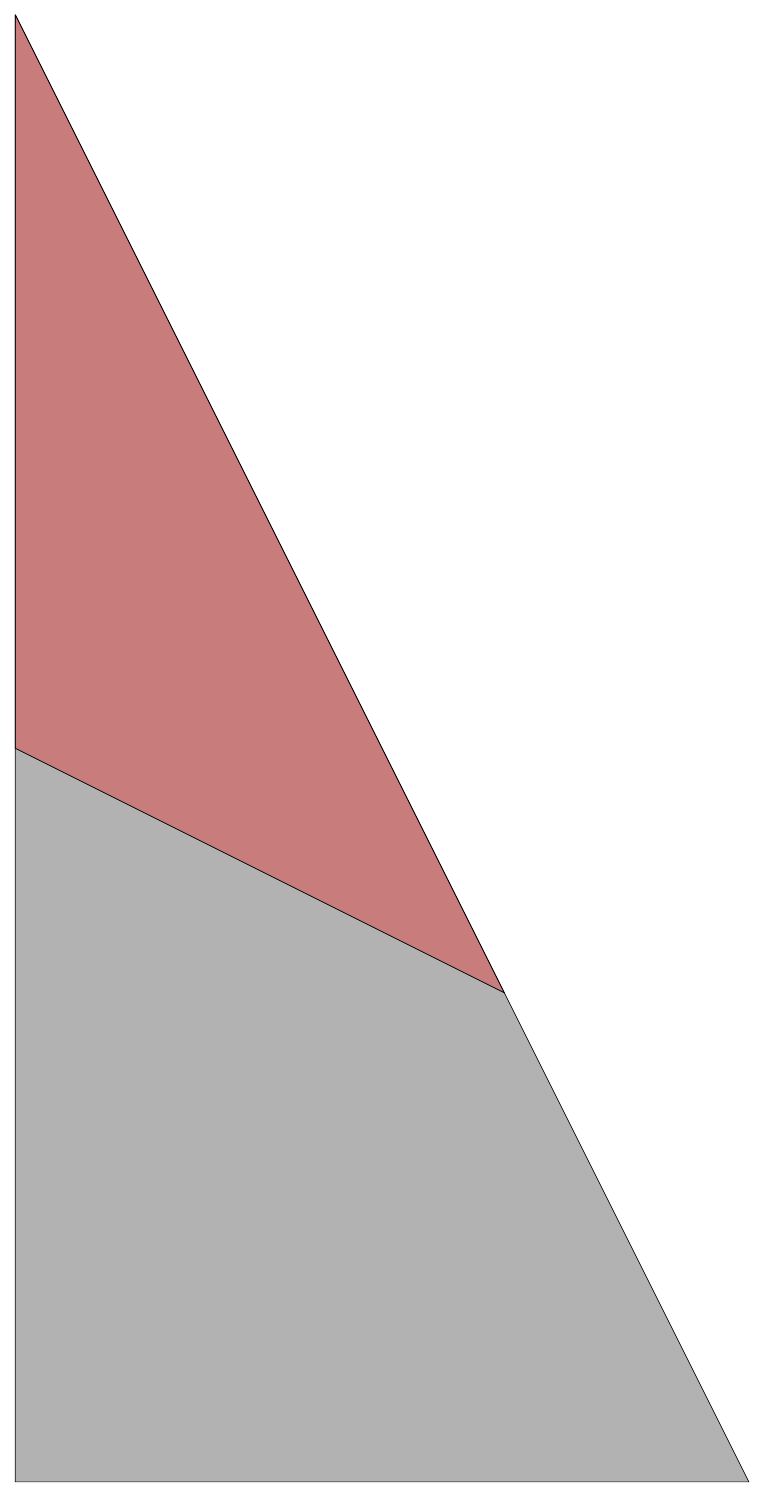}}
    \vspace{-.3cm}
    \caption{Example triangulations of convex polytope (quadrilateral) in $\mathbb{P}^2$ in terms of projective simplices in $\mathbb{P}^2$ (triangles). The last two triangulations involve additional points in $\mathbb{P}^2$ that are either inside, or outside the original quadrilateral.}
    \label{fig:poly4_p2_triangulations}
    \vspace{-.7cm}
\end{figure}

One feature that is clearly visible even for the simple projective polytopes and that is going to play an important role in the remainder of our discussion is the fact that we can build up (or \emph{triangulate}) more complicated spaces from the ``fundamental'' simplices. As we will discuss in greater detail in the physical examples of interest, such triangulations are not unique. For example, the quadrilateral in $\mathbb{P}^2$ can be obtained from the simplicial triangles in multiple ways, some of which are given in Fig.~\ref{fig:poly4_p2_triangulations}.
%

\newpage
\subsection{Canonical Forms}
\label{subsec:canon_forms}

From the above definition of positive geometries, it is clear that \emph{canonical forms} play a major role, both for the positive geometry as a mathematical object, as well as from a physics point of view. It is therefore of paramount importance to understand the properties of canonical forms and how to determine them. Two of the main tools to obtain canonical forms of more complicated positive geometries are \emph{triangulations} and \emph{push forwards}. For particular examples, like the projective convex polytopes of Sec.~\ref{subsec:proj_polytopes}, it is also possible to dualize the geometry in a way that the canonical forms are given in terms of the volume of the dual geometry. So far, such a dualization has been elusive for the curvy amplituhedron geometries despite some evidence for its existence \cite{Arkani-Hamed:2014dca}. 

\subsubsection{Triangulations of Positive Geometries}
\label{subsubsec:canon_forms}
Currently, one of the most common approaches to determining the canonical forms for positive geometries is via \emph{triangulation}. In this setup, one subdivides a more complicated positive geometry $\mathcal{A}$ into a set of simpler positive geometries $\{\mathcal{A}_a\}$ whose forms $\Omega(\mathcal{A}_a)$ are known. In physics, we often refer to any such subdivision of the full geometry $\mathcal{A}$ as a \emph{triangulation}, even if the components $\mathcal{A}_a$ are not simplices. The canonical form on $\mathcal{A}$ is given by
\begin{equation}
\hspace{3cm}
    \Omega(\mathcal{A}) = \sum_a \Omega(\mathcal{A}_a)\,.
\hspace{-1cm}    
\end{equation}
Again, in order to build some intuition, we turn to the convex polytopes of Section \ref{subsec:proj_polytopes} for guidance. In Fig.~\ref{fig:poly4_p2_triangulations}, we exemplified possible triangulations of a quadrilateral in $\mathbb{P}^2$ in terms of the simplicial triangles, which immediately shows that such triangulations are by no means unique. However, since the full positive geometry $\mathcal{A}$ is independent of the choice of triangulation, so is its canonical form.  

Since the basic building blocks for more general projective polytopes are the simplices, we ought to discuss how to obtain their canonical forms. Since projective geometry might be somewhat unfamiliar, we start our discussion in the de-projectivized setting and generalize from there. Crucial to the very definition of a positive geometry is the logarithmically singular behavior of the canonical form as one approaches a boundary of the space, see Eq.~(\ref{eq:pos_geom_omega_boundary}). The starting point are zero-dimensional geometries, i.e. points for which the associated canonical zero-forms are $\pm 1$ by the above definition. The first nontrivial case is one-dimensional and we consider the line-segment $[a,b] \subset \mathbb{R}$. This one-dimensional geometry has two boundaries, the points $x=a$ and $x=b$, and we would like to find a one-form with logarithmic singularities only at these two locations. It is not hard to see that correct form is
\begin{equation}
\label{eq:line_segment_form}
\hspace{-3.7cm}
    \raisebox{-2pt}{ \includegraphics[scale=.45,trim={2cm 6.6cm 2cm 5.3cm },clip]{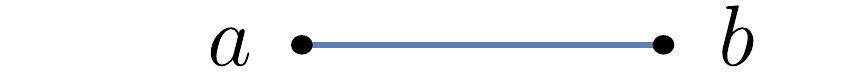}
    }
    \hspace{-.5cm}
    \leftrightarrow
    \Omega([a,b]) {=} \frac{dx}{(x-a)} - \frac{dx}{(x-b)} 
    {=} d\log(x-a) - d\log(x-b) 
    {=} d\log\frac{x-a}{x-b}\,,
    \hspace{.1cm} \label{dlog}
\end{equation}
where the relative signs in $\Omega([a,b])$ are such that the poles at $x\to \infty$ cancel. Note that we performed the change of variables in the form to rewrite $dx/(x-a)=d{\rm log}(x-a)$ making the logarithmic singularities manifest. While this was a very trivial example, in general this change of variables can be quite challenging. At the level of the form, the only singularities of (\ref{dlog}) are at $x=a$ and $x=b$ as required. Restricting the form to the two boundaries via the residue operation gives,
\begin{equation}
    {\rm Res}_{x=a}\, \Omega([a,b]) = +1\,, 
    \qquad 
    {\rm Res}_{x=b}\, \Omega([a,b]) = -1\,,
\end{equation}
which are the correct zero forms on the boundary points and the relative signs reflect the orientation of the line segment. 

Next, we consider the two-dimensional simplex, i.e.~a triangle. Working in the de-projectivized setting, a triangle is bounded by three one-dimensional lines (faces) or determined by three points (vertices). We would like to associate a canonical from to such a triangle, subject to the requirements of a positive geometry. Say, our triangle has boundaries at $x=0, y=0$ and $x+y=1$\footnote{As an exercise, check the equivalence between the $d\log$ form and the rational form.}
\begin{equation}
\hspace{-2cm}
    \raisebox{-40pt}{ 
    \includegraphics[scale=.2]{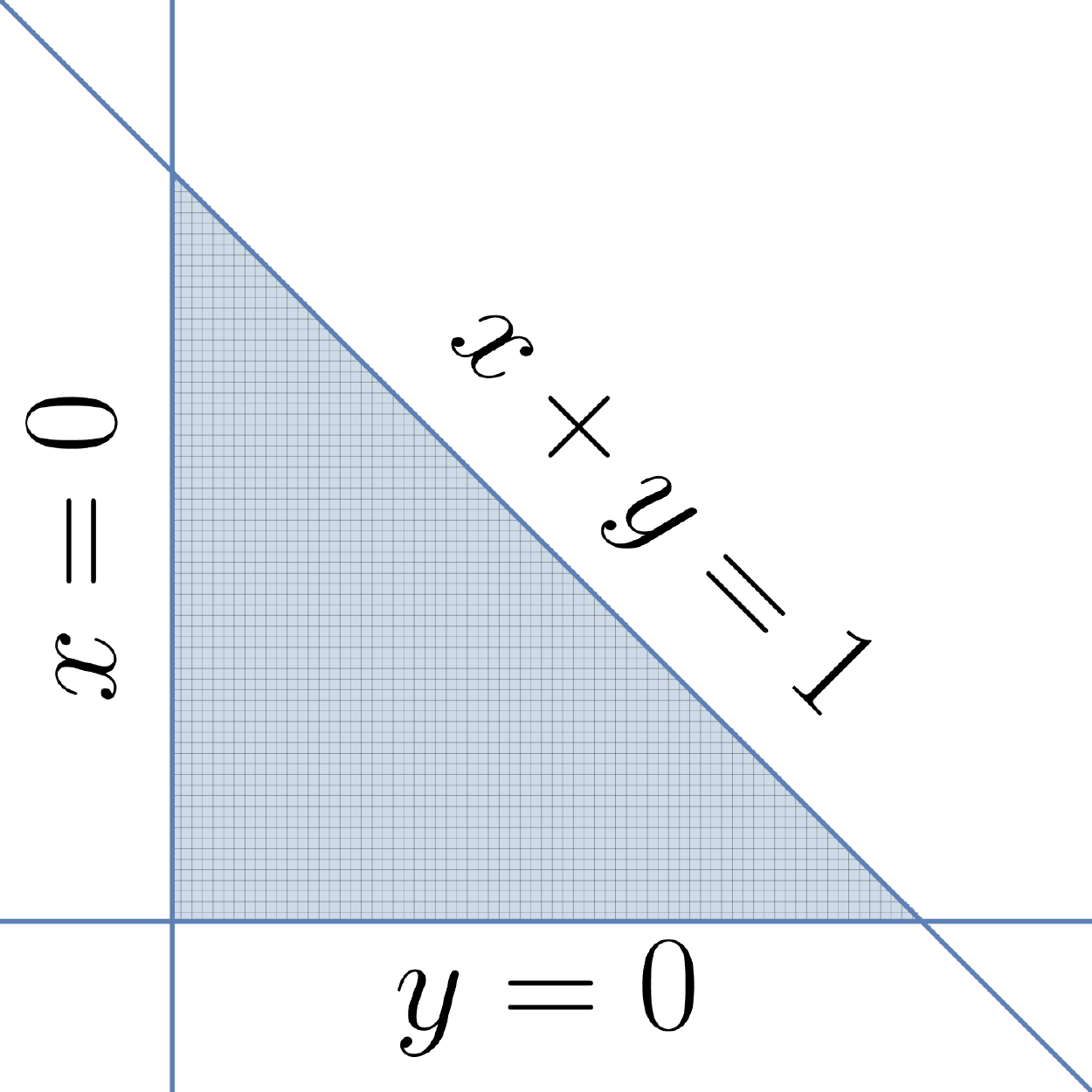}
    }
    \hspace{-.4cm}
    \leftrightarrow
    \Omega(\blacktriangle) = \frac{dx\wedge dy}{x\, y\, (1-x-y)} 
    = d\log \frac{x}{1-x-y} \wedge\, d\log \frac{y}{1-x-y}
    \,,
\end{equation}
which has the property that any codimension-one boundary of the triangle yields a one-dimensional positive geometry of a line segment discussed above. For example, the boundary at $x=0$ gives the correct canonical one-form on the line segment $y\in[0,1]$
\begin{equation}
   \Omega(y\in[0,1]) = 
   {\rm Res}_{x=0}\, \Omega(\blacktriangle) = \frac{dy}{y(1-y)} = d\log \frac{y}{1-y}\,. 
\end{equation}
The two examples above immediately suggest the generalization to higher dimensional simplices. The canonical form of a $D$-dimensional simplex is a $D$-form, where the poles are determined by $D+1$ linear equations in $D$ coordinates that specify the $D+1$ boundaries of the simplex. It is more natural to phrase such statements projectively. A simplex $\Delta$ in $\mathbb{P}^N$ is either determined by the $N{+}1$ vertices $Z_i \in \mathbb{R}^{N{+}1}/\{0\}$ or equivalently in terms of its $N{+}1$ facets. The protectively invariant form is~\cite{Arkani-Hamed:2010wgm,Enciso:2014cta,Enciso:2016cif}
\begin{equation}
\label{eq:proj_simplex_form}
\hspace{-2cm}
    \Omega(\Delta) = \frac{\ab{Z_1\,Z_2\, \ldots\, Z_{N+1}}^N \ab{Y d^NY}}
    {N!\, \underbrace{\ab{Y\, Z_1\, Z_2\cdots Z_N} \ab{Y\, Z_2 \, Z_3\, \cdots Z_{N+1}} \cdots \ab{Y\, Z_{N+1}\, Z_1\, \cdots Z_{N-1}}}_{(N+1)\ {\rm  factors}}}\,,
\end{equation}
where $ \frac{1}{N!} \ab{Y d^N Y} = \sum^{N+1}_{A=1} (-1)^A\, Y^A \, dY^1 \wedge\cdots \wedge d\widehat{Y}^A \wedge \cdots \wedge dY^{N+1},$ is the natural measure on $\mathbb{P}^N$ and the factor of $\ab{Z_1\,Z_2\, \ldots\, Z_{N+1}}^N$ in the numerator of (\ref{eq:proj_simplex_form}) ensures the projective invariance of the form under rescaling of $Z_i \to t_i\, Z_i$. 

Having described the canonical forms for the simplices, we go back to the original question of triangulating more complicated convex projective polytopes. It is our goal to find their associated canoncial forms. It is easiest to visualize a triangulation in $\mathbb{P}^2$. The first nontrivial case is the quadrilateral depicted in Fig.~\ref{fig:poly4_p2_triangulations}. Labeling the external vertices of the quadrilateral in $\mathbb{P}^2$ by $Z_1,\ldots,Z_4$, the first triangulation
\vspace{-.1cm}
\begin{equation}
\label{eq:poly4_p2_triang_1}
\hspace{-1cm}
    \raisebox{-35pt}{\includegraphics[trim={0cm 0cm 4.8cm 5cm},clip,scale=.3]{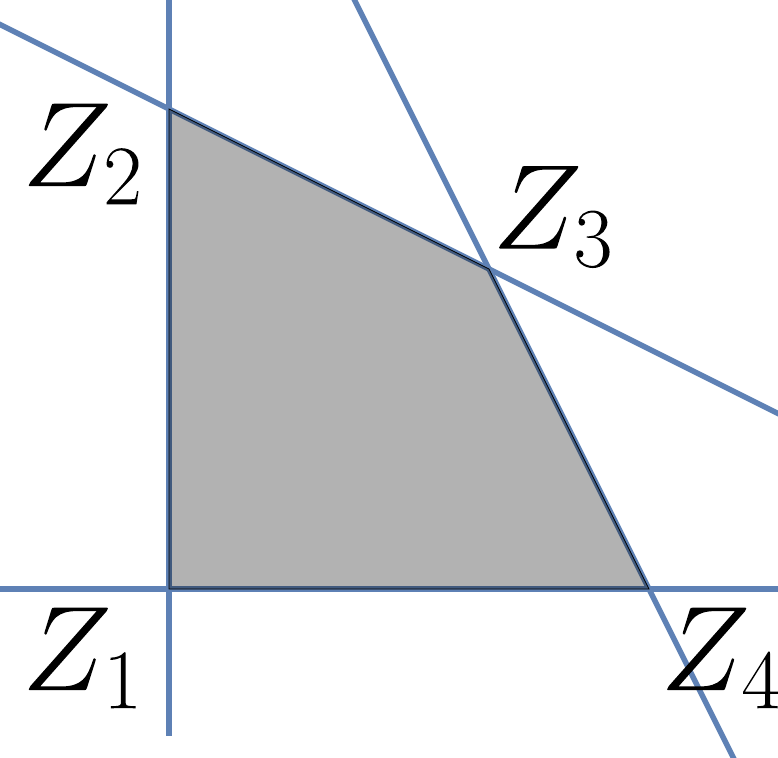}}
    =
    \quad
    \raisebox{-35pt}{\includegraphics[scale=.3]{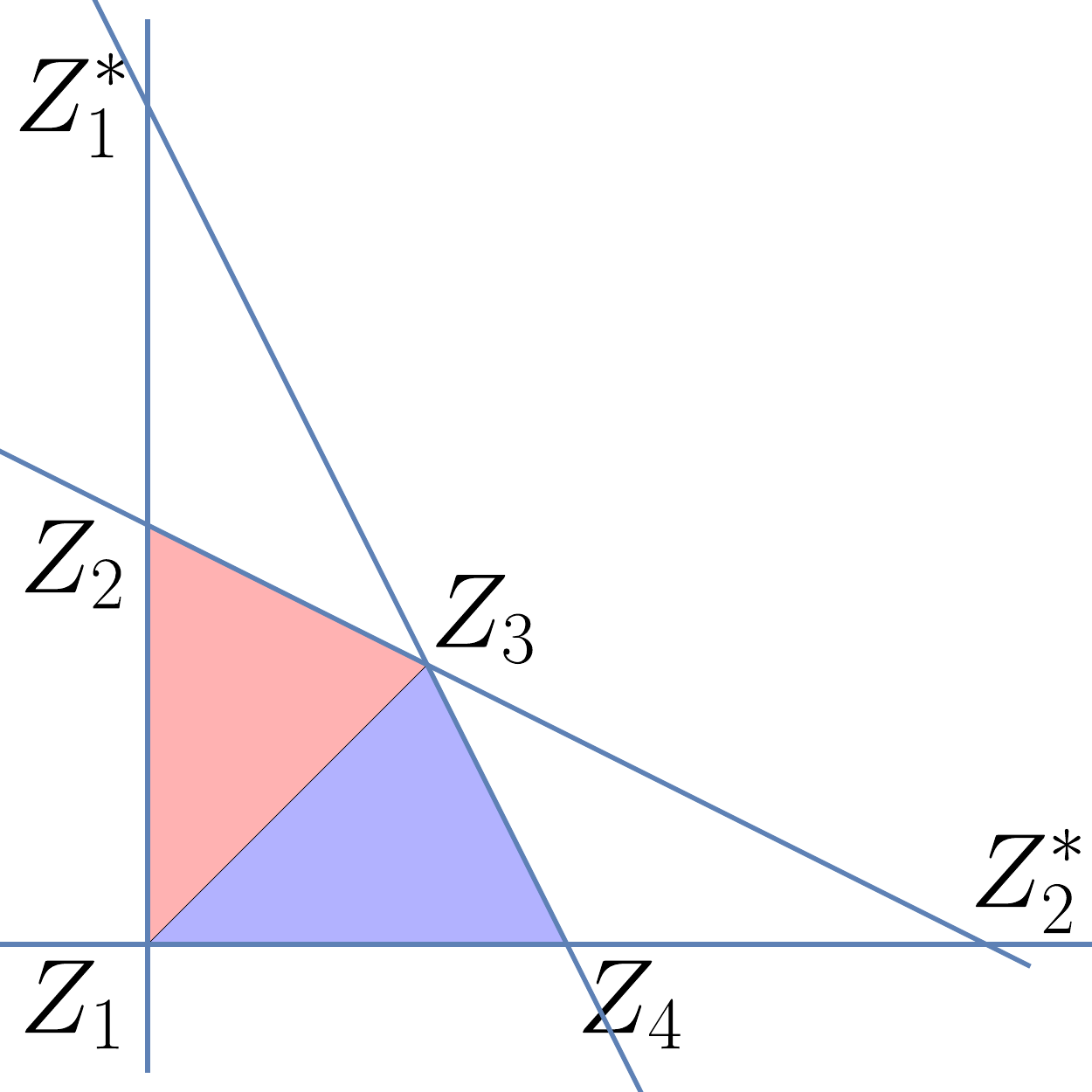}}
    \,,
\end{equation}
in terms of two simplices $\Delta_1\equiv [Z_1,Z_2,Z_3]$ and $\Delta_2\equiv [Z_3,Z_4,Z_1]$ labeled by the vertex coordinates $Z_i$. Specializing the generic simplex form in Eq.~(\ref{eq:proj_simplex_form}) to $N=2$, we find\footnote{We often use the shorthand notation $\ab{Z_iZ_jZ_k} \equiv \ab{ijk}$, et cetera. }
\begin{equation}
\label{eq:poly4_p2_triang_1_form}
\hspace{-2cm}
    \Omega(\mathcal{P}(Z_1,\ldots,Z_4)) = \Omega(\Delta_1) + \Omega(\Delta_2) 
    = \frac{\ab{123}^2\ab{Yd^2Y}}{\ab{Y12}\ab{Y23}\ab{Y31}}
     +\frac{\ab{134}^2\ab{Yd^2Y}}{\ab{Y13}\ab{Y34}\ab{Y41}}\,. 
\end{equation}
Interestingly, the triangulation (\ref{eq:poly4_p2_triang_1}) introduces a spurious one-dimensional boundary, the line $(Z_1Z_3)$, which is absent in the quadrilateral and an artifact of the way we ``chopped'' the polytope into simplices. At the level of the canonical form, this spurious boundary is encoded in the pole $\ab{Y31} = - \ab{Y13}$ which cancels in the sum
\begin{equation}
\label{eq:poly4_p2_form}
\hspace{-1.0cm}
    \Omega(\mathcal{P}(Z_1,\ldots,Z_4)) = \frac{\ab{Yd^2Y} \left[\ab{123}\ab{234}\ab{Y41}+\ab{124}\ab{134}\ab{Y23}\right]}
    {\ab{Y12}\ab{Y23}\ab{Y34}\ab{Y41}}
\end{equation}
by using the Shouten identity in $\mathbb{P}^2$\footnote{In (momentum) twistor space $\mathbb{P}^3$ \cite{Hodges:2009hk}, the Shouten identity is a $5$-term identity $Z_a \ab{b,c,d,e} + \cdots =0$. Generally, these identities are due to the fact that $N+2$ vectors in $\mathbb{P}^N$ satisfy a Gram relation~\cite{Arkani-Hamed:2010pyv}.}
\begin{equation}
    Z_a \ab{b,c,d} + Z_b \ab{c,d,a} + Z_c \ab{d,a,b} + Z_d \ab{a,b,c} = 0\,.
\end{equation}
It is left as an exercise to check the equivalence of Eqs.~(\ref{eq:poly4_p2_triang_1_form}) and (\ref{eq:poly4_p2_form}). Crucially, the pole $\ab{Y31}$ is absent in the form (\ref{eq:poly4_p2_form}) which holds more generally. A given triangulation of a bigger space introduces spurious boundaries that cancel in the full sum. In (\ref{eq:poly4_p2_triang_1}), we discussed the simplest possible case of a pairwise cancellation between two simplices. More generally, it is necessary to combine all sub-geometries $\{\mathcal{A}_a\}$ in order to achieve the spurious-pole cancellation, see e.g.~the third triangulation in Fig.~\ref{fig:poly4_p2_triangulations}. 

Another recurring feature that already appears in the quadrilateral example in Eq.~(\ref{eq:poly4_p2_form}) is the presence of a nontrivial numerator involving $Y$. The purpose of this numerator is to guarantee that the form has support on the correct boundaries. In the quadrilateral example, all codimension-one boundaries are encoded in the $\ab{Yi\,i{+}1}$ poles  (cyclically identified $5\equiv 1$). However, as we can see in (\ref{eq:poly4_p2_triang_1}), there are spurious codimension-two points that do not belong to the geometry where two codimension-one boundaries intersect $Z^\ast_1 = (Z_1 Z_2) \cap (Z_3Z_4)$ and  $Z^\ast_2 = (Z_2 Z_3) \cap (Z_4Z_1)$. The numerator guarantees, that the form (\ref{eq:poly4_p2_form}) does not have support on the corresponding residues where $\ab{Y12}=\ab{Y34}=0$ and $\ab{Y23}=\ab{Y41}=0$. In fact, in this simple case the numerator can be rewritten as 
\begin{equation}
  \ab{Y 1^\ast 2^\ast} \equiv  \left[\ab{123}\ab{234}\ab{Y41}+\ab{124}\ab{134}\ab{Y23}\right]
\end{equation}
As stated multiple times before, triangulations are not unique and different ways of chopping the geometric space $\mathcal{A}$ into simpler building blocks leads to different representations of the canonical form $\Omega(\mathcal{A})$, which, at the end of the day is, however, triangulation independent. As we will see in the later sections, specific triangulations are important for physical applications, such as the Feynman-diagram triangulation of the associahedron for bi-adjoint $\phi^3$ theory or BCFW triangulations of amplituhedra. 

\vspace{-.4cm}
\subsubsection{Other Constructions of Canonical Forms}
\label{subsubsec:other_form_construction}
Besides triangulating positive geometries in terms of simpler geometric building blocks for which the canonical forms are known, for certain examples, there exist alternative methods to determine canonical forms \cite{Arkani-Hamed:2017mur,Ferro:2020ygk}. 
\begin{itemize}
    \item \textbf{Integral representation from dual geometry:} Using projective duality for convex projective polytopes (see e.g.~Appendix A.4 of \cite{Arkani-Hamed:2017mur}), it is possible to determine the canonical differential from as the actual volume of the dual polytope \cite{Hodges:2009hk,Arkani-Hamed:2010wgm} leading to early notions of positivity of amplitudes \cite{Arkani-Hamed:2014dca}. Generalizing the dual geometry of polytopes to Grassmannian and amplituhedron spaces is an interesting open problem. 
    \item \textbf{Push forward:} In certain instances, it is possible to find simpler positive geometries of the same dimension with a map $\phi: \mathcal{A} \to \mathcal{A}'$ between them. Given a form $\omega$ on $\mathcal{A}$, the \emph{push forward} gives a new form $\omega'$ on $\mathcal{A}'$. One reason that the push forward is useful is the conjecture (proven in examples but not in generality) that it preserves canonical forms. For details see e.g.~Appendix A.3 of~\cite{Arkani-Hamed:2017mur}.
    \item \textbf{Direct construction:} Demanding that the only poles of the canonical form are logarithmic in nature and located at the boundaries of the positive geometry under consideration, one can write down an ansatz for the form where the denominator contains all codimension-one boundaries. The numerator is subject to the defining residue constraints of a positive geometry including the fact that the canonical form vanishes on loci that do not correspond to singularities of the space. This approach was pursued e.g.~in \cite{Arkani-Hamed:2014dca}. A toy problem is the construction of the canonical form of the quadrilateral in Eq.~(\ref{eq:poly4_p2_form}) starting with the four codimension-one boundaries in the denominator and an ansatz for the numerator. Demanding that only physical poles survive fixes the form (modulo one overall sign). Similar reasoning was used for amplitudes where no geometric picture has yet emerged which hinted to the more general existence of geometric structures underlying particle scattering \cite{Arkani-Hamed:2014via,Bern:2014kca,Bern:2015ple}.
\end{itemize}

%
\vspace{-.3cm}
\section{ABHY Associahedron and $\phi^3$ Amplitudes}
\label{sec:associahedron}
%
Upon describing some of the mathematical background on positive geometries, canonical forms, and their simplest incarnation in terms of convex polytopes, we begin our study of their applications to physical particle scattering. We first consider the associahedron polytope (which is well known to mathematicians \cite{Stasheff:1963I,Ziegler:1995polytopes,Ceballos:2014aaa}) for bi-adjoint scalar $\phi^3$ theory (ABHY associahedron \cite{Arkani-Hamed:2017mur}), before diving into the ``curvy'' amplituhedron construction of planar $\mathcal{N}=4$ SYM in Section \ref{sec:amplituhedron}. Here, we only discuss some of the fundamental properties of the amplitudes-associahedron relation, which have been understood and generalized in much greater detail: 
$\bullet$ ABHY associahedron and tree-level amplitudes of bi-adjoint $\phi^3$:~\cite{Arkani-Hamed:2017mur}; 
$\bullet$ causal diamonds (in 2d kinematic `spacetime') and cluster polytopes:~\cite{Arkani-Hamed:2019vag,Bazier-Matte:2018rat};
$\bullet$ positive geometry of $\phi^n$ interactions at tree-level:~\cite{Banerjee:2018tun,Salvatori:2019phs,Aneesh:2019cvt,Srivastava:2020dly,Raman:2019utu,Kojima:2020tox,Aneesh:2019ddi,Barmeier:2021iyq};
$\bullet$ positive geometry of $\phi^3\,\&\,\phi^4$ at one-loop level:
    \cite{Salvatori:2018fjp,Salvatori:2018aha,Arkani-Hamed:2019vag,Jagadale:2020qfa,Abhishek:2020sdr};
$\bullet$ mathematical references for associahedra and polytopes:~\cite{Stasheff:1963I,Ziegler:1995polytopes,Ceballos:2014aaa}.

As applicable from the definition of positive geometries in the previous section, one of the crucial ingredients linked to the geometry is a \emph{canonical} differential form with properties summarized in section \ref{sec:pos_geom_primer}. One of the achievements in the work of Arkani-Hamed, Bai, He, and Yan \cite{Arkani-Hamed:2017mur} is the connection of the classical associahedron polytope to ``scattering forms'' for bi-adjoint scalar theory directly in kinematic space.  

\subsection{An Amplitude Invitation}
\label{subsec:associahedron_invite}

We are interested in the tree-level scattering amplitudes of (massless) bi-adjoint scalar $\phi^3$ theory. At the Lagrangian level, this model is specified by
\begin{equation}
    \mathcal{L} = \frac{1}{2} (\partial_\mu \phi_{a A})(\partial^\mu \phi^{a A}) - \lambda \, f^{abc}\tilde{f}^{ABC} \phi_{aA}\, \phi_{bB}\, \phi_{cC}
\end{equation}
where $f^{abc}$ and $\tilde{f}^{ABC}$ are the structure constants of two flavor groups and $a, A$, etc. are their respective adjoint flavor indices. At tree-level, we can perform a color-like-decomposition (see e.g.~\cite{Dixon:1996wi}) into two (single-)trace structures of the respective groups labeled by two permutations $\alpha,\beta \in S_n$ of $n$ legs and amplitudes decompose as
\begin{equation}
\hspace{-1.5cm}
    \mathcal{M}_n = \sum_{\alpha, \beta \in S_n} M_n^{\alpha|\beta}(p_1,\cdots, p_n) \, 
    {\rm tr}[T^{a_{\alpha(1)}}\cdots T^{a_{\alpha(n)}}]\, 
    {\rm tr}[T^{A_{\beta(1)}}\cdots T^{A_{\beta(n)}}]
\end{equation}
where the (color-stripped) double partial amplitudes $M_n^{\alpha|\beta}(p_1,\cdots, p_n)$ depend on the two permutations but are independent of color factors. From a Feynman diagrammatic point of view, only diagrams with compatible color orderings in $\alpha$ and $\beta$ contribute to the answer.  For our discussion, we focus on the case where $\alpha = \beta= (1,2,3,\ldots, n-1,n)$ and leave the generalization to \cite{Arkani-Hamed:2017mur}. 

First, let us give a brief invitation by considering the simplest $n=4$ four-particle amplitude which is given by the sum of two planar Feynman diagrams
\begin{equation}
\hspace{-1.5cm}
    M_4 \equiv M^{(1234)|(1234)}_4 = 
    \vcenter{\hbox{\scalebox{1}{\trees}}}
    + 
    \vcenter{\hbox{\scalebox{1}{\treet}}}
    = \frac{1}{s} + \frac{1}{t}\,.
\end{equation}
Here and in the following, we suppress powers of the three-point coupling constant $\lambda$ and introduced the standard Mandelstam invariants $s\equiv s_{12}=(p_1{+}p_2)^2=s_{34} = (p_3{+}p_4)^2$ and $t\equiv s_{14}=(p_1{+}p_4)^2= s_{23} = (p_2{+}p_3)^2$ in an all-incoming (or all-outgoing) momentum convention. Instead of the amplitude $M_4$, consider the logarithmic one-form 
\begin{equation}
\label{eq:1form_n_4_scattering}
    \hspace{2cm}
    \Omega^{(1)}_{n{=}4} \equiv \frac{ds}{s} - \frac{dt}{t}\,,
\end{equation}
obtained by replacing inverse propagators by $d\log$s of the inverse propagators with the crucial extra ingredient of a relative sign. In our positive geometry primer in Section \ref{sec:pos_geom_primer} we stressed that the relevant geometric objects naturally live in projective space. Indeed, the relative sign in Eq.~(\ref{eq:1form_n_4_scattering}) guarantees that $\Omega^{(1)}_{n{=}4}$ is not only well-defined on the two-dimensional $(s,t)$-space, but also on a projectivized version where the form is invariant under local rescalings $(s,t) \mapsto \Lambda(s,t)\, (s,t)$, where $\Lambda(s,t)\in GL(1)$. Effectively, this means that our one-form can only depend on the ratio $s/t$. At this point, we should ask how we can extract the actual amplitude $M_4$ from the one-form $\Omega^{(1)}_4$ and how exactly this form is related to positive geometry. Both questions can be answered by first identifying some natural regions in kinematic space. If the poles in the amplitude correspond to the boundary of some geometric region, we ought to impose the positivity conditions on $s\geq 0$ and $t\geq0$, which still leaves us with the two-dimensional positive quadrant. The positivity constraints on $s$ and $t$ are not quite enough to describe the relevant geometric space, since our four-particle scattering form $\Omega^{(1)}_{n{=}4}$ is a one-form. We are therefore led to consider a particular restriction of the positive quadrant to a one-dimensional subspace: A natural way for such a restriction is to impose 
\begin{equation}
    s+t = -u = c >0  
    \qquad 
    \longrightarrow
    \qquad
    \raisebox{-40pt}{\includegraphics[scale=.2]{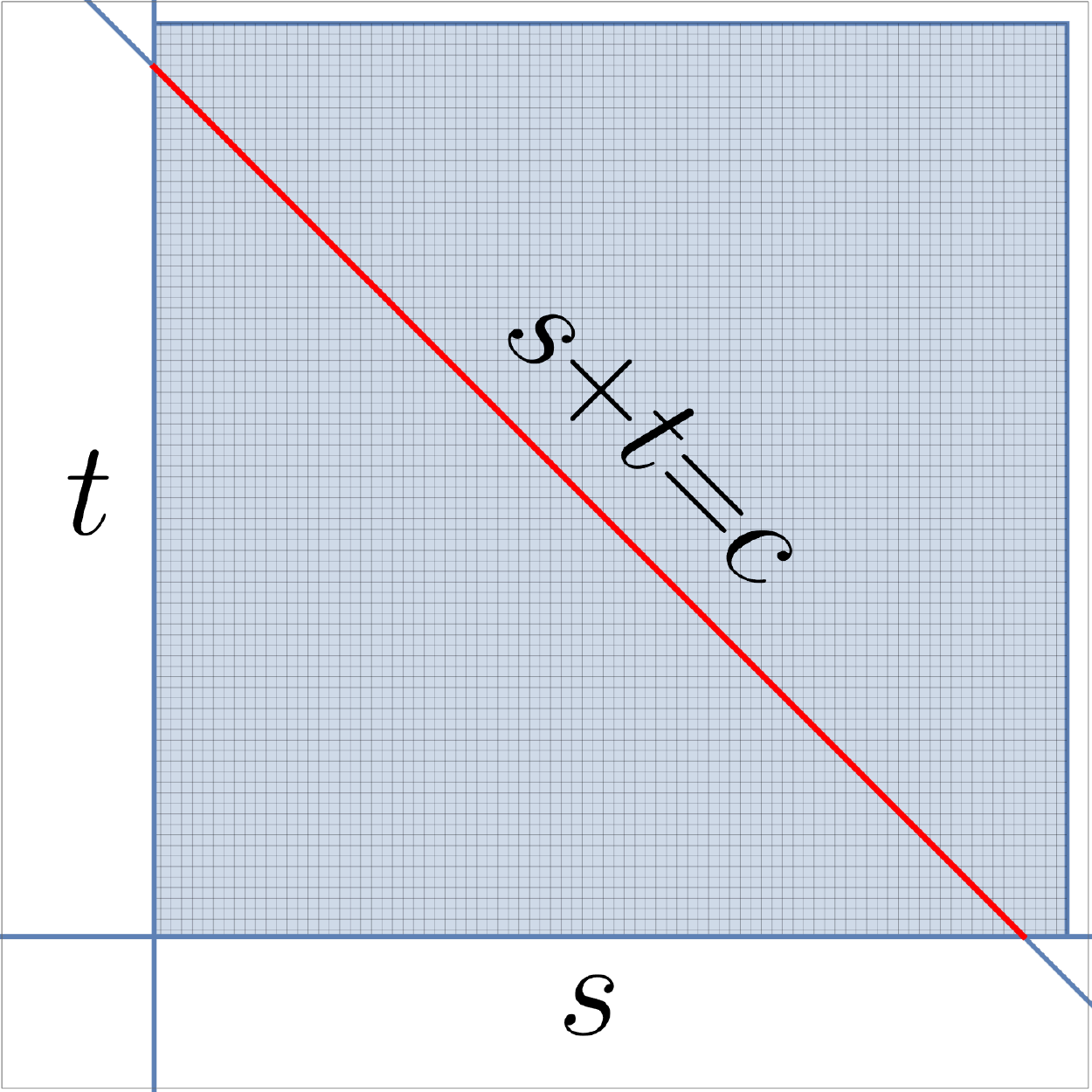}}\,,
\end{equation}
for some positive constant $c$ which sets the nonplanar Mandelstam invariant $u<0$. This additional constraint restricts the two-dimensional space $(s,t)\geq 0$ to the line-segment bounded by $s=0$ and $t=0$. Restricting (``pulling back'') $\Omega^{(1)}_{n{=}4}$ to this subspace yields the canonical form of the line-segment, discussed abstractly around Eq.~(\ref{eq:line_segment_form}). Taking into account\footnote{The Mandelstam variable $u{=}s_{13}$ is dependent and satisfies $s{+}t{+}u{=}0$ for massless external states.} $-u=s+t=c$, we have $ds+dt=0$. The constant $c$ drops out and we obtain
\begin{equation}
\hspace{2cm}
    \Omega^{(1)}_{n{=}4}\Big|_{s+t=c} = ds \left[
    \frac{1}{s} + \frac{1}{t}
    \right] = ds \, M_4\,.
\end{equation}
Factoring out $ds$ (top form on line segment), we exactly find the 4-particle amplitude.

\subsection{Planar Scattering Form on Kinematic Space}
\label{subsec:scattering_form_kinematic_space}

At $n$-points, the simple four-point example generalizes. Each tree-level Feynman diagram has $n-3$ propagators from which we obtain a $(n-3)$-form by wedging together the $d\log$'s of the inverse propagators. The amplitude is obtained by summing over all graphs and the relative signs of the forms are given by demanding projectivity. This $(n-3)$-dimensional space is, again, to be viewed as a particular subspace of $\mathcal{K}_n$---the $n(n-3)/2$-dimensional space of Mandelstam invariants\footnote{We assume a sufficiently large spacetime dimension $D\geq (n-1)$. For $D<(n-1)$ there are additional Gram constraints on the particle momenta.}.  A particularly nice basis for $\mathcal{K}_n$ is given by the set of \emph{all} planar (inverse) propagators $s_{i,i{+}1,\cdots j{-}1}$ for any pair $1\leq i <j\leq n$. One way of representing the kinematic setup is by introducing planar variables (for the standard ordering ($1,2,\ldots,n$) of the particles). We draw the $n$ momenta as a closed convex polygon and label the vertices of the polygon by associate new coordinates, $i={1,\ldots,n}$, and associate $X_{i,j}$ variables to pairs of vertices
\begin{equation}
\hspace{-2.5cm}
    \raisebox{-37pt}{\includegraphics[scale=.22]{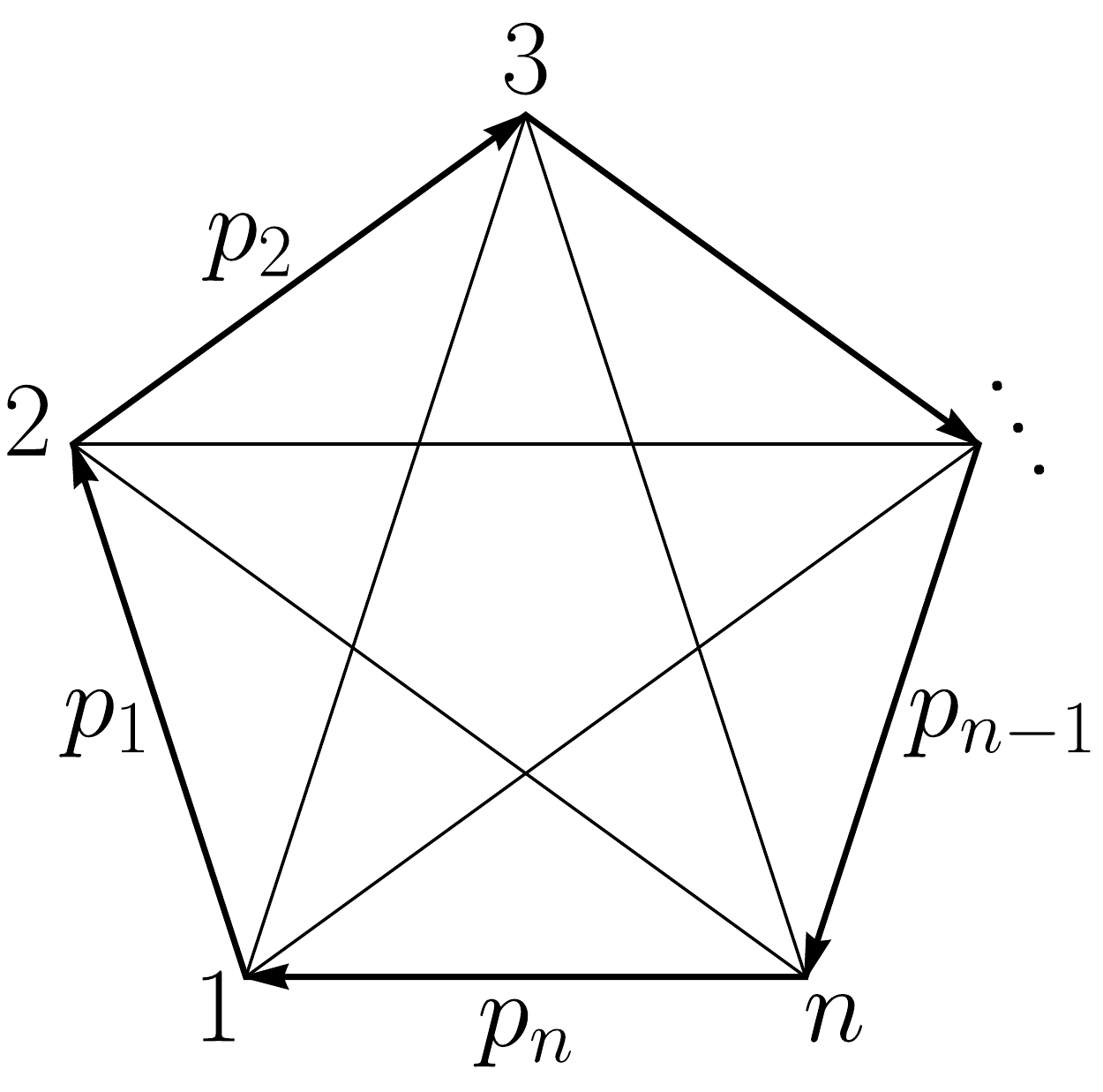}} 
    \hspace{-.3cm}
    \leftrightarrow
    X_{i,j} \equiv (p_i + p_{i+1}+{\cdots}+p_{j-1})^2 = s_{i,i{+}1,\cdots j{-}1}\,,
    {\rm for }\ 1\leq i <j\leq n
\end{equation}
Generalizing the 4-particle example, there is a ``positive region'', $\Delta_n{\subset}\mathcal{K}_n$, within kinematic space where all $n(n{-}3)/2$ nonzero $X_{i,j}$ are positive ($X_{i,i{+}1}{=}0$ and $X_{1,n}{=}0$).

Having defined the kinematic space, let us discuss the \emph{scattering form} in terms of (planar) Feynman diagrams of $\phi^3$ theory. Each cubic $n$-particle Feynman graph $\Gamma_n$ has $(n{-}3)$ inverse propagators $X_{i_a,j_a}$ so that we can assign a $(n{-}3)$ form to each $\Gamma_n$ 
\begin{equation}
\hspace{2cm}
    \Omega^{(n-3)}_{\Gamma_n} \equiv
    {\rm sign}(\Gamma_n) \bigwedge^{n-3}_{a=1} d\log X_{i_a,j_a}\,,
\end{equation}
where ${\rm sign}(\Gamma_n){=}\pm 1$ and flips under the swap of two propagators. The planar scattering form is then given by the sum of all Feynman diagrams
\begin{equation}
\label{eq:amp_scattering_form}
\hspace{2cm}
    \Omega^{(n-3)}_n \equiv \sum_{{\rm planar}\ \Gamma_n} \Omega^{(n-3)}_{\Gamma_n}\,.
\end{equation}
For $n{=}3$, we define the zero-form $\Omega^{(0)}_{n{=}3} \equiv \pm 1$. For general $n$, there are exactly $C_{n{-}2}$ planar cubic tree-level diagrams, where $C_k {=} 1/(k{+}1)$Binomial$(2k,k)$ is the k$^{{\rm th}}$  Catalan number (e.g.~$C_{n-2}=\{1,2,5,14,42\}$ diagrams for $n=\{3,4,5,6,7\}$). In principle, there are two sign choices for each $\Gamma_n$, and therefore a priori many possible scattering forms. To select a unique (up to an overall sign) form, one can demand that $\Omega^{(n-3)}_n$ is projective under local $GL(1)$ rescaling of the $\{X_{i,j}\} \mapsto \Lambda(X) \{X_{i,j}\}$. Summing the Feynman-diagram forms for $n=4,5$ and demanding projectivity, one finds~\cite{Arkani-Hamed:2017mur}
\begin{equation}
\hspace{0cm}
    \Omega^{(1)}_{n{=}4} = d\log \frac{X_{1,3}}{X_{2,4}} = d\log \frac{s}{t}\,,
\end{equation}
\begin{equation}
\hspace{0cm}
    \Omega^{(2)}_{n{=}5} = 
    d\log \frac{X_{1,3}}{X_{2,4}} \wedge d\log \frac{X_{1,3}}{X_{1,4}}
   +d\log \frac{X_{1,3}}{X_{2,5}} \wedge d\log \frac{X_{3,5}}{X_{2,4}}\,.
\end{equation}
It is important to note that projectivity of the form is a highly nontrivial feature which is not true for any subset of Feynman diagrams. 

\subsection{The Associahedron from the Combinatorics of Planar Cubic Graphs}
\label{subsec:associahedron_combinatoric}

After the kinematic warm-up and the discussion of projective logarithmic forms of bi-adjoint $\phi^3$ theory, we would like to discuss the associated positive geometry. As we will see momentarily, the relevant geometric object is well known to mathematicans as the associahedron polytope \cite{Stasheff:1963I}. Since this is a rather old topic in mathematics, there exist a number of combinatorial constructions of associahedra that are summarized in \cite{Ziegler:1995polytopes}. From the point of view of scattering amplitdues, one of the most natural descriptions of the associahedron is given in terms of triangulations of polygons. 
\begin{figure}[ht!]
\vspace{-.5cm}
\begin{subfigure}{.5\textwidth}
\centering
\begin{overpic}[width=5.6cm]
{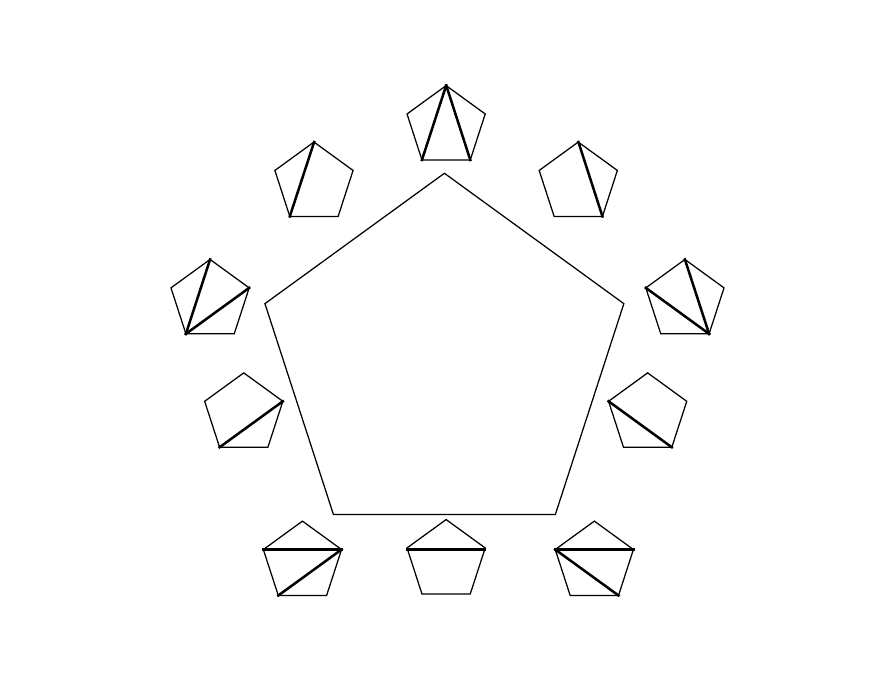}
\end{overpic}
\end{subfigure}
\hspace{-1cm}
\begin{subfigure}{.5\textwidth}
\centering
\begin{overpic}[width=4.4cm]
{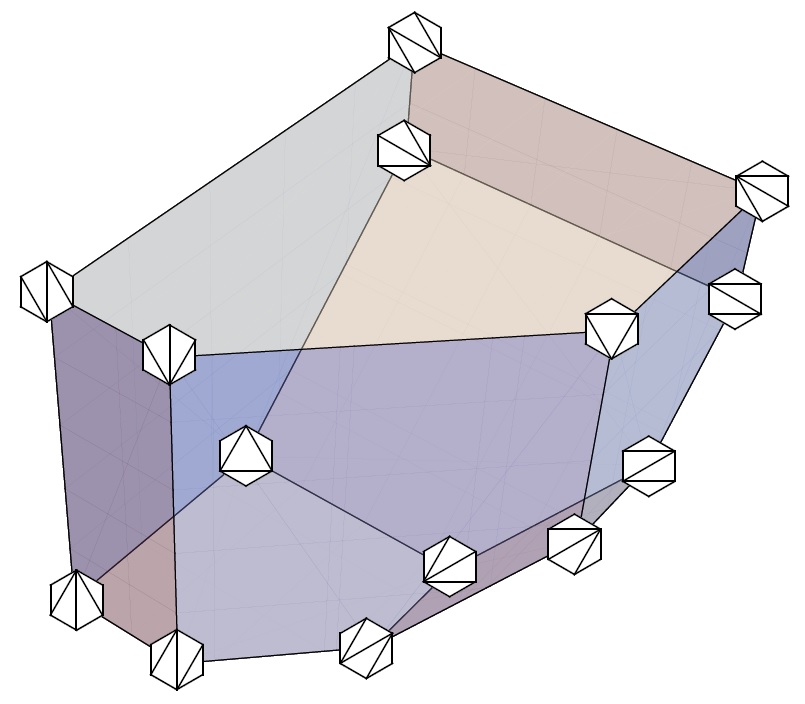}
\end{overpic}
\end{subfigure}
\vspace{-.4cm}
\caption{
\label{fig:associahedron_combinatorial}
Combinatorial structure of the $n{=}5$ (left) and the $n{=}6$ (right) associahedron in terms of (partial) triangulations of $n$-gons adapted from \cite{Arkani-Hamed:2017mur}.} 
\end{figure}

\begin{figure}[ht!]
\centering
\includegraphics[scale=.22]{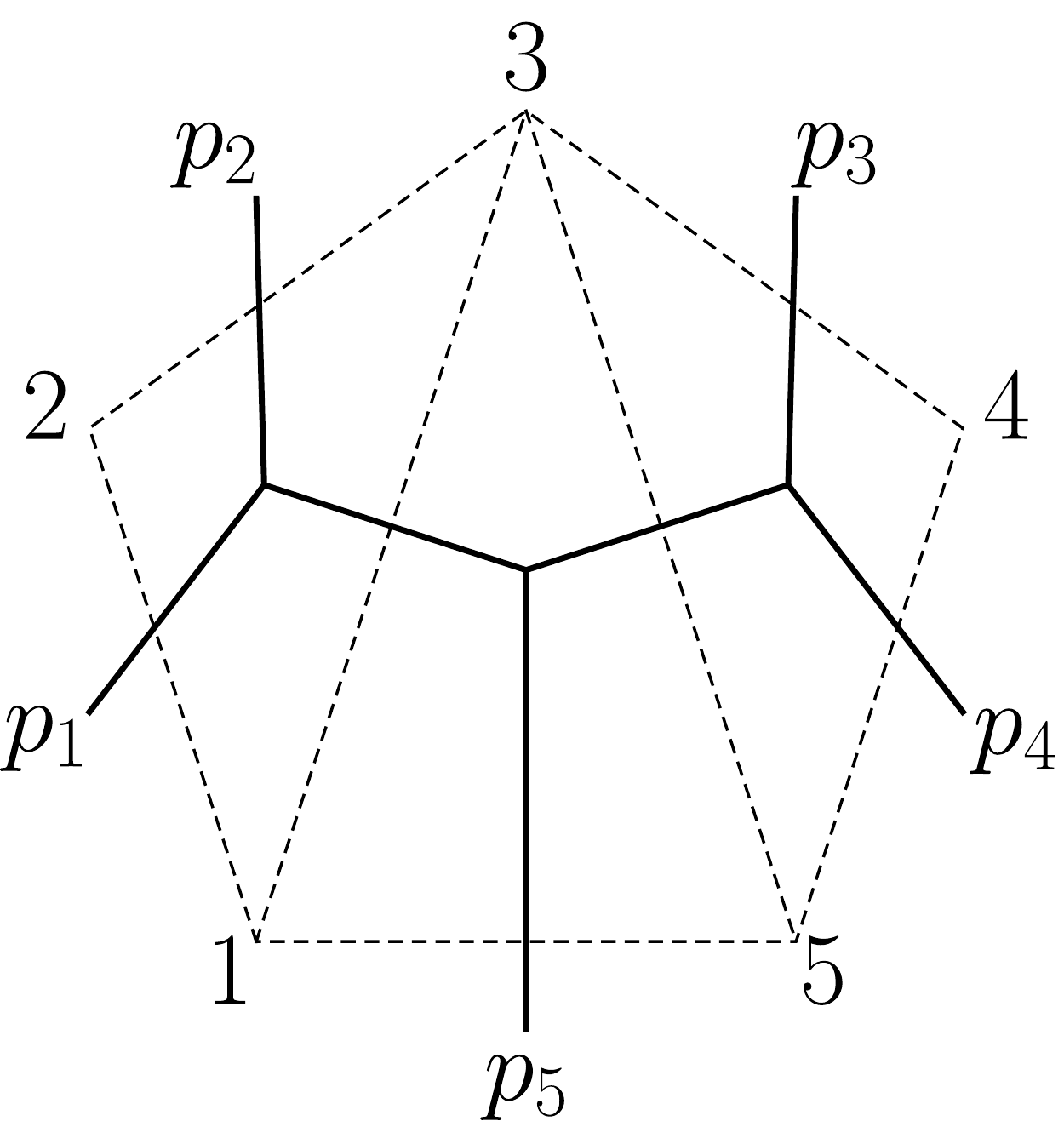}
\quad
\includegraphics[scale=.22]{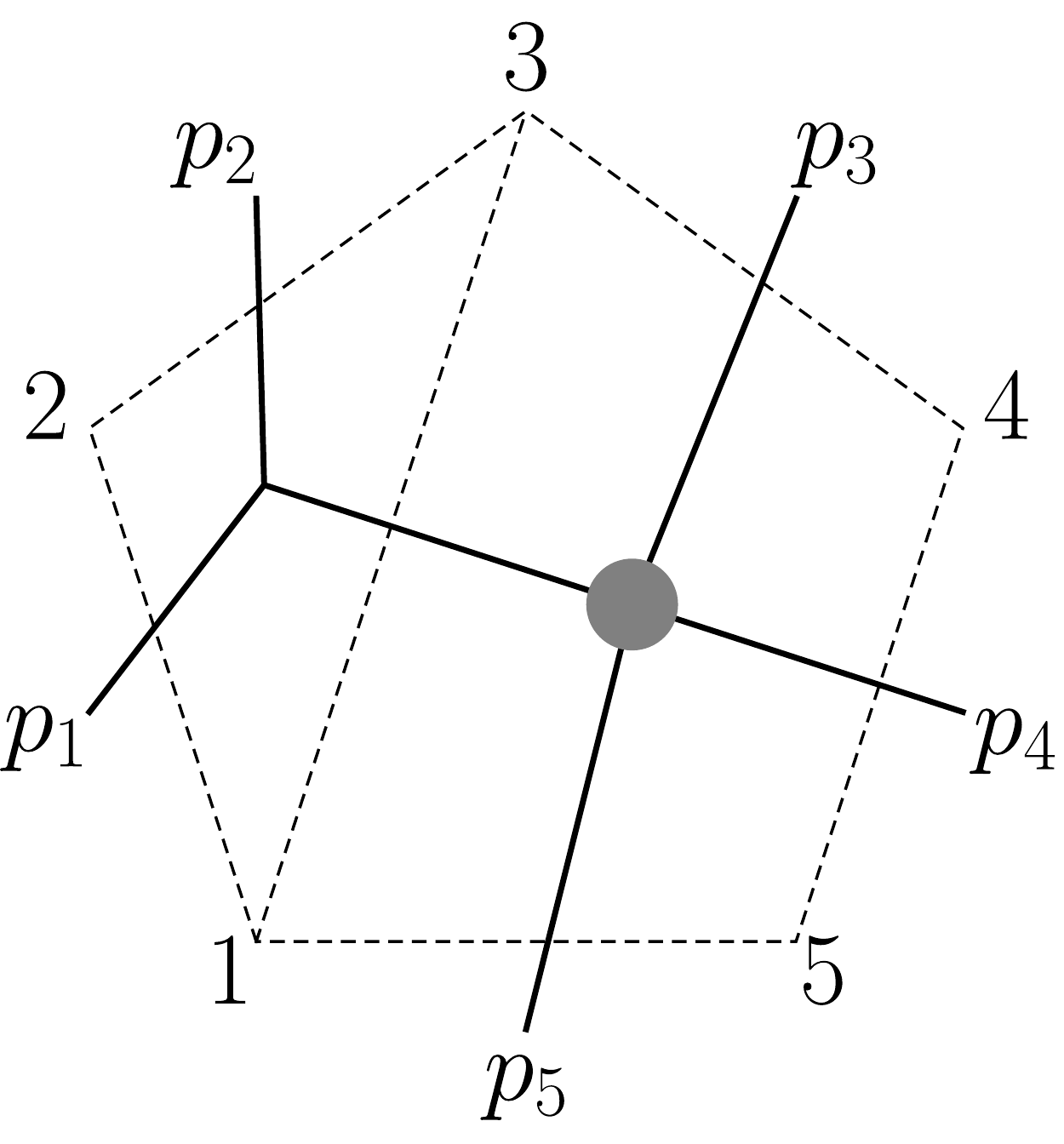}
\quad
\includegraphics[scale=.22]{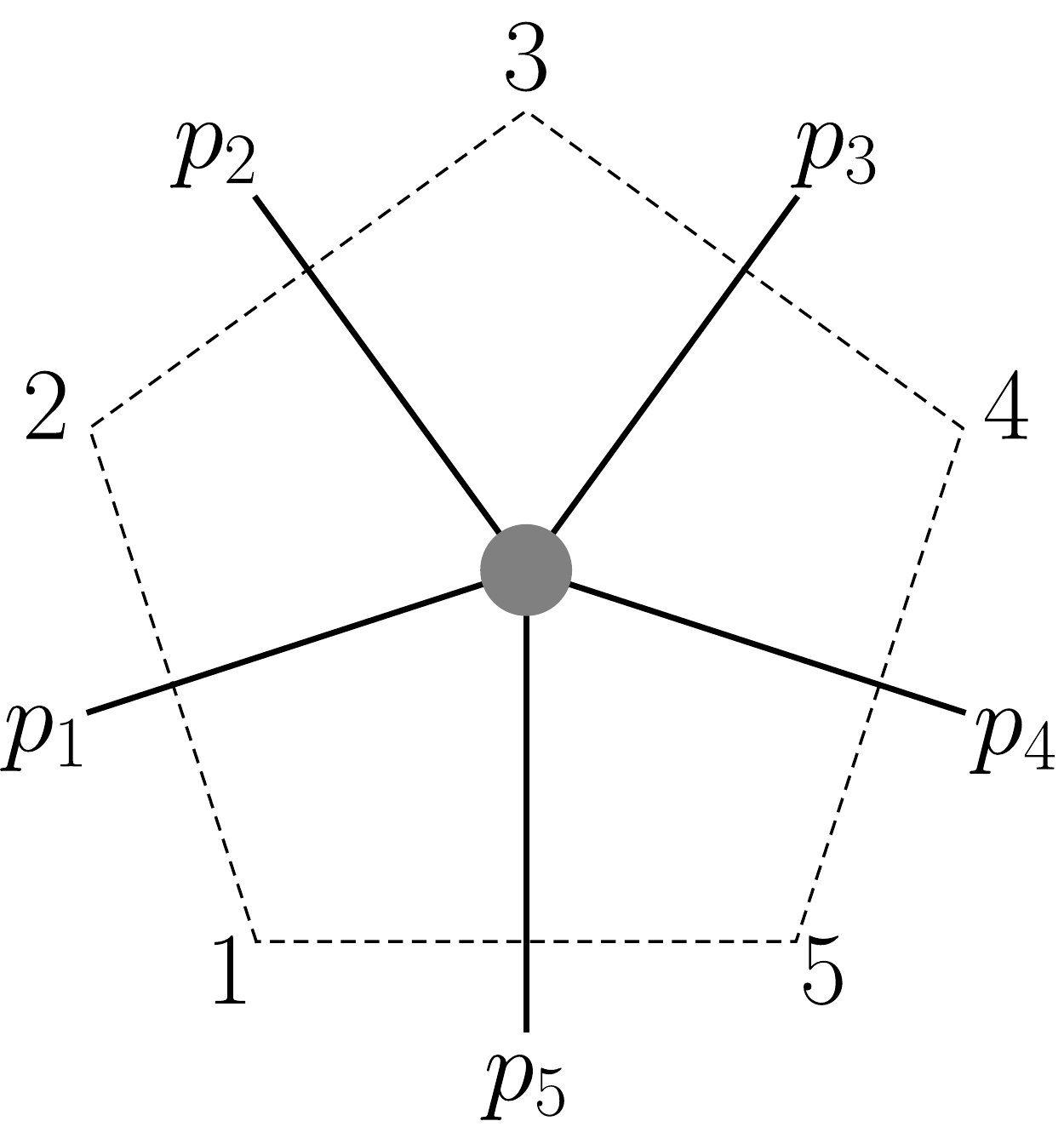} 
\vspace{-.4cm}
\caption{\label{fig:triangulations_5pt} Sample of complete (left) and partial (middle, right) triangulations of a pentagon and their dual Feynman diagrams.}
\end{figure}

In this setup, different triangulations of the polygon correspond to different boundary structures of the associahedron polytope. Here, we require that the ``diagonals'' that make up the (partial) triangulations do not cross (see Sec.3.1 of \cite{Arkani-Hamed:2017mur} for a precise definition). In this language, any convex polytope $\mathcal{A}_n$ of dimension $n-3$ that satisfies the following properties is an \emph{associahedron}~\cite{Arkani-Hamed:2017mur}:
\begin{enumerate}
    \item For every $d=0,1,\ldots,n{-}3$, there exists a one-to-one correspondence between the codimension $d$ boundaries and $d$-diagonal (partial) triangulations of a convex $n$-gon. (Partial triangulations with $d$ chords that chop the $n$-gon into $d+1$ pieces.)
    \item A codimension $d$ boundary $F_1$ and a codimension-$(d+k)$ boundary are adjacent, iff the partial triangulation corresponding to $F_2$ can be obtained from the ones for $F_1$ by adding $k$ additional diagonals 
\end{enumerate}
Particular examples of associahedra for $n=5,6$ are given in Fig.~\ref{fig:associahedron_combinatorial}. The extreme cases of ``triangulations'' without any diagonal correspond to the polytope's interior, whereas (complete) triangulations with $d=(n-3)$ diagonals correspond to the vertices of the associahedron polytope. It is a classic combinatorial result, that the number of complete triangulations (and therefore the number of vertices of the polytope) of an $n$-gon is the Catalan number $C_{n-2}$. Recall, that we encountered the same number when we counted the planar, cubic Feynman diagrams that contribute to tree-level $n$-particle scattering amplitudes in bi-adjoiont $\phi^3$ theory. This is not surprising, once we realize that planar cubic graphs are in one-to-one correspondence to (complete) triangulations of a polygon (see e.g.~LHS of Fig.~\ref{fig:triangulations_5pt}).
Partial triangulations can be drawn by collapsing propagators in the cubic graphs (middle and RHS of Fig.~\ref{fig:triangulations_5pt}). The boundary compatibility of the associahedron polytope corresponds to mutually consistent (planar) factorization channels in the scattering amplitude setup. More generally, factorization of the associahedron is purely combinatorial. Consider some codimension-one boundary (``facet'' $F$) of the associahedron which corresponds to a case where the $n$-gon is divided into an $m$-gon and an $(n-m+2)$-gon (e.g.~middle Fig.~\ref{fig:triangulations_5pt}) which provide the lower-point associahedra $\mathcal{A}_m$ and $\mathcal{A}_{n{-}m{+}2}$. Therefore, $F$ is combinatorically identical to their product $F\simeq \mathcal{A}_m \times \mathcal{A}_{n{-}m{+}2}$. Combinatorially, it follows that scattering amplitudes are not allowed to have iterative singularities in overlapping channels. For example, the five-particle amplitude does not have overlapping poles in $s_{12}$ and $s_{23}$ simultaneously (the diagonals would cross). Similar exclusions of overlapping singularities have played a role in the scattering-amplitudes community via the so-called ``Steinmann'' conditions \cite{Bartels:2008ce,Caron-Huot:2016owq,Dixon:2016nkn}. Historically, these were derived within algebraic QFT in the 1960s \cite{Steinmann} and it would be interesting to study the connection to geometry in more detail.  

\subsection{The Kinematic Associahedron}
\label{subsec:associahedron_kinematic_space}

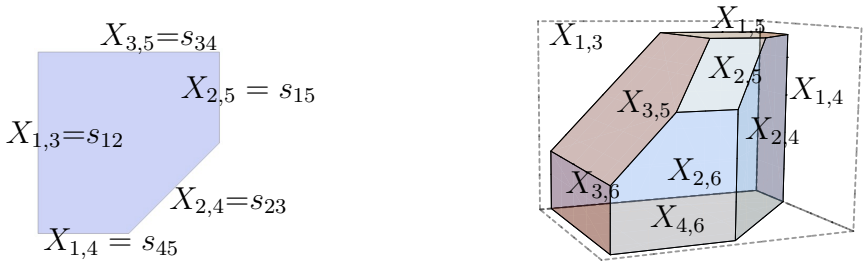
\begin{figure}[hb!]
\begin{subfigure}{.45\textwidth}
\centering
\hspace{-1cm}
\begin{overpic}[width=2.5cm]
{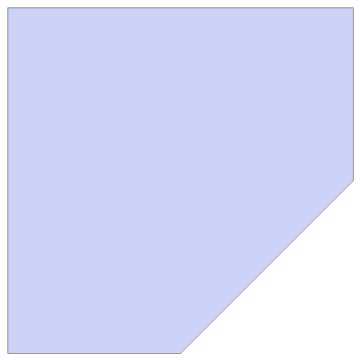}
\put(5,-7){$X_{1,4}=s_{45}$}
\put(-15,50){$X_{1,3}{=}s_{12}$}
\put(35,102){$X_{3,5}{=}s_{34}$}
\put(78,74){$X_{2,5}=s_{15}$}
\put(72,15){$X_{2,4}{=}s_{23}$}
\end{overpic}
\end{subfigure}
\begin{subfigure}{0.45\textwidth}
\centering
\begin{overpic}[width=4.5cm]
{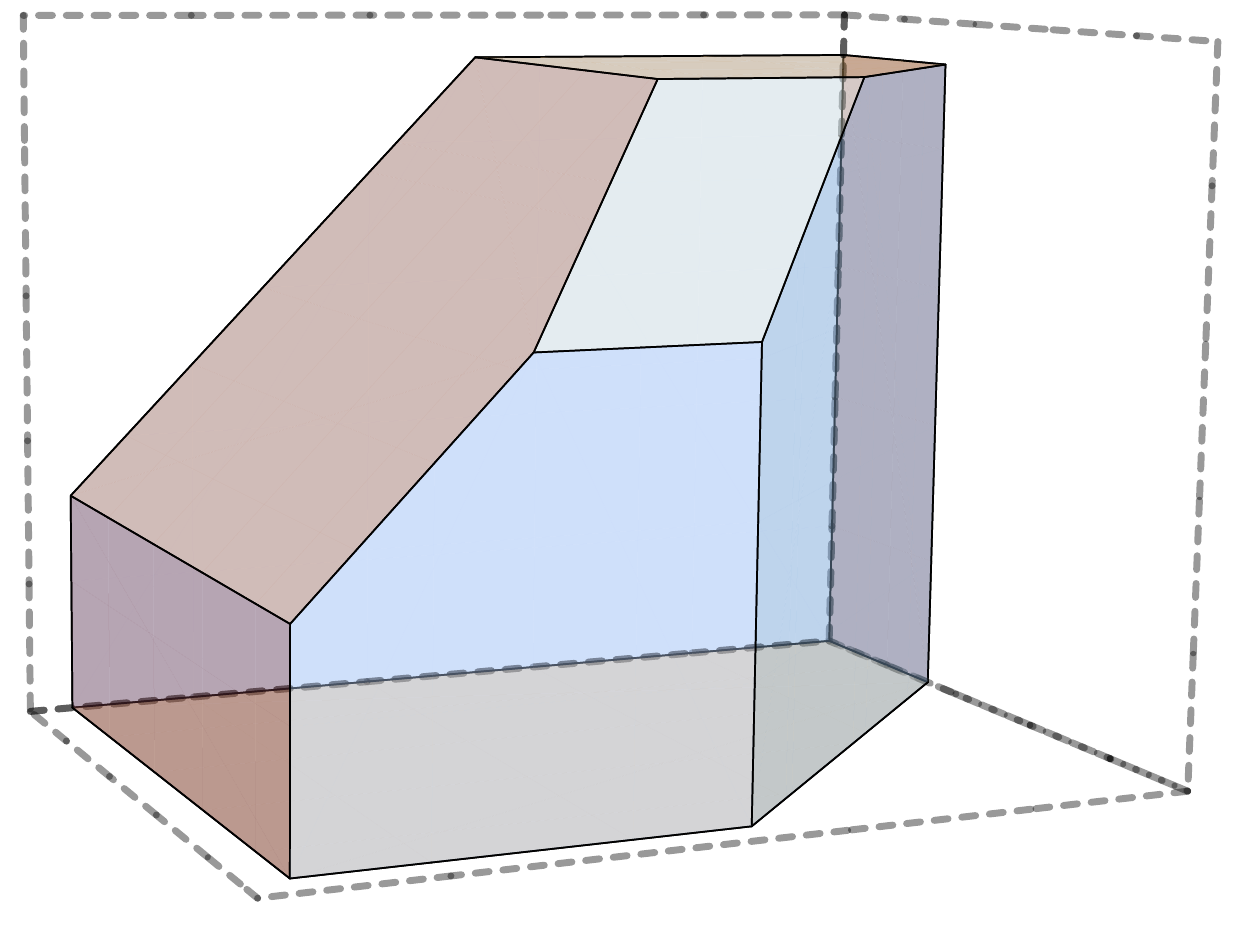}
\put(5,66){$X_{1,3}$}
\put(76,50){$X_{1,4}$}
\put(35,12){$X_{4,6}$}
\put(53,70.9){$X_{1,5}$}
\put(25,46){$X_{3,5}$}
\put(52,56){$X_{2,5}$}
\put(40,26){$X_{2,6}$}
\put(10,22){$X_{3,6}$}
\put(63,38){$X_{2,4}$}
\end{overpic}
\end{subfigure}
\vspace{-.4cm}
\caption{
\label{fig:eg_associahedra}
Adaption of $n{=}5$ (left) and $n{=}6$ (right) associahedra from Ref.~\cite{Arkani-Hamed:2017mur}, where every facet is labeled by the corresponding vanishing planar Mandelstam invariant.}
\end{figure}

We are now introducing the ABHY construction \cite{Arkani-Hamed:2017mur} of the associahedron in kinematic space. We have seen in Section \ref{subsec:scattering_form_kinematic_space}, that the $n(n{-}3)/2$-dimensional positive kinematic space $\Delta_n\subset \mathcal{K}_n$ is too big and we are looking for some $(n-3)$-dimensional subspace that determines the scattering forms (\ref{eq:amp_scattering_form}). Interestingly, there is a natural $(n-3)$ dimensional subspace that is carved out by the following equations \cite{Arkani-Hamed:2017mur} 
\begin{equation}
\label{eq:associahedron_restriction}
   - s_{ij} = c_{ij} > 0\,, \quad {\rm for }\ i,j\ \neq n\,, {\rm non-adjacent}\,,  
\end{equation}
for positive constants $c_{ij}$ (which enforces the negativity of the associated nonplanar Mandelstam invariants) as a generalization of $-u=-s_{13} > c$. In terms of the dual $X_{i,j}$ variables, we have the relation
\begin{equation}
\label{eq:sij_via_xij}
    s_{ij} = X_{i,j+1} + X_{i+1,j} - X_{i,j} - X_{i+1,j+1}\,.
\end{equation}
The $(n-2)(n-3)/2$ equalities in Eq.~(\ref{eq:associahedron_restriction}) select a hyperplane in kinematic space such that the $n(n{-}3)/2-(n{-}2)(n{-}3)/2 = \mathbf{(n{-}3)}$-dimensional intersection with the positive region is the kinematic \emph{associahedron polytope} $\mathcal{A}_n$ \cite{Arkani-Hamed:2017mur}. (The explicit values of the $c_{ij}$ influence the parametric representation of the resulting space but do not influence the boundary structure.) Examples for the five and six-particle associahedra are given in Fig.~\ref{fig:eg_associahedra}. In order to guarantee the positivity of the non-adjacent $-s_{ij}=c_{ij}$, there is a surprising interpretation in terms of solving wave equations in a two-dimensional kinematic space of $X_{i,j}$ \cite{Arkani-Hamed:2019vag} which gives rise to certain cluster polytopes. For more details of this connection, we refer the interested reader directly to Ref.~\cite{Arkani-Hamed:2019vag}.
As noted in Section \ref{subsec:associahedron_combinatoric}, one key feature of the associahedron is the combinatorial factorization on each facet into the product of two lower-dimensional associahedra which also holds for the kinematic associahedron $\mathcal{A}_n$ (see Sec.~4.1 of \cite{Arkani-Hamed:2017mur} for details). One important point to show the factorization structure of $\mathcal{A}_n$ is to realize that the positivity conditions on the planar variables $X_{i,j}\geq 0$ allow us to reach any codimension-one boundary by setting a planar variable to zero, $X_{i,j}\to 0$. However, to reach a lower-dimensional boundary, we can not set any $X_{k,l}$ to zero at the same time for which the corresponding diagonals in the polygon triangulation cross (see e.g.~(\ref{eq:dissallowed_eg})). Let us see this property from the definition of the kinematic associahedron. Starting from $c_{a,b}>0$ in Eq.~(\ref{eq:associahedron_restriction}) and plugging in the definition of the Mandelstam invariants in terms of $X_{i,j}$ (\ref{eq:sij_via_xij}) and summing over $a,b$ in the range $i\leq a<j$ and $k\leq b<l$, several terms cancel telescopically and one finds
\begin{equation}
    X_{j,k}+X_{i,l} = X_{i,k}+X_{j,l} - \sum_{i\leq a < j \,,k\leq b <l} c_{a,b}\,.
\end{equation}
In a situation where we set crossing diagonals $X_{i,k}=0 = X_{j,l}$ to zero
\begin{equation}
\label{eq:dissallowed_eg}
\hspace{-1cm}
\raisebox{-40pt}{\includegraphics[scale=.2]{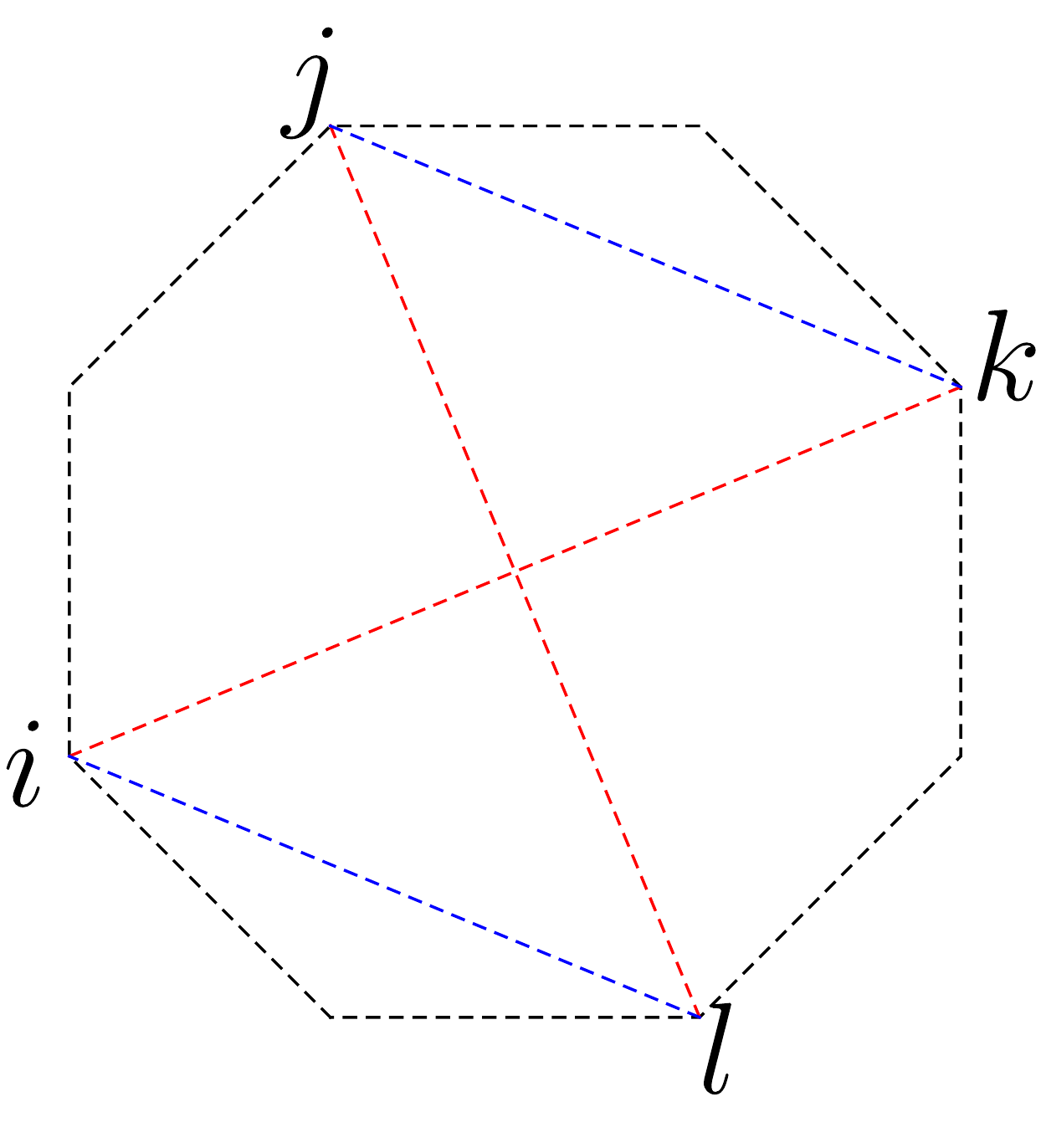}}
\quad \leftrightarrow \quad
   X_{j,k}+X_{i,l} = 0 + 0 - \sum_{i\leq a < j ,\  k\leq b <l} c_{a,b}\,, 
\end{equation}
we derive a contradiction where the LHS is non-negative and the RHS is strictly negative. We therefore conclude that every boundary of $\mathcal{A}_n$ is labeled by a set of non-crossing diagonals, just what was ordered for the associahedron. To understand these kinematic associahedra more explicitly, let's look at the $n=5$ example, depicted on the LHS of Fig.~\ref{fig:eg_associahedra}. The kinematic space $\mathcal{K}_5$ is 5-dimensional, but there are three constraints on the non-planar Mandelstam invariants $-s_{13}=c_{13}>0,-s_{1,4}=c_{14}>0$ and $-s_{24}=c_{24}>0$. If we parameterize the resulting two-dimensional space by $(X_{1,3},X_{1,4})$, the kinematic associahedron is the space $\mathcal{A}_5$ carved out by the inequalities
\begin{equation}
    X_{1,3} \geq 0 \,, \quad 
    X_{1,4} \geq 0 \,, \quad 
    X_{2,4} = X_{1,4} -X_{1,3} + c_{13} \geq 0\,,
\end{equation}
\begin{equation}
    X_{2,5} = -X_{1,3} + c_{13} + c_{14} \geq 0\,,\quad
    X_{3,5} = -X_{1,4} + c_{14} + c_{24} \geq 0\,.
\end{equation}
The last remaining step to complete the connection between the kinematic associahedron $\mathcal{A}_n$ and scattering amplitudes in bi-adjoint $\phi^3$ theory is to show how the scattering amplitude is related to the canonical form on $\mathcal{A}_n$. Since the associahedron is a (simple) polytope (See appendix A of \cite{Arkani-Hamed:2017mur} for details.) its canonical form is given by a sum over vertices, where for each vertex $Z$, there are exactly $(n-3)$ adjacent faces corresponding to $X_{i_a,j_a}{=}0$ so that the canonical form becomes
\begin{equation}
\label{eq:associahedron_form}
    \Omega(\mathcal{A}_n) = \sum_{{\rm vertex\ } Z} {\rm sign} (Z) \bigwedge^{n-3}_{a=1} d\log X_{i_a,j_a}
\end{equation}
where the sign$(Z)=\pm 1$ is determined geometrically from the relative orientation of the facets. Since $\Omega(\mathcal{A}_n)$ is defined on the $n-3$ dimensional subspace $H_n$ obtained by imposing the constraints (\ref{eq:associahedron_restriction}). To see the equivalence between the pullback of the scattering from (\ref{eq:amp_scattering_form}) to the subspace $H_n$ and the canonical form on the associahedron (\ref{eq:associahedron_form}), we have to remember the one-to-one correspondence between planar cubic graphs and vertices $Z$ of the associahedron (complete triangulations of the $n$-gon) outlined in Section \ref{subsec:associahedron_combinatoric}. One immediately realizes that the graph $\Gamma_n$ and the associated vertex $Z$ contain the same propagators $X_{i_a,j_a}$ and the only subtlety concerns the relative signs sign$(\Gamma_n)$ and sign$(Z)$ in Eqs.~(\ref{eq:amp_scattering_form}) and (\ref{eq:associahedron_form}), respectively. It is possible to show that both the projective sign choice in (\ref{eq:amp_scattering_form}) and the geometric one in (\ref{eq:associahedron_form}) are equivalent by studying certain mutation rules. For details, see Sec.~3.3 in \cite{Arkani-Hamed:2017mur}. One finds
\begin{equation}
    \Omega(\mathcal{A}_n) = \left[\sum_{{\rm planar\ }\Gamma_n} \frac{1}{\prod^{n-3}_{a=1}X_{i_a,j_a}}\right] d^{n-3}X = M_{n}\  d^{n-3}X\,.
\end{equation}
%

\section{Amplituhedron and Planar $\mathcal{N}=4$ SYM Amplitudes}
\label{sec:amplituhedron}

Historically, in the context of scattering amplitudes, the first example of a positive geometry in the literature is actually the more sophisticated ``curvy'' realization of the amplituhedron for planar ${\cal N}=4$ SYM theory. The original definition was discovered in some auxiliary space in Refs.~\cite{Arkani-Hamed:2013jha,Arkani-Hamed:2013kca}, which built on earlier work on positive Grassmannians \cite{Postnikov:2006kva,Arkani-Hamed:2009nll,Arkani-Hamed:2009kmp,Arkani-Hamed:2009pfk,Arkani-Hamed:2012zlh}, on-shell diagrams \cite{Arkani-Hamed:2012zlh}, and the realization of the BCFW recursion relation \cite{Britto:2004ap,Britto:2005fq,Arkani-Hamed:2010zjl}. In this framework, the amplituhedron space is a certain projection from the top cell of the positive Grassmannian $G_+(k,n)$, and the explicit expressions for tree-level amplitudes and loop integrands can be obtained from the $d\log$ form on this space. (The twistor Wilson loop also provides a $d\log$ form, as shown in \cite{Lipstein:2012vs}, but this description appears to be inequivalent to the amplituhedron beyond NMHV, as shown in \cite{Heslop:2018nht}.) This picture was later reformulated using topological winding properties \cite{Arkani-Hamed:2017vfh}, directly in momentum twistor space \cite{Hodges:2009hk}, the relevant kinematic space of planar scattering amplitudes in ${\cal N}=4$ SYM. The amplituhedron is a direct generalization of the simpler associahedron for bi-adjoint $\phi^3$ theory discussed in the previous section. At the moment, the amplituhedron construction is our only realization of positive geometry to all loop orders and has been used to obtain a number of all-loop results either by explicit triangulations or other methods.

Scattering amplitudes in ${\cal N}{=}4$ SYM theory can be packaged into superamplitudes for on-shell chiral superfields $\Phi$ (see e.g.~\cite{Nair:1988bq,Elvang:2013cua,Henn:2014yza} for more details),
\begin{equation}
\hspace{-2cm}
    \Phi = G^+ 
        + \twEta_A \Gamma^A 
        + \frac{1}{2}\twEta_A\twEta_B\, S^{AB} 
        + \frac{1}{6}\twEta_A\twEta_B\twEta_C\, \epsilon^{ABCD}\, \overline{\Gamma}_D 
        + \frac{1}{24} \twEta_A\twEta_B\twEta_C\twEta_D\, \epsilon^{ABCD}\, G^-
\end{equation}
where $\twEta_A$ are Grassmann odd variables, $G^\pm$ are positive and negative helicity gluons, $\Gamma$, $\overline{\Gamma}$ are fermions and $S$ are scalars. The so-called $R$-symmetry indices $A,B,C,D$ run from $1,\cdots, 4$ and label different states in the supermultiplet. We calculate $n$-particle superamplitudes ${\cal A}_n(\Phi_1,\dots,\Phi_n)$ which contain component amplitudes of all particles in the multiplet, including the purely gluonic sector, that can be extracted by simple manipulations (see e.g.~\cite{Bourjaily:2010wh,Bourjaily:2012gy}). The amplitude $\mathcal{A}_n$ breaks into helicity components,  $ {\cal A}_n = \sum_{k=2}^{n{-}2} {\cal A}_{n,k},$ where $\mathcal{A}_{n,k}$ has Grassmann degree $4k$ and is multiplied by $\eta^{4k}$---we refer to it as N$^{k-2}$MHV superamplitude. In the purely gluonic sector, $A_{n,k}$ is an amplitude of $k$ negative helicity and $n{-}k$ positive helicity gluons. 

At tree-level, we can decompose ${\cal A}_{n,k}$ into single trace color sectors,
\begin{equation}
\hspace{-1.5cm}
    {\cal A}_{n,k} = \sum_{\sigma \in S_n/Z_n}
    A_{n,k}(\sigma(1),\sigma(2),\dots,\sigma(n))\, 
    {\rm tr}(T^{a_{\sigma(1)}}T^{a_{\sigma(2)}}\dots T^{a_{\sigma(n)}})
\label{color}
\end{equation}
where $A_{n,k}$ is a \emph{color-ordered} amplitude. The sum is over all permutations of the $n$ legs modulo cyclic shifts. At loop-level, this decomposition is modified due to the presence of multiple trace terms. Nonetheless, in the planar limit for $N\rightarrow\infty$ the single trace dominates and we can expand the $L$-loop amplitude ${\cal A}_{n,k}^{(L-{\rm loop})}$ in the same way as (\ref{color}), see e.g.~\cite{Dixon:1996wi}. The planar limit allows us to define global loop variables, thereby removing the usual ambiguity in assigning loop momenta in Feynman integrals. Hence, we can define a unique \emph{loop integrand} ${\cal I}_{n,k}^{(L)}$ \cite{Arkani-Hamed:2010zjl},
\begin{equation}
\hspace{.5cm}
    A_{n,k}^{(L-{\rm loop})} = \int d^4\ell_1\dots d^4\ell_L\, {\cal I}_{n,k}^{(L-{\rm loop})}.
\end{equation}
The planar loop integrand ${\cal I}_{n,k}^{(L)}$ is a rational function of external and loop momenta, and can be most conveniently written using momentum twistor variables. 

The general $n$-point kinematics is captured by $n$ ordered momentum twistors $z_i$\footnote{We reserve lower-case $z_i$ for ordinary momentum twistors. Later on, we also introduce momentum supertwistors $\mathcal{Z}_i$ which include Grassmann variables, and in the context of the amplituhedron, we have bosonized momentum twistors $Z_i$ whose size depends on the $k$-charge.} which are points in the projective space $\mathbb{P}^3$. The $k^{{\rm th}}$ loop momentum $\ell_k$ is represented by a line ${\cal L}_k=(AB)_k$ in $\mathbb{P}^3$, where $A,B$ are two arbitrary points on this line. Momentum twistor variables are unconstrained, this means that for any choice of the $z_i$ the corresponding spinor helicity variables $\lambda_i$, $\widetilde{\lambda}_i$ satisfy momentum conservation (see e.g.~\cite{Bourjaily:2010wh} for a detailed discussion). We can define \emph{four-brackets} which are $(4\times4)$ determinants build from momentum twistors,
\begin{equation}
\hspace{-1cm}
    \la ijkl\ra \equiv \epsilon_{ABCD}\,  z_i^A z_j^B z_k^C z_l^D\,,
    \quad {\rm where, }\quad  A,B,C,D=\{1,2,3,4\}.
\end{equation}
For tree-level amplitudes and loop integrands we can then write,
\begin{equation}
    A_{n,k} = \frac{\delta^4(P)\delta^8(\Q)}{\la12\ra\la23\ra\dots \la n1\ra} \times R_{n,k}
\end{equation}
where $R_{n,k}$ is a rational function of four-brackets, $\delta^4(P)$ represents overall momentum conservation, and $\delta^8(\Q)$ is a super-momentum conserving delta function, see e.g.~\cite{Elvang:2013cua,Henn:2014yza}. Explicit scattering amplitudes in planar $\mathcal{N}=4$ SYM, revealed that the theory possesses several special features and a novel hidden symmetry---dual conformal invariance (DCI)~\cite{Drummond:2006rz,Alday:2007hr,Drummond:2008vq}. The dual conformal symmetry acts as $SL(4)$ on momentum twistors under which $R_{n,k}$ is invariant. In fact, the symmetry enlarges to the dual \emph{super}conformal symmetry and together with the ordinary superconformal symmetry of ${\cal N}=4$ SYM it closes to the infinite dimensional Yangian symmetry \cite{Drummond:2009fd}. 

\subsection{On-Shell Diagrams and Positive Grassmannian}

As is taught in virtually all introductory QFT textbooks, perturbative scattering amplitudes can be calculated using Feynman diagrams. This is a very universal method but there also is a significant drawback---Feynman diagrams do not manifestly preserve gauge invariance. Furthermore, they can also obscure non-trivial features and symmetries of amplitudes in particular theories. For example, both the simplicity and the manifest dual conformal symmetry of planar ${\cal N}=4$ SYM amplitudes are not visible in the Feynman diagram expansion. There are alternative methods to Feynman diagrams which are not only computationally efficient, but also exhibit underlying mathematical structures and symmetries. The most prominent method is the Britto-Cachazo-Feng (BCFW) recursion relation \cite{Britto:2004ap,Britto:2005fq}, which reconstructs tree-level amplitudes from lower-point amplitudes using tree-level unitarity. In planar ${\cal N}=4$ SYM, the BCFW recursion relations were extended to all-loop integrands \cite{Arkani-Hamed:2010zjl}.

The individual terms in the BCFW recursion relation can be represented as \emph{on-shell diagrams}---gauge-invariant objects built from (on-shell) three-point amplitudes \cite{Arkani-Hamed:2012zlh}. For massless particles the kinematics is governed by spinor-helicity variables, $p_i^\mu = \sigma^\mu_{\alpha\dot{\alpha}}\lambda^\alpha_i\widetilde{\lambda}^{\dot{\alpha}}_i$, where the null-momentum vector is related to two-dimensional spinors ($\alpha,\dot \alpha = 1,2$) via the Pauli-matrices $\sigma$ \cite{Xu:1986xb,Elvang:2013cua,Henn:2014yza}. The three-point kinematics is very restrictive and allows for two non-trivial solutions to the on-shell conditions and momentum conservation. For the gluon-amplitudes in Yang-Mills theory we have two elementary amplitudes with MHV $(+--)$ or $\MHVbar$ $(-++)$ helicity configuration. In the maximally supersymmetric case, these gluonic amplitudes are embedded in the MHV, resp. $\MHVbar$ superamplitudes (see e.g.~\cite{Drummond:2008vq}) 
\begin{equation}
\hspace{-1cm}
	\raisebox{-26pt}{\includegraphics[scale=0.6]{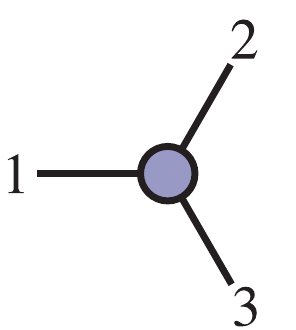}}
	\hskip -.5cm
	=\frac{\delta^4(P)\delta^8(\Q)}{\ab{12}\ab{23}\ab{31}}\,,
	\qquad\qquad
	\raisebox{-26pt}{\includegraphics[scale=0.6]{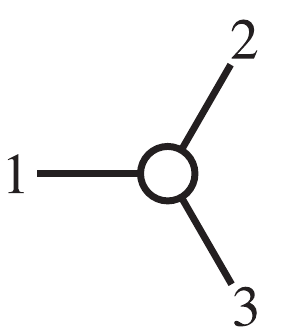}} 
	\hskip -.5cm 
	= \frac{\delta^4(P)\delta^4(\tw{\Q})}{[12][23][31]}\,,
\label{ThreeP}
\end{equation}
where $\ab{ij} = \epsilon_{\alpha\beta} \lam{i}^{\alpha}\lam{j}^{\beta}$ and $[ij] = \epsilon_{\dot{\alpha}\dot{\beta}}\lamt{i}^{\dot{\alpha}}\lamt{j}^{\dot{\beta}}$. The arguments of the respective delta-functions in (\ref{ThreeP}) are given by (neglecting all spinor- and $SU(4)\ R$-symmetry indices),
\begin{equation}
\hspace{-2cm}
P 
= \lam{1}\lamt{1} + \lam{2}\lamt{2} + \lam{3}\lamt{3},
\ 
\Q 
= \lam{1}\tw{\eta}_1 + \lam{2}\tw{\eta}_2 + \lam{3}\tw{\eta}_3, 
\  
\widetilde{\Q} = [12]\twEta_3 + [23]\twEta_1 +[31]\twEta_2\,.
\end{equation}
The general on-shell diagram is given by the product of three-point amplitudes where all legs, both external and internal, are on-shell. Any on-shell diagram has the interpretation as the cut of $L$-loop $n$-point N$^{k{-}2}$MHV loop integrand. The helicity index $k$ is $k = 2B + W - 4L$, where $B$ is the number of blue (MHV) vertices, $W$ is the number of white ($\overline{\rm MHV}$) vertices, and $L$ is the number of loops inside the on-shell diagram. 

Planar On-shell diagrams are known in mathematics as planar bi-partite (plabic) graphs \cite{Postnikov:2006kva}, and have a number of very interesting properties. In physics, we can also generalize beyond the planar limit, see e.g.~\cite{Arkani-Hamed:2014bca,Franco:2015rma,Frassek:2016wlg,Herrmann:2016qea}. Each diagram represents a cell in the Grassmannian $G(k,n)$ and describes a $k$-plane in $n$ dimensions which can be represented by a $(k\times n)$ matrix $C$
\begin{equation}
\hspace{-2cm}
    C = \left(
    \vec{C}_1 \, \vec{C}_2 \, \vec{C}_3 \, \cdots \, \vec{C}_{n-1} \, \vec{C}_n
    \right) 
     = \left(
    	\begin{array}{cccccc} 
        		C_{11} & C_{12} & C_{13} & \dots & C_{1\, n{-}1} & C_{1n} \\
        		C_{21} & C_{22} & C_{23} & \dots & C_{2\, n{-}1} & C_{2n} \\
        		\vdots & \vdots & \vdots & \ddots & \vdots & \vdots \\
        		C_{k1} & C_{k2} & C_{k3} & \dots &  C_{k\, n{-}1} & C_{kn}
     	\end{array}
     \right)
     \,,
\end{equation}
modulo $GL(k)$ symmetry. Planar on-shell diagrams are related to the positive Grassmannian $G_+(k,n)$ \cite{Arkani-Hamed:2012zlh}. The positive part, $G_+(k,n)$, corresponds to a subspace where all ordered $(k\times k)$ minors of $C$ are positive: $\left|\vec{C}_{i_1}\vec{C}_{i_2}\cdots \vec{C}_{i_k}\right|{\equiv}(i_1 i_2 \cdots i_k) {\geq} 0$ for $1\leq i_1<i_2<\cdots<i_k \leq n$. Without imposing further conditions, this is the $k(n{-}k)$ dimensional \emph{top cell}. Lower-dimensional cells can be approached by sending some \emph{consecutive} minors to zero while keeping all other ordered minors positive. 

A particular cell in $G_+(k,n)$ is a positive geometry, and can be parameterized by a set of positive parameters. For the simplest case with $k{=}1$, $G_+(1,n)\equiv \mathbb{P}^{n{-}1}$
\begin{equation}
    C = \left(\begin{array}{ccccc} 1 & x_1 & x_2 & \dots & x_{n{-}1}\end{array}\right)\,,
\end{equation}
is a set of positive numbers $x_k\geq0$. For a $d$-dimensional cell, only $d$ parameters are positive, the rest being zero. This yields a simple positive geometry with $d\log$ form,
\begin{equation}
\Omega = \frac{dx_1}{x_1}\frac{dx_2}{x_2}\dots \frac{dx_d}{x_d}\,.
\end{equation}
Even in more complicated situations, a general cell in $G_+(k,n)$ is a positive geometry, defined by a set of vanishing consecutive $(k\times k)$ minors, while all other ordered minors remain positive. The collection of all cells of $G_+(k,n)$---starting from the top cell all the way down to zero-dimensional cells---is called a \emph{stratification} and has a beautiful characterization via the geometry of $n$ points in $\mathbb{P}^{k{-}1}$ and permutations of $n$ labels. 

Unlike projective space, the positive geometry of $G_+(k,n)$ is more intriguing and it is not easy to see how to parameterize a particular cell using a positive parametrization. \emph{Plabic graphs} (i.e.~planar on-shell diagrams) do exactly that and they do provide a parametrization of the $C$-matrix for which all minors are non-negative. Many details can be found in \cite{Arkani-Hamed:2012zlh}.

First, we assign a \emph{perfect orientation} to a graph and add arrows to all edges in the graph according to the following rule: There are two incoming and one outgoing arrow at each black vertex, one incoming and two outgoing arrow in each white vertex, while there are $k$ incoming and $n{-}k$ outgoing arrows on external edges. Then we assign \emph{edge variables} $\alpha_e$ to each leg while fixing one of the variables to $1$ in each vertex\footnote{Alternatively, we can parameterize on-shell diagrams by \emph{face} -variables, for details see \cite{Arkani-Hamed:2012zlh}.}. Upon this assignment, the entries of the $C$ matrix are given by \cite{Arkani-Hamed:2012zlh}
\begin{equation}
    C_{iJ} = - \sum_{\Gamma_{i{\to}J}} \prod_{e\in \Gamma_{i{\to}J}} \alpha_e,
\end{equation}
where we sum over all (directed) paths $\Gamma_{i{\to}J}$ from the source $i$ (external incoming edge) to sink $J$ (external outgoing edge) by following the arrows in the graph. For each path we take the product of all edge variables along the way. By convention $C_{ii} = 1$ and $C_{iJ}=0$ if $J$ is an edge with incoming arrow. 

Let us exemplify the rules with the top cell of $G_+(2,4)$, where the $GL(2)$ gauge-fixed $C$ matrix
\begin{equation}
\hspace{-1cm}
\raisebox{-38pt}{
\includegraphics[trim={0cm .1cm 0cm 0cm},clip,scale=.3]{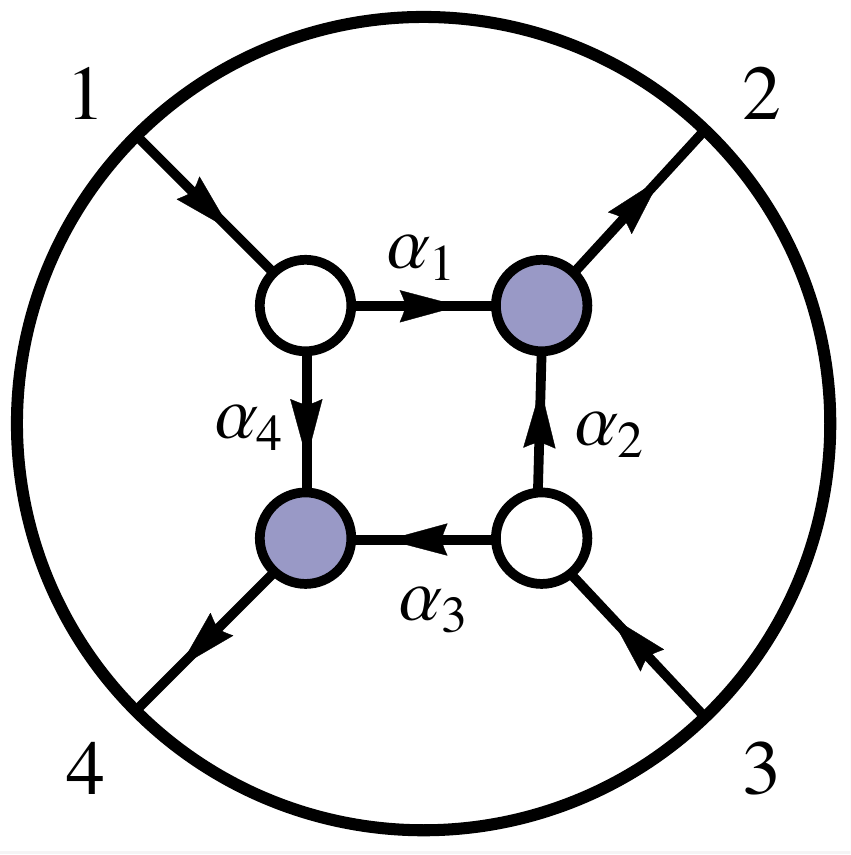}
}
\qquad 
\leftrightarrow
\qquad
    C = \left(
    \begin{array}{cccc}
     1 & -\alpha_1 & 0 & -\alpha_4 \\
     0 & -\alpha_2 & 1 & -\alpha_3
    \end{array}
    \right)\,,
\end{equation}
is parameterized by edge variables $\alpha_e$ and all ordered minors are positive for fixed signs of $\alpha_e$. Note that the minors $(13),(24)>0$ but they can not be set to zero without sending to zero two of the consecutive minors $(12),(23),(34),(14)$. However, we still need to impose $(13),(24)>0$ to define the cell. This is the main difference from the earlier examples in projective space where the positive geometry was entirely defined by the inequalities for boundaries. 

Generally, for the positive Grassmannian $G_+(k,n)$ in addition to boundary minors being positive, we need to impose a collection of extra inequalities which do not correspond to boundaries, but nevertheless are necessary to define the cell. The (redundant) collection of all inequalities is the list of all $(k\times k)$ minors which are zero in a given cell, while all other ordered $(k\times k)$ minors are positive. Once we parameterize the cell in $G_+(k,n)$ using the edge (or face) variables of a plabic graph, the $d\log$ form is simply,
\begin{equation}
\hspace{2cm}
    \Omega = \frac{d\alpha_1}{\alpha_1}\frac{d\alpha_2}{\alpha_2}\dots \frac{d\alpha_d}{\alpha_d}
\end{equation}
where $d$ is the dimensionality of the cell.

\subsection{BCFW Recursion Relations via On-Shell Diagrams}

The BCFW recursion relations allow us to express the scattering amplitudes in planar ${\cal N}=4$ SYM as the sum of on-shell diagrams \cite{Arkani-Hamed:2012zlh}
\begin{equation}
\label{eq:bcfw_os_diag_loop_recursion}
\hspace{-1cm}
    \raisebox{-40pt}{
    \includegraphics[scale=.7]{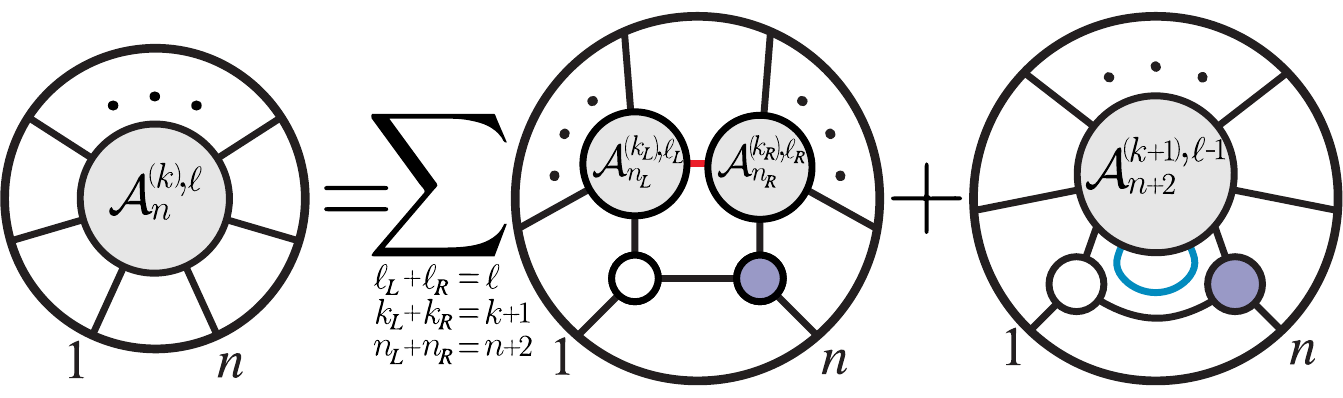}
    }\,,
\end{equation}
where each blob is recursed down to three-point amplitudes. At tree-level, the forward limit term (second term on the RHS of (\ref{eq:bcfw_os_diag_loop_recursion})) is absent. For the $n$ point N$^{k{-}2}$MHV amplitude the diagrams are $2n{-}4$ dimensional cells in $G_+(k,n)$. For MHV tree-level amplitudes, this is the top cell of $G_+(2,n)$ and there is only a single on-shell diagram for any $n$. In contrast, for non-MHV amplitudes we get a sum of on-shell diagrams which are lower dimensional cells. The link to the plabic graphs is not just superficial, in fact we can calculate the on-shell diagrams in ${\cal N}=4$ SYM theory using the edge variables and the $d\log$ form,
\begin{equation}
\label{eq:os_diags_spinor_hel}
    {\cal F}_\Gamma = \int \frac{d\alpha_1}{\alpha_1}\frac{d\alpha_2}{\alpha_2}\dots\frac{d\alpha_{2n{-}4}}{\alpha_{2n{-}4}}\delta(C\cdot\widetilde{\lambda})\delta(C^\perp\cdot\lambda)\delta(C\cdot\eta)
\end{equation}
where $C^\perp$ is the orthogonal complement of $C$, s.t. $C^\perp\cdot C=0$, see~\cite{Arkani-Hamed:2012zlh} Eq.~(4.6). The first two delta functions solve for all $\alpha_e$ in terms of $\lambda$, $\widetilde{\lambda}$ and reconstruct the momentum conserving delta function $\delta^4(P)$. The last delta function reproduces supermomentum conservation $\delta^8(\Q)$ and (possibly) additional Grassmann delta functions. 

Besides the spinor-helicity expression in Eq.~(\ref{eq:os_diags_spinor_hel}), there is a momentum twistor version for on-shell diagrams. First, for any on-shell diagram we can factor out the Parke-Taylor factor and the momentum and supermomentum delta functions,
\begin{equation}
    {\cal F}_\Gamma = \frac{\delta^4(P)\delta^8(\Q)}{\la12\ra\la23\ra\dots\la n1\ra} \times {\cal R}_\Gamma
\end{equation}
where the $n$-point N$^k$MHV Yangian invariant ${\cal R}_{\Gamma}$ can be calculated from the $d\log$ form on the $4k$-dimensional cell in $G_+(k,n)$, (Note that there is a $k\rightarrow k{-}2$ shift when going from ordinary spinor helicity variables to momentum twistors.)
\begin{equation}
    {\cal R}_\Gamma = \int \frac{d\alpha_1}{\alpha_1}\frac{d\alpha_2}{\alpha_2}\dots\frac{d\alpha_{4k}}{\alpha_{4k}}\delta^{(4|4)}(C\cdot {\cal Z})
\end{equation}
where ${\cal Z}_a = (z_a\,,\eta_a)$ consists of the bosonic momentum twistors $z_a$ and momentum twistor Grassmann variables $\eta_a$, related to the $\twEta_a$ defined above. (See e.g.~\cite{Bourjaily:2010wh} for the precise relation.) Notably, the cells are always $4k$-dimensional, independent of $n$.

\vspace{-.4cm}
\subsection{Tree-Level Amplituhedron}

When written in terms of momentum (super-)twistor variables, the BCFW recursion relations represent the $n$-point N$^k$MHV tree-level (super-)amplitudes as the sum of $d\log$ forms on $4k$-dimensional cells inside $G_+(k,n)$. Generalizing our notion of simplices and convex polytopes in projective space of Sec.~\ref{sec:pos_geom_primer}, we might ask whether there exists a positive geometry associated to the whole tree-level amplitude. We could glue together all contributing cells in $G_+(k,n)$ and try to associate the tree-level amplitude with that region. This naive answer turns out not to work: the combination of different cells does not give a nice region inside $G(k,n)$ (note that all cells are $4k$-dimensional rather than $k(n-k)$ dimensional top cell) and furthermore, the particular region depends on the choice of the BCFW shift, ie. it is not triangulation-independent. Therefore, we should strive for an intrinsic definition of the positive geometry for any tree-level amplitude in ${\cal N}=4$ SYM, independent of any particular triangulation.

\newpage
\subsubsection{Original Tree-Level Amplituhedron}

The tree-level amplituhedron $\mathbb{A}_{n,k}$  is a certain projection of the \emph{top cell} of $G_+(k,n)$ to a certain $m\,k$ dimensional space, which is a subspace of $G(k,k{+}m)$,
\begin{equation}
  \Phi: G_+(k,n) \rightarrow G(k,k{+}m)\,. 
\end{equation}
The parameter $m$ denotes the ``dimensionality" of external momentum twistors, where $m=4$ is relevant for physics, but other values of $m$ might be interesting for studying certain mathematical aspects of the amplituhedron. The map $\Phi$ is defined by the positive $((k{+}m) \times n)$ matrix $Z=M_+(k{+}m,n)$, where all ordered $(k{+}m \times k{+}m)$ minors are positive (this is a fixed matrix, so we do not call it a positive Grassmannian),
\begin{equation}
    Y_\alpha^I = C_{\alpha a}Z_a^I,\qquad \alpha=1,{\dots},k;\quad a = 1,{\dots},n;\quad I = 1,{\dots}k{+}m \,,
\label{Ydef}
\end{equation}
and $Y\in G(k,k{+}m)$ describes a region of maximal $km$ dimensionality in the non-positive part of $G(k,k{+}m)$. In this space $Y$ represents a $k$-plane and the $Z_a$ are $k{+}m$ dimensional \emph{bosonized} momentum twistors. In particular, the $Z_a$ are given in terms of the ordinary $m$-dimensional momentum twistors $z_a$ and $k$ extra components which involve Grassmann variables $\eta_a^A$ with $A=1,{\dots},m$,
\begin{equation}
\label{eq:bosonized_mom_twistors}
    Z_a = \left(
    \begin{array}{ccccc} z_a & 
        (\phi_1\cdot\eta_a) & 
        (\phi_2\cdot\eta_a) & \dots & 
        (\phi_k\cdot\eta_a)
    \end{array}\right)\,,
\end{equation}
from which we can build natural $SL(k{+}m)$ invariants $\la Z_{a_1}Z_{a_2}{\cdots}Z_{a_{k{+}m}}\ra,\, \la Y Z_{a_1}{\cdots}Z_{a_m}\ra$. 
We can define the canonical volume form of this positive geometry,
\begin{equation}
    \Omega_{n,k}^{(m)} = {\cal F}_{n,k}^{(m)}\, d\mu \quad
    \mbox{with measure }\,\,
    d\mu = \prod_{\alpha = 1}^k \la Y_1\dots Y_k\, d^mY_\alpha\ra\,,
\end{equation}
where ${\cal F}_{n,k}^{(m)}$ is the volume function. The tree-level amplitude ${\cal A}_{n,k}$ can be extracted from ${\cal F}_{n,k}^{(m=4)}$ by integrating out the $\phi_1,\ldots, \phi_k$ from the $Z_a$ in Eq.~(\ref{eq:bosonized_mom_twistors}).

\vspace{-.3cm}
\subsubsection{Topological Definition of Tree-Level Amplituhedron}

There is an equivalent definition of the tree-level amplituhedron $\mathbb{A}_{n,k}^{(m)}$ which uses a topological description \cite{Arkani-Hamed:2017vfh}. The definition (\ref{Ydef}) implies that for any $Y$ inside the amplituhedron:
\begin{equation}
    \ab{ Y\,i_1\,i_1{+}1\,\dots\,i_{\frac{m}{2}}\,i_{\frac{m}{2}}+1} > 0 \,,
    \qquad \mbox{ $m$ even,}
\label{pole}
\end{equation}
and there are $k$ sign flips in the sequence of brackets
\begin{equation}
\hspace{-1.5cm}
    \{
    \ab{Y123\dots m{-}1\,m},
    \ab{Y123\dots m{-}1\,m{+}1},\dots,
    \ab{Y123\dots m{-}1\,n}
    \}\,. 
\label{seq}
\end{equation}
A general bracket $\la Y a_1 a_2 \dots a_m\ra$ does not have a fixed sign, but for a fixed $Y$ inside the amplituhedron, the signs of all brackets are correlated such that the sequence (\ref{seq}) has exactly $k$ sign flips. For the physical case, $m=4$, Eqs.~(\ref{pole}) and (\ref{seq}) reduce to $\la Y\,i\,i{+}1\,j\,j{+}1\ra>0$ and $k$ sign flips in the sequence $\{\la Y1234\ra,\la Y1235\ra,\dots,\la Y123n\ra\}$. 

The form $\Omega_{n,k}^{(m)}$ on the amplituhedron is not a simple $d\log$ form as the positive geometry is generally complicated. We do have a single $d\log$ form on the $k(n{-}k)$-dimensional top cell of $G_+(k,n)$ but after the $Z$-map the dimensionality of the space drops to $mk$, and the canonical form is more complicated. To calculate $\Omega_{n,k}^{(m)}$, we triangulate the amplituhedron $\mathbb{A}_{n,k}$ in terms of elementary regions with a simple $d\log$ form, and then sum all contributions. Note that the BCFW recursion relations also provide such a triangulation. Each BCFW term originates as a $4k$-dimensional cell in $G_+(k,n)$ (rather than a top cell), and under the $Z$-map it maps to a particular region in $G(k,k{+}4)$ which is also $4k$-dimensional. There is no dimensionality drop, and hence the $d\log$ form on $G_+(k,n)$ maps to the $d\log$ form on $G(k,k{+}4)$. 

As an example, consider the simplest $k=1$ case for (unphysical) $m=2$ kinematics. Here, the amplituhedron is a map from $G_+(1,n)$ to $G(1,3)\equiv \mathbb{P}^2$, and the positive geometry is a region in $\mathbb{P}^2$, see Sec.~\ref{sec:pos_geom_primer}. The $Z$-matrix is a $(3\times n)$ matrix with positive ordered $(3\times3)$ minors. For any point $Y$ inside this region we have,
\begin{equation}
\hspace{-2cm}
    \la Y\,i\,i{+}1\ra>0\,\, \mbox{and sequence}\,\,
    \{\la Y12\ra,\la Y13\ra,\dots,\la Y1n\ra\}\,\,\mbox{has one sign flips.} \label{poly1}
\end{equation}
As $\la Y12\ra{>}0$, $\la Y1n\ra{<}0$, the $n{-}2$ sign patterns depend on the position of the $+{\to}-$ flip
\begin{equation}
\hspace{-1.5cm}
    \left(
        \begin{array}{cccccc} 
        \ab{Y12} & \dots & \ab{Y1\,i} & \ab{Y1\,i{+}1} & \dots & \ab{Y1n}
        \\
        + & \dots & + & - & \dots & -
        \end{array}
    \right)\quad 
    \mbox{where $i=2,{\dots},n{-}1$}\,.
    \quad 
\label{poly2}
\end{equation}
For a fixed sign pattern we can express $Y=Z_1+\alpha_i Z_i+\alpha_{i{+}1}Z_{i{+}1}$ and the inequalities (\ref{poly1}), (\ref{poly2}) imply $\alpha_i,\alpha_{i{+}1}>0$. This is the interior of a triangle in $\mathbb{P}^2$ with vertices $Z_1$, $Z_i$, $Z_{i{+}1}$. The collection of all such triangles forms a polygon,
\begin{equation}
\hspace{2cm}
\mathbb{A}^{(m{=}2)}_{n,k{=}1} = 
\sum^{n-1}_{i=2}
    \raisebox{-32pt}{
    \includegraphics[scale=.2]{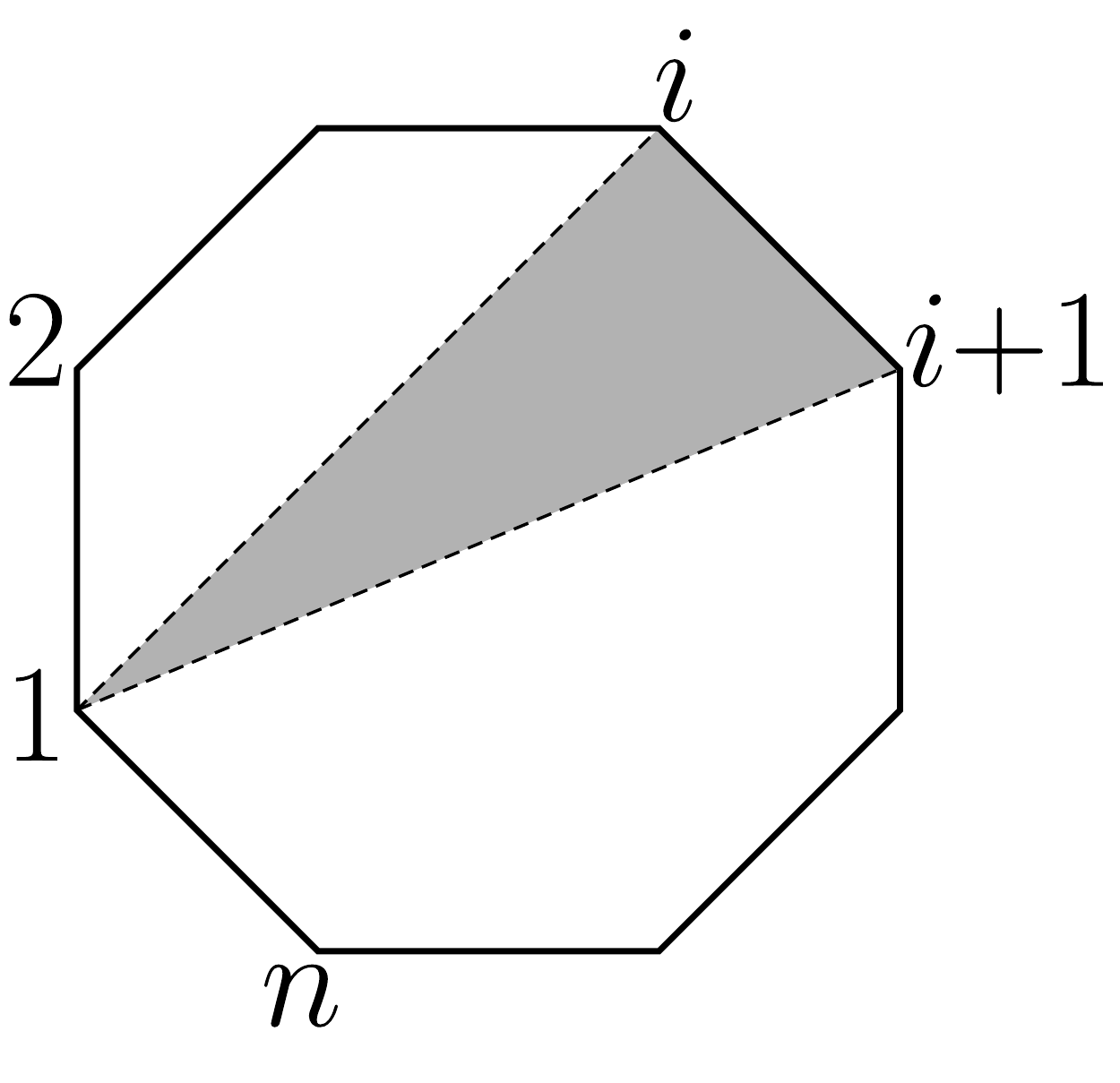}}
\label{triang}    
\end{equation}
which is the amplituhedron geometry for $k{=}1$, $m{=}2$. Its canonical form is known from our discussion in Section \ref{sec:pos_geom_primer}. Note that each triangle can also be thought of as a map from the 2-dimensional cell in $G_+(1,n)$ where all entries except $C_1$, $C_i$, $C_{i{+}1}$ are zero,
\begin{equation}
    C = \left(\begin{array}{ccccccccc} 1 & 0 & \dots & 0 & C_i & C_{i{+}1} & 0 & \dots & 0\end{array}\right)\,.
\end{equation}
The logarithmic form of each triangle in (\ref{triang}) is $    \Omega_{\blacktriangle} = \frac{d\alpha_i}{\alpha_i}\frac{d\alpha_{i{+}1}}{\alpha_{i{+}1}} = \frac{\la Y\,d^2Y\ra\la 1\,i\,i{+}1\ra^2}{\la Y\,1\,i\ra\la Y\,i\,i{+}1\ra\la Y\,i{+}1\,1\ra},$ where we expressed $\alpha_i$, $\alpha_{i{+}1}$ via $Y$ and $Z$'s. Summing all contributing triangle forms,
\begin{equation}
    \Omega_{n,k=1}^{(m=2)} = \sum_{i=2}^{n{-}1} \frac{\la Y\,d^2Y\ra\la 1\,i\,i{+}1\ra^2}{\la Y\,1\,i\ra\la Y\,i\,i{+}1\ra\la Y\,i{+}1\,1\ra}\,,
\end{equation}
we obtain the amplitude, where all spurious $\la Y1i\ra$ poles cancel in the sum leaving only the $\la Yii{+}1\ra$ ones that correspond to physical boundaries. 

From the differential form, we can extract the on-shell superfunction for the scattering amplitude (for $m=2$ it would be just a toy amplitude).  We fix $Y$ such that $\la Y\,a_1\,a_2\,{\dots}\,a_m\ra = \la z_{a_1}z_{a_2}\dots z_{a_m}\ra$, i.e.~any bracket with $Y$ depends only on the bosonic momentum twistors $z_a$. Evaluating the superfunction ${\cal F}^{(m)}_{n,k}$, and integrating over all $\phi_j$ we obtain an $SL(m|m)$ invariant superfunction,
\begin{equation}
    F_{n,k}^{(m)} = \int d^m\phi_1\dots d^m\phi_k\,{\cal F}_{n,k}^{(m)}
\end{equation}
For $m=4$ this is equal to $n$-point N$^k$MHV tree amplitude in ${\cal N}=4$ SYM. For our $m=2$, $k=1$ toy example, we get
\begin{equation}
\hspace{-1.8cm}
   F_{n,k=1}^{(m=2)} = \sum_{i=2}^{n{-}1} \frac{(\la 1i\ra\eta_{i{+}1} + \la i\,i{+}1\ra\eta_1 + \la i{+}1\,1\ra\eta_i)^2}{\la Y\,1\,i\ra\la Y\,i\,i{+}1\ra\la Y\,i{+}1\,1\ra},
   \quad
   {\rm where}\ 
   \ab{ab} = \epsilon_{IJ} z_a^Iz_b^J . 
\end{equation}
In fact, the entire picture can be re-formulated directly in the $m$-dimensional momentum twistor space. We define an amplituhedron space $\mathbb{A}_{n,k}^{(m)}$ as the space of all $m$-dimensional momentum twistor variables $z_i$ which satisfy following constraints: 
\begin{itemize}
    \item They are obtained by the projection from the $m{+}k$-dimensional positive momentum twistors $Z_i\in M_+(m{+}k,n)$ through a $k$-plane $Y$: $Z_i \rightarrow z_i$.
    \item For all brackets $\la a_1\dots a_m\ra = \epsilon_{I_1\dots I_m}z_{a_1}^{I_1}\dots z_{a_m}^{I_m}$,
    \begin{equation}
    \la i_1\,i_1{+}1{\dots} i_{\frac{m}{2}}i_{\frac{m}{2}}+1\ra > 0 \label{tree1}
    \end{equation}
    \item And the following sequence must have $k$ sign flips:
    \begin{equation}
        \{\la 123{\dots}m{-}1m\ra,{\dots},\la 123{\dots}m{-}1\,n\} 
        \label{tree2}
    \end{equation}
\end{itemize}
We refer to these $z_i$ as N$^k$MHV external data, and the space of all of them is the tree-level N$^k$MHV amplituhedron geometry. There is a form with logarithmic singularities on the boundaries of this space $\omega_{n,k}^{(m)}$. The on-shell superfunction can be obtained by replacing $dz_i \rightarrow \eta_i$ in $\omega_{n,k}^{(m)}$. For our toy example, we would get for $\omega_{n,k{=}1}^{(m{=}2)}$,
\begin{equation}
    \omega_{n,k=1}^{(m=2)} = \sum_{i=2}^{n{-}1} \frac{(\la 1\,i\ra dz_{i{+}1} + \la i\,i{+}1\ra dz_1 + \la i{+}1\,1\ra dz_i)^2}{\la 1\,i\ra\la i\,i{+}1\ra\la i{+}1\,1\ra}
\end{equation}
and for $dz_i\rightarrow\eta_i$ we would recover $F_{n,k=1}^{(m=2)}$. Note that these two definitions of $\mathbb{A}_{n,k}^{(m)}$ are equivalent, but the latter one defines the amplituhedron directly in the kinematic space rather than in the abstract $Y$-space.
%
\hspace{-.4cm}
\subsection{Loop Amplituhedron}

The tree-level amplituhedron construction has an extension to loops. We focus on the physically relevant value $m=4$ for the external data and label the amplituhedron geometry $\mathbb{A}^{\ell}_{n,k}$ by the loop order $\ell$.

To define the one-loop amplituhedron topologically \cite{Arkani-Hamed:2017vfh}, we start with the N$^k$MHV external data $z_i$, satisfying the tree-level amplituhedron conditions (\ref{tree1}), (\ref{tree2}). At one-loop, the positive geometry also involves configurations of the line $(AB)\equiv (AB)^{IJ}$ (where $z_A^I$, $z_B^J$ are two arbitrary points on that line) subject to the following conditions:
\begin{equation} 
\hspace{-2cm}
\la AB\,i\,i{+}1\ra>0 \,, \quad 
\mbox{sequence} \,\,\{\ab{AB12},\la AB13\ra,{\dots},\la AB1n\ra\} 
\mbox{ has $k{+}2$ sign flips.}
\label{oneloop}
\end{equation}
This is the maximal number of sign flips we can have in this sequence for N$^k$MHV external data. We define the form $\omega^{\ell}_{n,k}$ with logarithmic singularities on the boundaries of this space. The loop integrand ${\cal I}^{\ell}_{n,k}$ of the $n$-point N$^k$MHV one-loop amplitude can be obtained from $\omega^{\ell}_{n,k}$ by replacing $dz_i\rightarrow \eta_i$. Note that the line $(AB)$ is the momentum twistor representation of the loop momentum, and hence ${\cal I}^{\ell}_{n,k}$ is in fact a 4-form in $(AB)$ (momentum twistor analogue of $d^4\ell$).

The standard way to obtain $\omega^{\ell}_{n,k}$ is by triangulating the amplituhedron geometry and summing the $d\log$ forms for elementary regions. We focus on the simple case of the $n$-point MHV one-loop amplitude, corresponding to $k=0$ where the fixed positive external data $\{Z_i\}$ is the same as $\{z_i\}$. (The projection plane $Y$ is zero dimensional.) Therefore, the only object in our positive geometry is the line $(AB)$. The sequence of brackets (\ref{oneloop}) has two sign flips occurring at positions $i$ and $j$,
\begin{equation}
\hspace{-2.5cm}
    \left(\!\!
    \begin{array}{ccccccccc} \ab{AB12} & {\dots} & \la AB1\,i\ra & \la AB1\,i{+}1\ra  & {\dots} & \la AB1\,j\ra & \la AB1\,j{+}1\ra & {\dots} & \la AB1n\ra \\
    + & {\dots} & + & - & {\dots} & - & + & {\dots} & + 
    \end{array}
    \!\!\right).
\label{oneloop2}
\end{equation}
Together with the positivity of all $\la AB\,l\,l{+}1\ra>0$, each sign flip pattern defines one elementary positive geometry. Inside each geometry, we parameterize $(AB)$ as
\begin{equation}
    z_A = z_1 + \alpha\, z_i + \beta\, z_{i{+}1}\,,
    \qquad 
    z_B = z_1 + \gamma\, z_j + \delta\, z_{j{+}1}\,, 
\label{oneloop3}
\end{equation}
by putting $A$ in the plane $(1\,i\,i{+}1)$ and $B$ in the plane $(1\,j\,j{+}1)$. Imposing all conditions of Eq.~(\ref{oneloop}) on $(AB)$ we get a simple region $\alpha,\beta,\gamma,\delta>0$ in parameter space. The form with logarithmic singularities on the boundaries of this region is
\begin{equation}
\hspace{-2.4cm}
    \omega =    \frac{d\alpha}{\alpha}
                \frac{d\beta}{\beta}
                \frac{d\gamma}{\gamma}
                \frac{d\delta}{\delta} 
           = \frac{\la AB\,d^2A\ra\la AB\,d^2B\ra\,\la AB\,(1\,i\,i{+}1){\cap}(1\,j\,j{+}1)\ra^2}
           {\la AB\,1\,i\ra\la AB\,1\,i{+}1\ra\la AB\,i\,i{+}1\ra\la AB\,1\,j\ra\la AB\,1\,j{+}1\ra\la AB\,j\,j{+}1\ra}        
\label{oneloop4}
\end{equation}
expressed both in parameter space and projectively using Eq.~(\ref{oneloop3}) and the definition $\la AB\,(abc){\cap}(def)\ra = \la Aabc\ra\la Bdef\ra - \la Babc\ra\la Adef\ra$. The complete one-loop amplituhedron $\mathbb{A}^{\ell=1}_{n,k=0}$ is the union of all regions in (\ref{oneloop2}) with associated canonical form
\begin{equation}
\hspace{-2.5cm}
    \omega^{\ell=1}_{n,k=0} {=} \sum_{i=2}^{n{-}2}\sum_{j=i{+}1}^{n{-}1}
    \frac{\la AB\,d^2A\ra\la AB\,d^2B\ra\,\la AB\,(1\,i\,i{+}1){\cap}(1\,j\,j{+}1)\ra^2}{\la AB\,1\,i\ra\la AB\,1\,i{+}1\ra\la AB\,i\,i{+}1\ra\la AB\,1\,j\ra\la AB\,1\,j{+}1\ra\la AB\,j\,j{+}1\ra}\,,
\end{equation}
reproducing the ``kermit'' expansion of the one-loop MHV integrand in ${\cal N}
{=}4$ SYM~\cite{Arkani-Hamed:2010zjl}. 

Likewise, we define the $\ell$-loop $n$-point N$^k$MHV amplituhedron $\mathbb{A}^{\ell}_{n,k}$ in terms of N$^k$MHV external data $z_i$ and $\ell$ loop lines $(AB)_j$. The external data must lie in the N$^k$MHV tree-level amplituhedron $\mathbb{A}^{(m=4),\ell=0}_{n,k}$, i.e.~satisfying the conditions (\ref{tree1}) and (\ref{tree2}). Each line $(AB)_j$ is inside the one-loop amplituhedron $\mathbb{A}^{\ell{=}1}_{n,k}$ subject to Eq.~(\ref{oneloop}). In addition, we impose the mutual positivity condition for any pair of lines~$\{(AB)_i$,~$(AB)_j\}$,
\begin{equation}
\hspace{2cm}
 \la (AB)_i(AB)_j\ra > 0\,., 
\end{equation}
The planar ${\cal N}=4$ SYM loop integrand can be then extracted from the form $\omega^{\ell}_{n,k}$ with logarithmic singularities on the boundaries of $\mathbb{A}^{\ell}_{n,k}$. 

As a simple example, we consider the 2-loop 4-point MHV amplituhedron $\mathbb{A}^{\ell=2}_{n=4,k=0}$. The space of all MHV external data $z_i$ is trivial (with $\la1234\ra>0$), and the geometry contains only two loop lines $(AB)$ and $(CD)$. Each line is inside the 1-loop 4-point MHV amplituhedron, which fixes the signs of all brackets $\la ABij\ra$ and $\la CDij\ra$,
\begin{equation}
\hspace{-2.2cm}
    \ab{AB12}, 
    \ab{AB23}, 
    \ab{AB34},\ab{AB14} > 0,\,\,\, 
    \la AB13\ra, \la AB24\ra < 0,\,\,\,  
    (AB)\leftrightarrow (CD)\,.
\end{equation}
Expanding $z_A = z_1 + \alpha_1 z_2 + \beta_1 z_4$, $z_B = z_3 - \gamma_1 z_2 + \delta_1 z_4$ and $z_C = z_1 + \alpha_2 z_2 + \beta_2 z_4$, $z_D = z_3 - \gamma_2 z_2 + \delta_2 z_4$, the above inequalities fix $\alpha_1,\beta_1,\gamma_1,\delta_1,\alpha_2,\beta_2,\gamma_2,\delta_2>0$. Finally we have to impose $\ab{ABCD}>0$. In this parametrization this reduces to 
\begin{equation}
    (\alpha_1-\alpha_2)(\delta_1-\delta_2) + (\beta_1-\beta_2)(\gamma_1-\gamma_2)<0\,. \label{twoloop}
\end{equation}
This space needs to be triangulated which gives definite bounds to all eight parameters. We first order three pairs of the variables, $\alpha_1>\alpha_2$ or $\alpha_2>\alpha_1$ et cetera, which breaks the whole space into eight regions. Fixing $\delta_1$ we find the bound for $\delta_2$ from (\ref{twoloop}). For example, for one of the regions, the parameter ranges are
\begin{equation}
\hspace{-1cm}
    \alpha_1>\alpha_2,\,\, \beta_1>\beta_2,\,\, \gamma_1>\gamma_2 \quad \mbox{and}\quad \delta_2 > \delta_1 + \frac{(\beta_1-\beta_2)(\gamma_1-\gamma_2)}{(\alpha_1-\alpha_2)}\,,
\end{equation}
leading to the form with logarithmic singularities on the boundaries of this region,
\begin{equation}
\hspace{-1.5cm}
    \omega = 	\frac{d\alpha_1\,d\alpha_2}{\alpha_2(\alpha_2-\alpha_1)}
    			\frac{d\beta_1\,d\beta_2}{\beta_2(\beta_2-\beta_1)}
			\frac{d\gamma_1\,d\gamma_2}{\gamma_2(\gamma_2-\gamma_1)}
			\frac{d\delta_1\,d\delta_2}{\delta_1\left(\delta_2-\delta_1-\frac{(\beta_1-\beta_2)(\gamma_1-\gamma_2)}{(\alpha_1-\alpha_2)}\right)}\,.
\end{equation}
We can re-write this form projectively and sum over all eight contributions in the triangulation. This cancels all spurious poles---these are $(\alpha_2-\alpha_1)$ type poles---which are introduced in the triangulation, leading to the final result
\begin{equation}
\hspace{-2.5cm}
    \omega^{\ell=2}_{n=4,k=0} {=} \frac{d\mu\, \ab{1234}^3
    [\ab{AB12} \ab{CD34} + \ab{AB23}\ab{CD14} + (AB)\leftrightarrow (CD)]}
    {\ab{AB12}\ab{AB23}\ab{AB34}\ab{AB14}\ab{ABCD}\ab{CD12}\ab{CD23}\ab{CD34}\ab{CD14}} 
    \label{twoloop2}
\end{equation}
where $d\mu = \ab{AB\,d^2A}\ab{AB\,d^2B}\ab{CD\,d^2C}\ab{CD\,d^2D}$. Expanding the numerator in (\ref{twoloop2}) we get a sum of four terms which are nothing else than double box integrands,
\begin{equation}
\hspace{-1.3cm}
 \omega^{\ell=2}_{n=4,k=0} = 
    \hspace{-.6cm}
    \vcenter{\hbox{\scalebox{.7}{\dblboxS{AB}{CD} }}}
    \hspace{-.2cm}
    +
    \hspace{-.2cm}
    \vcenter{\hbox{\scalebox{.7}{\dblboxS{CD}{AB} }}}
    \hspace{-.2cm}
    +
    \hspace{-.2cm}
    \vcenter{\hbox{\scalebox{.7}{\dblboxT{AB}{CD} }}}
    \hspace{-.2cm}
    +
    \hspace{-.2cm}
    \vcenter{\hbox{\scalebox{.7}{\dblboxT{CD}{AB} }}}
    \hspace{-.3cm}.
\end{equation}
At higher loops, the problem of finding integrands using unitarity methods becomes increasingly difficult because of a huge number of contributing terms. On the other hand, the definition of the amplituhedron geometry $\mathbb{A}^{\ell}_{n,k}$ is extremely simple and compact for any $\ell$. Nevertheless, the complexity of the integrand is present in the intriguing mutual inequalities $\ab{(AB)_i(AB)_j}>0$ between individual loop lines. In fact, the geometric problem of triangulations turns into solving a system of quadratic inequalities which is indeed very difficult. Even the classification of all boundaries is not known in general \cite{Franco:2014csa,Rao:2018uta}. Recently, some interesting results \cite{Arkani-Hamed:2018rsk,Langer:2019iuo} were obtained for certain ``internal" cuts of the $n$-point MHV integrand for arbitrary $\ell$, where all mutual positivity conditions trivialize, $\la (AB)_i(AB)_j\ra = 0$. In another work, certain amplitude-like objects (ratio of Wilson loops) were computed to all orders in a certain (geometric) limit by resuming integrated $d\log$ forms over positive geometries \cite{Arkani-Hamed:2021iya}. Despite all these advances, the triangulation of the full amplituhedron space for all $\ell$, even at four points, is still an important open problem. 
%
\newpage
\section{Recent Appearances of Positive Geometry in Physics and Outlook}
\label{sec:advances_and_outlook}

The goal of this program is to reformulate perturbative scattering amplitudes in a geometric language. The S-matrix in a particular theory would be defined by a positive geometry which captures all singularities as boundaries, together with a differential form or other mathematical object which extract the explicit expressions for amplitudes. A systematic approach how to find this picture for a larger class of QFTs is a big open question. Nevertheless, apart from bi-adjoint $\phi^3$ theory and planar ${\cal N}{=}4$ SYM, positive geometry has appeared in various contexts in the literature.

\vspace{-.3cm}
\subsubsection*{Worldsheet Associahedron}

In this review, we have discussed the relation between the kinematic associahedron and amplitudes in $\phi^3$ theory. The main reason for this connection is the fact that the boundary structure of the associahedron mimics the factorization channels of $\phi^3$ amplitudes. It has been long known that the same object appears in the context of open string theory. The boundary structure of the open string moduli space (ordered points on the boundary of a disk) factorizes in the same way and there is a worldsheet canonical form associated with this positive geometry,
\begin{equation}
    \omega_n^{WS} = \frac{1}{{\rm vol [SL(2)]}}\prod_{a=1}^n \frac{d\sigma_a}{\sigma_a - \sigma_{a{+}1}}
\end{equation}
The worldsheet and kinematic associahedra are connected via \emph{scattering equations} which relate the positions on the worldsheet $\sigma_i$ to Mandelstam variables $s_{ij}$,
\begin{equation}
    E_i = \sum_{j\neq i} \frac{s_{ij}}{\sigma_i - \sigma_j} = 0
\end{equation}
More details on this fascinating connection are described in \cite{Arkani-Hamed:2017mur}. 

\vspace{-.3cm}
\subsubsection*{Positive geometry of chiral integrals and the dual Amplituhedron}

In the context of positive geometry for planar ${\cal N}=4$ SYM theory, we have seen that all tree-level amplitudes and loop integrands could be calculated from the differential form on the amplituhedron geometry. We could triangulate the amplituhedron using on-shell diagrams (via BCFW recursion relations) which were projections of the cells in the positive Grassmannian $G_+(k,n)$. As it was pointed out in \cite{Arkani-Hamed:2010pyv}, the loop integrands in planar ${\cal N}=4$ SYM theory can also be expanded in terms of chiral integrals using generalized (or prescriptive) unitarity~\cite{Bern:1994zx,Bern:1994cg,Britto:2004nc,Bern:2007ct,Bourjaily:2017wjl}. These integrals are naturally written in momentum twistor space with special chiral numerators which ensure cancellations on various cuts and IR finiteness, e.g.~the one-loop chiral pentagon
\begin{equation}
\hspace{-2.5cm}
\Omega^{(n)}_{ij} = 
\raisebox{-35pt}{\includegraphics[scale=.55]{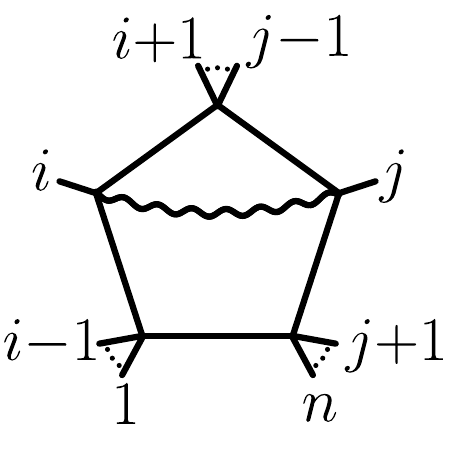}} 
\hspace{-.2cm}
= \frac{\ab{ABd^2A}\ab{ABd^2B}\ab{1ijn} \ab{AB(i{-}1 i i{+}1)\cap (j{-}1 j j{+}1)}}
       {\ab{ABi{-}1i}\ab{ABii{+}1}\ab{ABj{-}1j}\ab{ABjj{+}1}\ab{AB1n}}\,,
\label{eq:chi_pent_rat_form}       
\end{equation}        
utilized in \cite{Bourjaily:2013mma} for the expansion of $n$-point MHV one-loop integrands. There is a simple change of variables which allows us to write the integrand in $d\log$ form \cite{Herrmann:2020oud,Herrmann:2020qlt},
\begin{equation}
\hspace{-2.5cm}
   \Omega^{(n)}_{ij} {=} 
   d\log\frac{\la AB\,i{-}1i\ra}{\la AB\,i\,i{+}1\ra}\,
   d\log\frac{\la AB\,j{-}1\,j\ra}{\la AB\,j\,j{+}1\ra}\,
   d\log\frac{\la AB(n1i){\cap}(ij\ast)\ra}{\la ABn1\ra}\,
   d\log\frac{\la AB(n1j){\cap}(ij\ast)\ra}{\la ABn1\ra}
\label{eq:chi_pent_dlog}
\end{equation}
where $\ast$ is some arbitrary reference twistor. This $d\log$ form naturally defines a positive geometry by demanding definite signs for the $d\log$ arguments. It has been speculated that the expansion in terms of chiral pentagons provides a natural triangulation of the yet-to-be discovered \emph{dual} amplituhedron geometry. It was shown in \cite{Herrmann:2020qlt} that on certain boundaries the positive geometries defined by the $d\log$ form (\ref{eq:chi_pent_dlog}) indeed triangulate the dual amplituhedron geometry (in these cases the geometry reduces to simple polytope). The dual amplituhedron has been also studied from different perspectives \cite{Arkani-Hamed:2014dca} but the actual construction is still a big open question.

\vspace{-.3cm}
\subsubsection*{Momentum Amplituhedron}

The amplituhedron geometry is defined in momentum twistor space which requires a particular ordering of external momenta intricately tied to the planar limit. When trying to extend the geometric constructions to other theories and beyond the planar limit, it is natural to search for positive geometries in spinor helicity space. As a first step, the original amplituhedron construction was reformulated, at tree level, in spinor helicity variables, and called the \emph{momentum amplituhedron}~\cite{Damgaard:2019ztj}.

The external data $\lambda$, $\widetilde{\lambda}$ is embedded in two bosonized spinor helicity matrices, $\Lambda^A_a\in M(n{-}k{+}2,n)$ and $\widetilde{\Lambda}^{\dot{A}}_a\in M(k{+}2,n)$, where $\tw{\Lambda}$ and $\Lambda^\perp$ are positive. The momentum amplituhedron $\mathcal{M}_{n,k}$ is a map $\Phi$ from the top cell of $C_{\dot{\Gamma} a}\in G_+(k,n)$ to the product space $G_+(k,k{+}2) \times G_+(n{-}k,n{-}k{+}2)$. The coordinates in this space are given by a pair of Grassmannian elements $(\widetilde{Y},Y)$: $ \tw{Y}^{\dot{A}}_{\dot{\Gamma}} = C_{\dot{\Gamma} a} \tw{\Lambda}^{\dot{A}}_a\,,\ Y^{A}_{\Gamma} = C^{\perp}_{\Gamma a} \Lambda^{A}_a\,,$ and the momentum amplituhedron lives in a codimension-four subspace defined by the geometrized version of momentum conservation imposed on $Y,\widetilde{Y}$,
\begin{equation}
    P^{\alpha\dot{\alpha}} = \sum_{a{=}1}^n (Y^\perp\cdot\Lambda)_a^{\alpha} (\widetilde{Y}^\perp\cdot \widetilde{\Lambda})_a^{\dot{\alpha}} = 0\,.
\end{equation}
The positivity of the $\tw{\Lambda}$ and $\Lambda^\perp$ together with the $\Phi$ map leads to the correct sign-flip pattern in the external data similar to the topological construction of the momentum twistor amplituhedron.
\begin{equation}
\hspace{-2.5cm}
    \{\la Y12\ra,\la Y13\ra,{\dots},\la Y1n\ra\}: \mbox{$k{-}2$ sign flips,}
    \quad
    \{[\widetilde{Y}12],[\widetilde{Y}13],{\dots},[\widetilde{Y}1n]\}: \mbox{$k$ sign flips,}
\end{equation}
where $\la\cdots\ra$ and $[\cdots]$ are the usual brackets with $Y$, resp. $\widetilde{Y}$. We can also project $\Lambda$, $\widetilde{\Lambda}$ through the orthogonal complements of $Y$, $\widetilde{Y}$ to obtain the external spinors,
\begin{equation}
    \lambda_a^\alpha = (Y^\perp)^\alpha_A \Lambda_a^A,
    \qquad
    \widetilde{\lambda}_a^{\dot{\alpha}} = (\widetilde{Y}^\perp)^{\dot{\alpha}}_{\dot{A}} \widetilde{\Lambda}_a^{\dot{A}}\,.
\end{equation}
Besides the positivity in $\tw{\Lambda}$ and $\Lambda^\perp$, in order to specify the momentum amplituhedron space, one has to ensure the positivity of planar Mandelstam variables. These additional constraints are more subtle and discussed in Ref.\cite{Damgaard:2019ztj}. The resulting momentum amplituhedron space is a positive geometry with a canonical form with logarithmic singularities on its boundaries $\omega(\lambda,\widetilde{\lambda})$. The tree-level N$^{k-2}$MHV (super-)amplitude can be obtained by replacing $d\lambda_a \rightarrow \eta_a$, $d\widetilde{\lambda}_a\rightarrow \widetilde{\eta}_a$ where the function is written in the half-chiral superspace \cite{He:2018okq}. The momentum amplituhedron exhibits a lot of nice properties such as transparent singularity structure, KK relations or the connection to the kinematical associahedron \cite{Damgaard:2020eox,Damgaard:2021qbi,Ferro:2020lgp}.

\vspace{-.3cm}
\subsubsection*{Non-planar on-shell diagrams and positive geometry}

While the amplituhedron approach does not currently generalize to non-planar loop amplitudes, we do have interesting non-planar on-shell objects which are closely connected to positive geometry. These are \emph{non-planar on-shell diagrams} representing cuts of non-planar loop amplitudes. We can follow the same rules, and construct the $C$-matrix for any diagram and compute the canonical form. Each diagram represents a certain cell in the (non-positive) Grassmannian $G(k,n)$ with interesting boundary structure and possible combinatorial classification. At MHV level, the Grassmannian $G(2,n)$ can be tiled by positive parts $G_+(2,n)$ with different orderings. As a result, any non-planar MHV on-shell diagram evaluates to a linear combination of Parke-Taylor factors \cite{Arkani-Hamed:2014bca},
\begin{equation}
    \raisebox{-35pt}{\includegraphics[trim={0cm .1cm 7cm 0cm},clip,scale=.6]{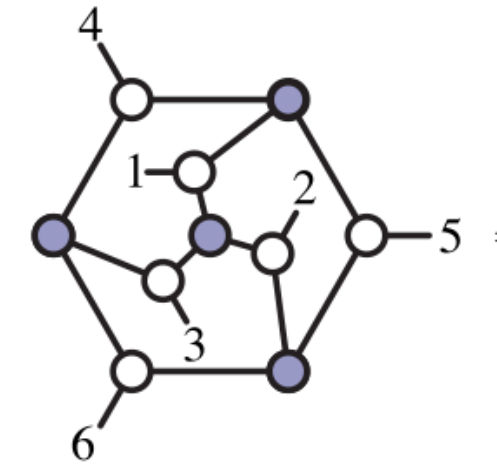}}
    \raisebox{-25pt}{\includegraphics[trim={5cm 0.5cm 0cm 2cm},clip,scale=.9]{./figures/Nonplanar.pdf}}
\end{equation}
For higher $k$, the positive geometries for on-shell diagrams in $G(k,n)$ are more complicated and can not be covered by a set of $G_+(k,n)$. For 6pt NMHV all on-shell diagrams were classified in \cite{Bourjaily:2016mnp}, and new types of on-shell superfunctions were identified. On the other hand, the subset of on-shell diagrams which arise from the BCFW recursion relations are expandable in terms of $d\log$ forms on cells in $G_+(k,n)$ \cite{new}, this is similar to the MHV case (where it was true for any on-shell diagram). The on-shell diagrams can be defined also in gravity \cite{Heslop:2016plj,Herrmann:2016qea} and a variety of other quantum field theories.

\vspace{-.3cm}
\subsubsection*{EFT-hedron and positivity bounds on low energy EFT Wilson}

One of the revolutionary ideas in our understanding of fundamental physics has been the development of effective field theories (EFTs). By now, they play a crucial role in virtually all areas of high energy physics ranging from the interpretation of collider physics experiments to understanding the quantum nature of gravity (see e.g.~\cite{Georgi:1993mps,Cohen:2019wxr}). One of the fundamental exploits of EFT is the fact that physics at different scales decouples and at low energies we can systematically parameterize unknown physics living at some high energy scale $\Lambda$. In EFT, this is accomplished by adding higher-dimensional operators with a priori undetermined \emph{Wilson} coefficients. Naively, from a low-energy perspective, one could think that these coefficients can assume arbitrary values. However, it has been understood \cite{Adams:2006sv} that ``not everything goes'' in QFT and that unitarity and causality in the UV impose (positivity) bounds on the low-energy coefficients. Recently, this has spurred a flurry of activity on bounding different EFTs in various contexts, including gravitational theories, see e.g.~\cite{deRham:2017zjm,Arkani-Hamed:2020blm,Caron-Huot:2021rmr,Bern:2021ppb,Chiang:2021ziz}.

In the context of positive geometry, we emphasize an analytic approach to such bounds---termed the EFThedron \cite{Arkani-Hamed:2020blm}---which has been further studied in e.g.~\cite{Chiang:2021ziz}. Roughly, the EFThedron places bounds on the low-energy expansion of the $2\to2$ scattering amplitude for small Mandelstam $s,t$
\begin{equation}
    \mathcal{M}_{{\rm IR}}(s,t) = \sum_{k,q\leq k} g_{k,q}\, s^{k-q}\, t^{q}\,,
\end{equation}
where the $g_{k,q}$ are related to the usual Wilson coefficients. The crucial insight of \cite{Arkani-Hamed:2020blm,Chiang:2021ziz} is based on the fact that high-scale unitarity places restrictions on the set of $\{g_{k,q}\}$ that can be organized in terms of the EFThedron positive geometry. Roughly, this is due to the fact that the positivity of the high-energy spectral density in a partial wave expansion is equivalent to certain moment problems that in turn are related to (generalizations) of cyclic polytopes. For details, c.f.~the excellent articles~\cite{Arkani-Hamed:2020blm,Chiang:2021ziz}. 

\vspace{-.3cm}
\subsubsection*{Positive Geometry for conformal bootstrap, cosmological correlation functions, and more}

Besides the applications of positive geometries mentioned above, they found their way into the bootstrap program for CFT correlators, where crossing-symmetry places nontrivial constraints on both the OPE coefficients and the spectrum of the theory that has been exploited non-perturbatively \cite{Rattazzi:2008pe} in a very active field of research (see e.g.~\cite{Simmons-Duffin:2016gjk} for an overview). Similar to the EFThedron, there are structures that can be organized in terms of a positive geometry, for details c.f.~\cite{Arkani-Hamed:2018ign}. 

Positive geometries also appeared in the context of cosmological correlation functions \cite{Arkani-Hamed:2017fdk,Arkani-Hamed:2018bjr,Benincasa:2018ssx,Benincasa:2020aoj,Benincasa:2021qcb} that are relevant to understand e.g.~the density perturbations at the end of inflation. Since these de Sitter observables have a connection to flat-space scattering on certain kinematic poles \cite{Arkani-Hamed:2015bza}, a relation to positive geometries is at least well motivated and has been indeed worked out in the above references.

Other fascinating work \cite{Arkani-Hamed:2019mrd} revealed that the string amplitudes naturally arise from a certain $\alpha'$-deformation of the $d\log$ forms (so called \emph{stringy canonical forms}). There are also ongoing attempts to extend the associahedron construction for $\phi^3$ theory to all loops \cite{new2}, and to find the positive geometry for gravity amplitudes (putative Gravituhedron geometry)  \cite{Armstrong:2020ljm,Trnka:2020dxl}. 

Finally, the amplituhedron and ABHY constructions have been a great source of inspiration for purely mathematical treatments, solidifying these constructions as well as bringing new important insights (see e.g. \cite{Lam:2014jda,Karp:2021uap,Dian:2021idl,Even-Zohar:2021sec,Lukowski:2020dpn,Parisi:2021oql,Moerman:2021cjg,Dian:2022tpf})
%
%
\newpage
\section{Summary}
\vspace{-.2cm}

In this review, we have highlighted the recent progress on reformulating Quantum Field Theory in terms of novel geometric concepts where all properties of scattering amplitudes are the answer to some natural questions in the context of \emph{positive geometry}. We briefly recalled the most important concepts of positive geometry, leaving a comprehensive definition to Ref.~\cite{Arkani-Hamed:2017tmz}. Our discussion continued with the simple and physically relevant example of the \emph{kinematic associahedron}, related to scattering amplitudes of bi-adjoint scalar $\phi^3$ theory \cite{Arkani-Hamed:2017mur}. The geometry of the associahedron is polytopal with linear boundaries which simplifies its geometric description. As second example, we discussed the ``curvy'' geometries of Grassmannians and on-shell diagrams that ultimately led to the description of scattering amplitudes in planar $\mathcal{N}{=}4$ SYM in terms of the \emph{amplituhedron} \cite{Arkani-Hamed:2013jha}. We closed with a very brief summary of further examples of positive geometries that entered in a physical context, such as the EFThedron, cosmological polytopes, stringy canonical forms, as well as positive geometry description of CFT correlation functions. Despite all these advances, we are still at the beginning of the long quest for a new geometric formulation for the perturbative $S$-matrix.

\vspace{-.5cm}
\section*{Acknowledgments}
\vspace{-.3cm}
This work  was supported  by the European Union's Horizon 2020 research and innovation programme under the Marie Sk\l{}odowska-Curie grant agreement No.~764850 {\it ``\href{https://sagex.org}{SAGEX}''}. We acknowledge support from the U.S. Department of Energy (DOE) under Award Numbers DE-SC0009937 and DE-SC0009999 and thank Nima Arkani-Hamed for comments on the manuscript.

\vspace{.2cm}
\bibliography{refs_geometry.bib}

\end{document}